\documentclass[a4paper,twoside,12pt]{article}
\usepackage{amsmath,amssymb}
\usepackage{bbm}
\usepackage{bm}
\usepackage{verbatim}
\usepackage{stmaryrd}
\usepackage{graphicx,color}
\usepackage{mathtools}
\usepackage{titlesec}
\usepackage{titletoc}

\titleformat{name=\section,numberless}{\large\bfseries}{}{0pt}{}
\titleformat*{\section}{\large\bfseries}
\titleformat*{\subsection}{\normalsize\bfseries}

\newcommand{\mostimportant}[1]{{\footnotesize #1\par}}

\newcommand{\grd}[2]{{\textmd{\bf G}_{#1}#2}}
\newcommand{\grdp}[2]{{\textmd{\bf G}_{#1}'#2}}
\newcommand{\grdpp}[2]{{\textmd{\bf G}_{#1}''#2}}
\newcommand{\fil}[2]{{\textmd{\bf F}_{#1}#2}}
\newcommand{\filp}[2]{{\textmd{\bf F}_{#1}'#2}}
\newcommand{\filpp}[2]{{\textmd{\bf F}_{#1}''#2}}

\newcommand{\sub}[1]{{\textmd{\bf S}#1}}

\newcommand{\odd}{\textmd{odd}}
\newcommand{\even}{\textmd{even}}

\newcommand{\dcube}{cube${}_{\D{}}$}
\newcommand{\dcubes}{cubes${}_{\D{}}$}
\newcommand{\Dcubes}{Cubes${}_{\D{}}$}
\newcommand{\ecube}{cube${}_{\Ein{}}$}
\newcommand{\ecubes}{cubes${}_{\Ein{}}$}
\newcommand{\Ecubes}{Cubes${}_{\Ein{}}$}

\newcommand{\CONJ}{\textmd{\bf C}}

\newcommand{\TWO}[4]{%
\mathllap{#1}&=\mathrlap{#2}\phantom{xxxxxxxxx}&
\mathllap{#3}&=\mathrlap{#4}}

\DeclareMathOperator{\Aut}{Aut}
\DeclareMathOperator{\RE}{Re}
\DeclareMathOperator{\IM}{Im}
\DeclareMathOperator{\image}{image}
\DeclareMathOperator{\rank}{rank}

\DeclareMathOperator{\Der}{Der}

\DeclareMathOperator{\maxcube}{max}
\DeclareMathOperator{\lcm}{lcm}
\DeclareMathOperator{\Sol}{Sol}

\newcommand{\overbar}[1]{\mkern 1mu\overline{\mkern-1mu#1\mkern-1mu}\mkern 1mu}

\newcommand{\cc}[1]{\overbar{#1}}

\newcommand{\dd}{\mathrm{d}}
\newcommand{\p}{\partial}

\renewcommand{\epsilon}{USE eps INSTEAD}

\newcommand{\C}{\mathbbm{C}}

\newcommand{\R}{\mathbbm{R}}
\newcommand{\Z}{\mathbbm{Z}}

\newcommand{\cre}{\texttt{e}}
\newcommand{\ann}{\texttt{i}}

\newcommand{\IWeyl}[1]{\mathcal{I}^{#1}}

\newcommand{\gauge}{\texttt{G}}

\newcommand{\Ein}[1]{\mathcal{E}^{#1}}
\newcommand{\EinG}[1]{\mathcal{E}^{#1}_{\smash{\gauge}}}
\newcommand{\D}[1]{\mathcal{L}^{#1}}
\newcommand{\DG}[1]{\mathcal{L}^{#1}_{\smash{\gauge}}}
\newcommand{\unk}{\gamma}
\newcommand{\KERN}{{\mathcal{J}}}
\newcommand{\DC}[1]{\D{#1}_{\C}}

\newcommand{\bkl}[2]{{\textmd{\bf f}_{#1}#2}}
\newcommand{\bklG}[1]{(\textmd{\bf f}_{#1}\Ein{})_{\smash{\gauge}}}

\newcommand{\EinC}[1]{\Ein{#1}_{\C}}

\newcommand{\IWeylC}[1]{\IWeyl{#1}_{\C}}
\newcommand{\IWeylX}[1]{\IWeyl{#1}_{\ast}}
\newcommand{\qq}[1]{\bm{\xi}_{#1}}
\newcommand{\QQ}[1]{\Xi_{#1}}
\newcommand{\qqpic}{\QQ}

\newcommand{\DERIVVERT}{\DERIV_{\textmd{vertical}}}

\newcommand{\DERIV}{\mathcal{D}}

\newcommand{\db}[2]{\llbracket #1,#2\rrbracket}
\newcommand{\eb}[2]{[#1,#2]}

\newcommand{\VV}{V\cc{V}}
\newcommand{\VVreal}{{(\VV)_{\textmd{real}}}}
\newcommand{\VVpos}{{(\VV)_{\textmd{positive}}}}

\newcommand{\ff}{finite-free}

\newcommand{\sref}[1]{\S \ref{sec:#1}}
\newcommand{\aref}[1]{\S \ref{app:#1}}

\newcommand{\rC}{\mathcal{C}}
\newcommand{\rR}{\mathcal{R}}

\newcommand{\Lang}{\mathcal{A}}

\titlecontents{section}[4.2em]{\addvspace{-4pt}}
{\contentslabel{1.5em}}
{}{\titlerule*[0.3pc]{.}\contentspage\rule{3.5em}{0pt}}

\newcommand{\modelspace}{model-space}

\newcommand{\eve}{vacuum Einstein equations}

\newcommand{\ixaux}[1]{$\bm{\mathrm{(i#1)}}$}
\newcommand{\ixherm}{\ixaux{1}}
\newcommand{\ixker}{\ixaux{2}}
\newcommand{\ixpos}{\ixaux{3}}

\newcommand{\suboxaux}[1]{$\sub{\bm{\mathrm{(o#1)}}}$}
\newcommand{\suboxsp}{\suboxaux{1}}
\newcommand{\suboxds}{\suboxaux{2}}
\newcommand{\suboxff}{\suboxaux{3}}

\newcommand{\kk}[2]{$\textmd{#1}_{#2}$}

\newcommand{\BG}[1]{\texttt{B}^{#1}}
\newcommand{\bil}[1]{\texttt{b}^{#1}}
\newcommand{\auxx}[1]{\texttt{a}^{#1}}
\newcommand{\Auxx}[1]{\texttt{A}^{#1}}

\newcommand{\je}{\texttt{j}}

\renewcommand{\subset}{\subseteq}
\renewcommand{\supset}{\supseteq}

\newcommand{\km}{\texttt{k}}

\newcommand{\twofootnotes}[3]{\begin{samepage}\mbox{#1\footnotemark\textsuperscript{,}\footnotemark}\addtocounter{footnote}{-2}\stepcounter{footnote}\footnotetext{#2}\stepcounter{footnote}\footnotetext{#3}\end{samepage}}

\newcommand{\filind}{p}
\newcommand{\otherfilind}{q}

\newcommand{\dec}[1]{#1_{(0)}}
\newcommand{\unkp}{\dec{\unk}}
\newcommand{\BKL}{one BKL-bounce}

\newcommand{\half}{\frac{1}{2}}
\newcommand{\thalf}{\frac{3}{2}}

\newcommand{\fundi}{\texttt{R}}
\newcommand{\pundi}{\texttt{P}}
\newcommand{\pundiaux}{\texttt{P}}
\newcommand{\pundievenall}{\pundiaux_{\even'}}
\newcommand{\pundialleven}{\pundiaux_{\even''}}
\newcommand{\pundioddall}{\pundiaux_{\odd'}}
\newcommand{\pundiallodd}{\pundiaux_{\odd''}}
\newcommand{\pundieveneven}{\pundiaux_{\even'\even''}}
\newcommand{\pundievenodd}{\pundiaux_{\even'\odd''}}
\newcommand{\pundioddeven}{\pundiaux_{\odd'\even''}}
\newcommand{\pundioddodd}{\pundiaux_{\odd'\odd''}}

\newcommand{\jj}[1]{\bm{\theta}_{#1}}
\newcommand{\JJ}[1]{\Theta_{#1}}
\newcommand{\jjpic}{\JJ}

\newcommand{\KERNcomp}{\cc{V}\grd{\textmd{any},0}{\DERIV}}

\newcommand{\DEF}{\stackrel{\text{def}}{=}}

\newcommand{\tju}{\bm{\zeta}}

\newcommand{\REES}{\mathcal{P}}
\newcommand{\GO}{\REES/s\REES}

\newcommand{\ax}{\xrightarrow{\,\dd_0\,}}
\newcommand{\sx}[1]{\REES^{#1}/s\REES^{#1}}
\newcommand{\ay}{\xrightarrow{\,\dec{\dd}\,}}
\newcommand{\sy}[1]{\fil{\filind}{\Ein{#1}}/\fil{\filind-1}{\Ein{#1}}}

\newcommand{\wwa}{$\mathclap{\subset}$}
\newcommand{\wwb}[1]{$\mathclap{\bkl{#1}{\Ein{}}}$}
\newcommand{\wwcdots}{$\cdots$}

\newcommand{\Dp}{\mathcal{L}'}
\newcommand{\Dpp}{\mathcal{L}''}
\newcommand{\Ep}{\mathcal{E}'}
\newcommand{\Epp}{\mathcal{E}''}

\newcommand{\reg}{\mathcal{X}}
\newcommand{\regperp}{\reg^{\perp}}
\newcommand{\pro}{\texttt{p}}

\newcommand{\fock}[2]{\langle #1 \rangle_{#2}}

\newcommand{\svac}{\sigma}

\newcommand{\gvac}[1]{\Omega_{#1}}

\newcommand{\mfd}{4-dim manifold}

\newcommand{\nextpartxx}[2]{%
\addtocontents{toc}{~\emph{#1}\par}
\section*{#2}}
\newcommand{\nextpart}[1]{\nextpartxx{#1}{#1}}

\newcommand{\gla}{\footnote{%
A real graded Lie algebra (see Nijenhuis, Richardson \cite{NR})
is a $\Z$-graded real vector space
with an $\R$-bilinear bracket 
$\eb{\,\cdot\,}{\,\cdot\,}: \Ein{k}\times \Ein{\ell} \to \Ein{k+\ell}$
such that
$\eb{x}{y} = -(-1)^{|x||y|} \eb{y}{x}$
and
$
(-1)^{|x||z|}
\eb{x}{\eb{y}{z}}
+
(-1)^{|y||x|}
\eb{y}{\eb{z}{x}}
+
(-1)^{|z||y|}
\eb{z}{\eb{x}{y}} = 0$
if $x \in \Ein{|x|}$,
$y\in\Ein{|y|}$,
$z\in\Ein{|z|}$.
}}

\newcommand{\differentialfootnote}{\footnote{%
In a graded Lie algebra,
$\eb{x}{\eb{x}{\,\cdot\,}} = \tfrac{1}{2}\eb{\eb{x}{x}}{\,\cdot\,}$
for all elements $x$ of degree one.
}}

\newcommand{\Gv}{v_-}
\newcommand{\Gw}{v_+}
\newcommand{\Gcv}{\cc{\Gv}}
\newcommand{\Gcw}{\cc{\Gw}}

\newcommand{\HALF}{\tfrac{1}{2}}
\newcommand{\imagunit}{i}
\newcommand{\squarethis}[1]{(#1)^2}

\newcommand{\aunkn}[1]{a_{#1}}
\newcommand{\bunkn}[1]{b_{#1}}
\newcommand{\cunkn}[1]{c_{#1}}
\newcommand{\dunkn}[1]{d_{}}
\newcommand{\eunkn}[1]{e_{#1}}
\newcommand{\funkn}[1]{f_{}}
\newcommand{\gunkn}[1]{g_{}}
\newcommand{\hunkn}[1]{h_{}}
\newcommand{\xunkn}[1]{x_{#1}}

\newcommand{\xiaux}[1]{\bm{\xi}_{#1}}
\newcommand{\axider}[1]{\ifx&#1&\xiaux{a}\else\xiaux{a}(#1)\fi}
\newcommand{\bxider}[1]{\ifx&#1&\xiaux{b}\else\xiaux{b}(#1)\fi}
\newcommand{\cxider}[1]{\ifx&#1&\xiaux{c}\else\xiaux{c}(#1)\fi}
\newcommand{\exider}[1]{\ifx&#1&\xiaux{e}\else\xiaux{e}(#1)\fi}
\newcommand{\xxider}[1]{\ifx&#1&\xiaux{}\else\xiaux{}(#1)\fi}

\newcommand{\commuttor}[2]{[#1,#2]}

\newcommand{\auxTHISdata}[4]{
\begin{align*}
#1|_{\rR} & \;=\; #2\rule{100mm}{0pt}\\
\rule{0pt}{13pt}
#1(\Gv) & \;=\; #3\\
\rule{0pt}{13pt}
#1(\Gw) & \;=\; \mathrlap{\text{(defined such that $\fundi''#1 = #4#1$)}}
\end{align*}
}

\newcommand{\auxA}{\unk^\texttt{0}}
\newcommand{\auxB}{\unk^\texttt{I}}
\newcommand{\auxD}{\unk^\texttt{II}}
\newcommand{\auxG}{\unk^\texttt{III}}

\newcommand{\auxAdata}{\auxTHISdata{
\auxA
}{
\mathrlap{(\Gw\Gcw+\Gv\Gcv){\xxider{}}}
}{
\mathrlap{\HALF(-\HALF \xunkn{2} - \HALF \xunkn{3}+\imagunit \xunkn{4}-\xunkn{5})\Gw\Gcw\Gv}\\
&\mathrlap{\qquad
+\HALF(-\xunkn{1}-\HALF \xunkn{2}-\HALF \xunkn{3}+\imagunit \xunkn{4}-\xunkn{5})\Gv\Gcv\Gv}\\
&\mathrlap{\qquad
+\HALF(-\HALF \xunkn{2}-\HALF \xunkn{3})\Gv\Gcw\Gw
+\HALF(\HALF \xunkn{2}-\HALF \xunkn{3})\Gw\Gcv\Gw}
}{
}}

\newcommand{\NAIVELEADINGTERM}{\auxTHISdata{
\unkp
}{
\mathrlap{(\Gw\Gcw+\Gv\Gcv){\axider{}}}
}{
\mathrlap{\HALF(-\HALF \aunkn{2} - \HALF \aunkn{3}+\imagunit \aunkn{4}-\aunkn{5})\Gw\Gcw\Gv}\\
&\mathrlap{\qquad
+\HALF(-\aunkn{1}-\HALF \aunkn{2}-\HALF \aunkn{3}+\imagunit \aunkn{4}-\aunkn{5} + \imagunit \aunkn{6})\Gv\Gcv\Gv}\\
&\mathrlap{\qquad
+\HALF(-\HALF \aunkn{2}-\HALF \aunkn{3})\Gv\Gcw\Gw
+\HALF(\HALF \aunkn{2}-\HALF \aunkn{3} + \imagunit \aunkn{6})\Gw\Gcv\Gw}
}{
}}

\newcommand{\auxDdata}{\auxTHISdata{
\auxD
}{
\mathrlap{0}
}{
\mathrlap{\HALF\imagunit\xunkn{} \Gw\Gcw\Gv
+\HALF\imagunit\xunkn{} \Gv\Gcw\Gw}
}{
}}

\newcommand{\auxBdata}{\auxTHISdata{
\auxB
}{
\mathrlap{(\Gw\Gcw-\Gv\Gcv){\xxider{}}}
}{
\mathrlap{\HALF(-\imagunit\xunkn{1}-\xunkn{6}+\imagunit\xunkn{7})\Gw\Gcw\Gv}\\
&\mathrlap{\qquad
+(\xunkn{4}+\imagunit\xunkn{5})\Gw\Gcv\Gw}\\
&\mathrlap{\qquad
+\HALF(\xunkn{2}-\xunkn{3})\Gv\Gcw\Gw}\\
&\mathrlap{\qquad
+\HALF(\imagunit\xunkn{1}-\xunkn{2}-\xunkn{3}-\xunkn{6}+\imagunit\xunkn{7})\Gv\Gcv\Gv}
}{
- 
}}

\newcommand{\auxGdata}{\auxTHISdata{
\auxG
}{
\mathrlap{0}
}{
\mathrlap{\imagunit\xunkn{}\Gv\Gcw\Gw}
}{
- 
}}

\newcommand{\BLSTR}{\rule{0pt}{14pt}}
\newcommand{\auxGOparametrization}{%
\begin{align*}
\unk_0\;=\;\;&
(\auxA_{\axider{},\aunkn{1},\ldots,\aunkn{5}} + \Exc'\auxD_{\aunkn{6}})\\
\BLSTR&+ s' \Exc\auxB_{\bxider{},\bunkn{1},\ldots,\bunkn{7}}
+ s'' \auxB_{\cxider{},\cunkn{1},\ldots,\cunkn{7}}\\
\BLSTR&+ (s')^2 \auxD_{\dunkn{}}
+ s's'' (\Exc'\auxB_{\exider{},\eunkn{1},\ldots,\eunkn{7}}
         + \Exc'\auxG_{\eunkn{8}})
+ (s'')^2 \Exc\auxD_{\funkn{}}\\
\BLSTR&+ (s')^2s'' \auxG_{\gunkn{}}
+ s'(s'')^2 \Exc\auxG_{\hunkn{}}
\end{align*}%
}

\newcommand{\twocomp}[4]{\rule{#1}{0pt}\mathllap{#3}\,\mathrlap{#4}\rule{#2}{0pt}}
\newcommand{\sixcomp}[6]{\twocomp{60pt}{40pt}{#1}{#2}
                         \twocomp{60pt}{40pt}{#3}{#4}
                         \twocomp{60pt}{40pt}{#5}{#6}}

\newcommand{\auxGOparametrizationFAKE}{%
\begin{align*}
\unk_0\;=\;\;&
\auxA_{\axider{},\aunkn{1},\ldots,\aunkn{5}}\\
\BLSTR&
\sixcomp{
+ s''s'''}{\auxB_{\cxider{},\cunkn{1},\ldots,\cunkn{7}}}{
+ s'''s'}{\Exc\auxB_{\bxider{},\bunkn{1},\ldots,\bunkn{7}}}{
+ s's''}{\Exc'\auxB_{\exider{},\eunkn{1},\ldots,\eunkn{7}}} \\
\BLSTR&
\sixcomp{+ (s')^2}{\auxD_{\dunkn{}}}{
+ (s'')^2}{\Exc\auxD_{\funkn{}}}{
+ (s''')^2}{\Exc'\auxD_{\aunkn{6}}} \\
\BLSTR&
\sixcomp{+ s''s'''(s')^2}{\auxG_{\gunkn{}}}{
+ s'''s'(s'')^2}{\Exc\auxG_{\hunkn{}}}{
+ s's''(s''')^2}{\Exc'\auxG_{\eunkn{8}}}
\end{align*}%
}

\newcommand{\listeqs}[1]{\begin{itemize}#1\end{itemize}}
\newcommand{\iitem}[1]{\item[$0=$] $#1$}

\newcommand{\aeqsnA}{\listeqs{%
\iitem{-\axider{\aunkn{1}}+\squarethis{\aunkn{6}}+\aunkn{5}\aunkn{1}}
\iitem{-\axider{\aunkn{2}}-\squarethis{\aunkn{6}}+\aunkn{5}\aunkn{2}}
\iitem{-\axider{\aunkn{3}}+\squarethis{\aunkn{6}}+\aunkn{5}\aunkn{3}}
\iitem{-\axider{\aunkn{6}}+\aunkn{2}\aunkn{6}+\aunkn{5}\aunkn{6}}
}}

\newcommand{\aeqsnB}{\listeqs{%
\iitem{\aunkn{2}\aunkn{3}
+\aunkn{3}\aunkn{1}
+\aunkn{1}\aunkn{2}
-\squarethis{\aunkn{6}}}
}}

\newcommand{\beqsnA}{\listeqs{%
\iitem{2\commuttor{\axider{}}{\bxider{}}+2\bunkn{6}\axider{}-\aunkn{1}\bxider{}-\aunkn{2}\bxider{}-2\aunkn{5}\bxider{}}
\iitem{-2\axider{\bunkn{1}}+\aunkn{1}\bunkn{1}+\aunkn{2}\bunkn{1}+2\aunkn{5}\bunkn{1}+2\aunkn{4}\bunkn{6}+\aunkn{6}\bunkn{6}-2\bxider{\aunkn{4}}-\bxider{\aunkn{6}}}
\iitem{-4\axider{\bunkn{2}}+2\aunkn{1}\bunkn{2}+2\aunkn{2}\bunkn{2}+4\aunkn{5}\bunkn{2}+3\aunkn{1}\bunkn{6}+3\aunkn{2}\bunkn{6}+2\aunkn{3}\bunkn{6}+4\aunkn{5}\bunkn{6}-3\bxider{\aunkn{1}}-3\bxider{\aunkn{2}}-2\bxider{\aunkn{3}}-4\bxider{\aunkn{5}}}
\iitem{4\axider{\bunkn{3}}-2\aunkn{1}\bunkn{3}-2\aunkn{2}\bunkn{3}-4\aunkn{5}\bunkn{3}-\aunkn{1}\bunkn{6}-\aunkn{2}\bunkn{6}-2\aunkn{3}\bunkn{6}-4\aunkn{5}\bunkn{6}+\bxider{\aunkn{1}}+\bxider{\aunkn{2}}+2\bxider{\aunkn{3}}+4\bxider{\aunkn{5}}}
\iitem{-4\axider{\bunkn{4}}+2\aunkn{1}\bunkn{4}+2\aunkn{2}\bunkn{4}+4\aunkn{5}\bunkn{4}-4\aunkn{6}\bunkn{5}+\aunkn{1}\bunkn{6}-\aunkn{2}\bunkn{6}-4\aunkn{6}\bunkn{7}-\bxider{\aunkn{1}}+\bxider{\aunkn{2}}}
\iitem{2\axider{\bunkn{5}}+\aunkn{6}\bunkn{2}-\aunkn{6}\bunkn{3}-2\aunkn{6}\bunkn{4}-2\aunkn{5}\bunkn{5}+\aunkn{6}\bunkn{6}+\aunkn{1}\bunkn{7}-\aunkn{2}\bunkn{7}-\bxider{\aunkn{6}}}
}}

\newcommand{\beqsnB}{\listeqs{%
\iitem{\aunkn{3}\bunkn{2}+\aunkn{1}\bunkn{3}+\aunkn{2}\bunkn{3}-\aunkn{3}\bunkn{3}-\aunkn{1}\bunkn{4}+\aunkn{2}\bunkn{4}+2\aunkn{6}\bunkn{5}+\bxider{\aunkn{1}}+\bxider{\aunkn{2}}}}}

\newcommand{\deqsnA}{\listeqs{%
\iitem{\axider{\dunkn{1}}-\aunkn{1}\dunkn{1}-\aunkn{5}\dunkn{1}+\bxider{\bunkn{5}}+\bxider{\bunkn{7}}+2\bunkn{4}\bunkn{5}-\bunkn{5}\bunkn{6}+2\bunkn{4}\bunkn{7}-\bunkn{6}\bunkn{7}}}}

\newcommand{\deqsnB}{}

\newcommand{\eeqsnA}{\listeqs{%
\iitem{2\commuttor{\axider{}}{\exider{}}+2\eunkn{6}\axider{}-\aunkn{1}\exider{}-\aunkn{3}\exider{}-2\aunkn{5}\exider{}+2\bunkn{5}\cxider{}+2\bunkn{7}\cxider{}-2\cunkn{5}\bxider{}-2\cunkn{7}\bxider{}}
\iitem{-2\axider{\eunkn{1}}+\aunkn{1}\eunkn{1}+\aunkn{3}\eunkn{1}+2\aunkn{5}\eunkn{1}+2\aunkn{4}\eunkn{6}+\aunkn{6}\eunkn{6}+2\cunkn{5}\bunkn{1}+2\cunkn{7}\bunkn{1}-2\cunkn{1}\bunkn{5}-2\cunkn{1}\bunkn{7}-2\exider{\aunkn{4}}-\exider{\aunkn{6}}}
\iitem{-4\axider{\eunkn{2}}+2\aunkn{1}\eunkn{2}+2\aunkn{3}\eunkn{2}+4\aunkn{5}\eunkn{2}+3\aunkn{1}\eunkn{6}+2\aunkn{2}\eunkn{6}+3\aunkn{3}\eunkn{6}+4\aunkn{5}\eunkn{6}-2\bxider{\cunkn{5}}-2\bxider{\cunkn{7}}+4\cunkn{5}\bunkn{2}+4\cunkn{7}\bunkn{2}-4\cunkn{2}\bunkn{5}-2\cunkn{6}\bunkn{5}+2\cunkn{5}\bunkn{6}+2\cunkn{7}\bunkn{6}-4\cunkn{2}\bunkn{7}-2\cunkn{6}\bunkn{7}+2\cxider{\bunkn{5}}+2\cxider{\bunkn{7}}-3\exider{\aunkn{1}}-2\exider{\aunkn{2}}-3\exider{\aunkn{3}}-4\exider{\aunkn{5}}}
\iitem{-4\axider{\eunkn{3}}+2\aunkn{1}\eunkn{3}+2\aunkn{3}\eunkn{3}+4\aunkn{5}\eunkn{3}+\aunkn{1}\eunkn{6}+2\aunkn{2}\eunkn{6}+\aunkn{3}\eunkn{6}+4\aunkn{5}\eunkn{6}+2\bxider{\cunkn{5}}+2\bxider{\cunkn{7}}+4\cunkn{5}\bunkn{3}+4\cunkn{7}\bunkn{3}-4\cunkn{3}\bunkn{5}+2\cunkn{6}\bunkn{5}-2\cunkn{5}\bunkn{6}-2\cunkn{7}\bunkn{6}-4\cunkn{3}\bunkn{7}+2\cunkn{6}\bunkn{7}-2\cxider{\bunkn{5}}-2\cxider{\bunkn{7}}-\exider{\aunkn{1}}-2\exider{\aunkn{2}}-\exider{\aunkn{3}}-4\exider{\aunkn{5}}}
\iitem{4\axider{\eunkn{4}}-2\aunkn{1}\eunkn{4}-2\aunkn{3}\eunkn{4}-4\aunkn{5}\eunkn{4}+\aunkn{1}\eunkn{6}-\aunkn{3}\eunkn{6}+2\bxider{\cunkn{5}}+2\bxider{\cunkn{7}}+4\cunkn{5}\bunkn{4}+4\cunkn{7}\bunkn{4}+4\cunkn{4}\bunkn{5}-2\cunkn{6}\bunkn{5}-2\cunkn{5}\bunkn{6}-2\cunkn{7}\bunkn{6}+4\cunkn{4}\bunkn{7}-2\cunkn{6}\bunkn{7}+2\cxider{\bunkn{5}}+2\cxider{\bunkn{7}}-\exider{\aunkn{1}}+\exider{\aunkn{3}}}
\iitem{2\axider{\eunkn{5}}+\aunkn{6}\eunkn{2}-3\aunkn{6}\eunkn{3}+2\aunkn{6}\eunkn{4}-2\aunkn{5}\eunkn{5}-\aunkn{1}\eunkn{7}+\aunkn{3}\eunkn{7}+2\aunkn{2}\eunkn{8}+2\bxider{\cunkn{2}}-\bxider{\cunkn{3}}-3\bxider{\cunkn{4}}+\bxider{\cunkn{6}}+\cunkn{2}\bunkn{2}-\cunkn{3}\bunkn{2}+2\cunkn{4}\bunkn{2}+\cunkn{2}\bunkn{3}-\cunkn{3}\bunkn{3}-4\cunkn{4}\bunkn{3}-\cunkn{6}\bunkn{3}+2\cunkn{3}\bunkn{4}-4\cunkn{4}\bunkn{4}+\cunkn{6}\bunkn{4}+2\cunkn{7}\bunkn{5}-\cunkn{3}\bunkn{6}+\cunkn{4}\bunkn{6}-2\cunkn{6}\bunkn{6}+2\cunkn{5}\bunkn{7}+\cxider{\bunkn{3}}+\cxider{\bunkn{4}}+\cxider{\bunkn{6}}-2\exider{\aunkn{6}}}
\iitem{-2\axider{\eunkn{8}}-\aunkn{6}\eunkn{2}+\aunkn{6}\eunkn{3}-\aunkn{6}\eunkn{6}+2\aunkn{5}\eunkn{8}-\bxider{\cunkn{2}}+\bxider{\cunkn{4}}-\bxider{\cunkn{6}}+\cunkn{3}\bunkn{2}-\cunkn{4}\bunkn{2}-\cunkn{2}\bunkn{3}+\cunkn{4}\bunkn{3}-\cunkn{6}\bunkn{3}+\cunkn{2}\bunkn{4}-\cunkn{3}\bunkn{4}+\cunkn{6}\bunkn{4}+\cunkn{3}\bunkn{6}-\cunkn{4}\bunkn{6}+\cxider{\bunkn{2}}-\cxider{\bunkn{4}}+\cxider{\bunkn{6}}+\exider{\aunkn{6}}}}}

\newcommand{\eeqsnB}{\listeqs{%
\iitem{-2\eunkn{8}\axider{}+\aunkn{6}\exider{}+\commuttor{\bxider{}}{\cxider{}}+\bunkn{3}\cxider{}-\bunkn{4}\cxider{}-\cunkn{3}\bxider{}+\cunkn{4}\bxider{}}
\iitem{-\aunkn{6}\eunkn{1}-2\aunkn{4}\eunkn{8}-\aunkn{6}\eunkn{8}-\bxider{\cunkn{1}}+\cunkn{3}\bunkn{1}-\cunkn{4}\bunkn{1}-\cunkn{1}\bunkn{3}+\cunkn{1}\bunkn{4}+\cxider{\bunkn{1}}}
\iitem{-\aunkn{6}\eunkn{2}-\aunkn{6}\eunkn{3}-2\aunkn{1}\eunkn{8}-2\aunkn{2}\eunkn{8}-2\aunkn{3}\eunkn{8}-4\aunkn{5}\eunkn{8}-\bxider{\cunkn{2}}-\bxider{\cunkn{3}}+\cunkn{3}\bunkn{2}-\cunkn{4}\bunkn{2}-\cunkn{2}\bunkn{3}-\cunkn{4}\bunkn{3}+\cunkn{2}\bunkn{4}+\cunkn{3}\bunkn{4}+\cxider{\bunkn{2}}+\cxider{\bunkn{3}}}
\iitem{-\aunkn{2}\eunkn{2}-\aunkn{1}\eunkn{3}+\aunkn{2}\eunkn{3}-\aunkn{3}\eunkn{3}-\aunkn{1}\eunkn{4}+\aunkn{3}\eunkn{4}+2\aunkn{6}\eunkn{8}-2\bxider{\cunkn{5}}+2\cunkn{5}\bunkn{2}-4\cunkn{5}\bunkn{3}-2\cunkn{5}\bunkn{4}-2\cunkn{2}\bunkn{5}+4\cunkn{3}\bunkn{5}+2\cunkn{4}\bunkn{5}+2\cxider{\bunkn{5}}-\exider{\aunkn{1}}-\exider{\aunkn{3}}}
\iitem{-\aunkn{6}\eunkn{2}+3\aunkn{6}\eunkn{3}-2\aunkn{2}\eunkn{8}-\bxider{\cunkn{2}}+\bxider{\cunkn{3}}+2\bxider{\cunkn{4}}+\cunkn{3}\bunkn{2}-\cunkn{4}\bunkn{2}-\cunkn{2}\bunkn{3}+3\cunkn{4}\bunkn{3}+\cunkn{2}\bunkn{4}-3\cunkn{3}\bunkn{4}+\cxider{\bunkn{2}}-\cxider{\bunkn{3}}-2\cxider{\bunkn{4}}+2\exider{\aunkn{6}}}}}

\newcommand{\geqsnA}{\listeqs{%
\iitem{-4\axider{\gunkn{1}}+2\aunkn{1}\gunkn{1}+4\aunkn{5}\gunkn{1}+\bxider{\eunkn{2}}+\bxider{\eunkn{3}}+2\bxider{\eunkn{6}}+\bunkn{3}\eunkn{2}+\bunkn{4}\eunkn{2}-\bunkn{2}\eunkn{3}+\bunkn{4}\eunkn{3}-2\bunkn{6}\eunkn{3}+\bunkn{2}\eunkn{4}+\bunkn{3}\eunkn{4}+2\bunkn{6}\eunkn{4}+2\bunkn{3}\eunkn{6}+2\bunkn{4}\eunkn{6}+4\bunkn{5}\eunkn{8}+4\bunkn{7}\eunkn{8}-\cunkn{2}\dunkn{1}-\cunkn{3}\dunkn{1}-2\cunkn{6}\dunkn{1}-\exider{\bunkn{2}}-\exider{\bunkn{3}}-2\exider{\bunkn{6}}}}}

\newcommand{\geqsnB}{\listeqs{%
\iitem{-2\aunkn{4}\gunkn{1}-\aunkn{6}\gunkn{1}+\bxider{\eunkn{1}}+\bunkn{3}\eunkn{1}+\bunkn{4}\eunkn{1}-\bunkn{1}\eunkn{3}+\bunkn{1}\eunkn{4}-\cunkn{1}\dunkn{1}-\exider{\bunkn{1}}}
\iitem{-2\aunkn{1}\gunkn{1}-2\aunkn{2}\gunkn{1}-2\aunkn{3}\gunkn{1}-4\aunkn{5}\gunkn{1}+\bxider{\eunkn{2}}+\bxider{\eunkn{3}}+\bunkn{3}\eunkn{2}+\bunkn{4}\eunkn{2}-\bunkn{2}\eunkn{3}+\bunkn{4}\eunkn{3}+\bunkn{2}\eunkn{4}+\bunkn{3}\eunkn{4}-\cunkn{2}\dunkn{1}-\cunkn{3}\dunkn{1}-\exider{\bunkn{2}}-\exider{\bunkn{3}}}
\iitem{-2\aunkn{1}\gunkn{1}+\bxider{\eunkn{2}}-\bxider{\eunkn{3}}-2\bxider{\eunkn{4}}+\bunkn{3}\eunkn{2}+\bunkn{4}\eunkn{2}-\bunkn{2}\eunkn{3}-3\bunkn{4}\eunkn{3}+\bunkn{2}\eunkn{4}-3\bunkn{3}\eunkn{4}-4\bunkn{5}\eunkn{8}-4\bunkn{7}\eunkn{8}+2\cxider{\dunkn{1}}-\cunkn{2}\dunkn{1}+3\cunkn{3}\dunkn{1}-\exider{\bunkn{2}}+\exider{\bunkn{3}}-2\exider{\bunkn{4}}}
\iitem{-2\gunkn{1}\axider{}+\commuttor{\exider{}}{\bxider{}}+\eunkn{3}\bxider{}-\eunkn{4}\bxider{}-\bunkn{3}\exider{}-\bunkn{4}\exider{}+\dunkn{1}\cxider{}}
}}

\newcommand{\ieqsnA}{}

\newcommand{\ieqsnB}{\listeqs{%
\iitem{\bxider{\hunkn{1}}+2\bunkn{3}\hunkn{1}+\bunkn{6}\hunkn{1}+\cxider{\gunkn{1}}+2\cunkn{3}\gunkn{1}+\cunkn{6}\gunkn{1}+\exider{\eunkn{8}}+2\eunkn{3}\eunkn{8}+\eunkn{6}\eunkn{8}}}}

\begin{document}

\noindent{\bf\Large
Filtered expansions in general relativity\\
\rule{0pt}{18pt}%
and \BKL
}\\
\noindent\rule{0pt}{30pt}{\bf
Michael Reiterer\footnote{School of Mathematics, IAS Princeton, NJ, USA},
Eugene Trubowitz\footnote{Department of Mathematics, ETH Zurich, Switzerland}}
\vskip 10mm
\noindent {\bf Abstract:}
When the \eve~are formulated in terms of a frame,
rather than a metric,
can one perturb solutions with a degenerate frame into ones
with a nondegenerate frame?
In examples
we point out that one can encounter issues
already at the level of formal perturbative expansions;
namely
the cohomological, so-called space of obstructions is nonzero.
In this paper we propose a
perturbative expansion based on filtrations.
We construct and prove properties
of a specific filtration, intended 
to make mathematical sense of \BKL,
a building block of a well-known but very heuristic conjecture due to
Belinskii, Khalatnikov and Lifshitz
(which would involve sticking together an infinite sequence of single bounces).
It seems possible that now the space of obstructions is zero,
but this question is left open.
\vskip 8mm

\setcounter{tocdepth}{1}
\tableofcontents



\newcommand{\degnon}{degenerate-to-nondegenerate}
\newcommand{\degnontech}{\degnon~perturbation}
\newcommand{\degnontechs}{\degnon~perturbations}

\newcommand{\comeswith}{comes with}
\section{Introduction}\label{sec:intro}

The \eve~of general relativity
are traditionally formulated in terms of a metric,
but they can also be formulated in terms of a frame.
It is then tempting to
start from a solution with a degenerate frame
and to try to perturb it into one with a nondegenerate frame.
The rationale is that degenerate
objects tend to be
simpler than nondegenerate ones.
We refer to this kind of perturbation as a \degnontech.

What is a degenerate frame?
Suppose we are on a manifold diffeomorphic to $\R^4$.
Then a frame is,
over each point of the manifold,
an invertible linear map
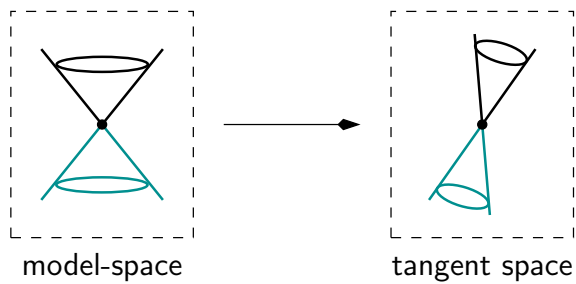
\begin{center}
\input{nocollapse.pstex_t}
\end{center}%

where the fixed \modelspace\footnote{%
Model-space: fiber of a real vector bundle of rank 4,
over the 4-dim manifold.}
has a fixed inner product
or conformal inner product of signature
${-}{+}{+}{+}$.
We would more specifically call this a nondegenerate frame.
By contrast,
a degenerate frame is a linear map that is not invertible:
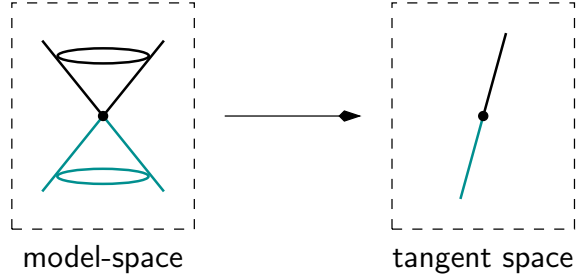
\begin{center}
\input{collapse.pstex_t}
\end{center}%

A solution with nondegenerate frame
yields a metric that solves the
\eve~(i.e.~a metric with vanishing Ricci curvature),
whereas one with degenerate frame does not
have an associated metric.

We do not actually carry out a \degnontech~in this paper.
Below we indicate some of the issues that one encounters.
We then 
make a proposal based on filtrations,
the main topic of this paper.

\subsection{Language}
At first it is enough to know just the bare
algebraic structure of the formalism
that we use;
more details are introduced as needed.
Let $\Ein{} = \bigoplus_{k \in \Z} \Ein{k}$ be a real graded Lie algebra\gla.
A natural object in a graded Lie algebra is
\[
\Sol(\Ein{})
\;\;=\;\;
\{
\unk \in \Ein{1} \mid \eb{\unk}{\unk} = 0
\}
\]
Conceptually the quotient
$\Sol(\Ein{})/\text{(automorphisms generated by $\Ein{0}$)}$
is deemed the more basic object,
but it is ill-defined at this general level
since it presumes that we can exponentiate elements
of the Lie algebra $\Ein{0}$.

General relativity is of this form:
one can choose $\Ein{}$ so that
$\Sol(\Ein{})$ is the set of solutions to the
\eve~\cite{RT}.
Here $\Ein{} = \Ein{0}\oplus \ldots \oplus \Ein{4}$.
Linearly associated to the unknown $\unk \in \Ein{1}$ is a frame\footnote{%
Associated to $\unk$ is also an affine connection,
and essentially the frame and the connection determine $\unk$.
However the map $\unk \mapsto \text{connection}$ is nonlinear,
and is not defined (is singular) when the frame is degenerate.
It would therefore be misleading and wrong, especially in this paper,
to say that $\unk$ consists of a frame and an affine connection.
}.
Essentially, $\Ein{0}$
generates
self-diffeomorphisms
of the \mfd~and rotations of \modelspace.

\subsection{Issues with naive \degnontechs}\label{sec:dkhflho4hire}

\newcommand{\toyex}{%
\footnote{%
Consider the toy example\begin{samepage}
\[
\begin{pmatrix}
\phantom{\varepsilon}\p_0 + \phantom{i}\varepsilon \p_3 & \varepsilon \p_1 + i\varepsilon \p_2\\
\varepsilon \p_1 - i\varepsilon \p_2 & \phantom{\varepsilon}\p_0 - \phantom{i}\varepsilon \p_3
\end{pmatrix}
\]
where\end{samepage} $\p_0,\p_1,\p_2,\p_3$ are partial derivatives,
$\varepsilon \in \R$ and $i = \sqrt{-1}$.
If $\varepsilon \neq 0$
then the matrix entries
make a (complex) nondegenerate frame on $\R^4$,
and the matrix is symmetric hyperbolic.
If $\varepsilon = 0$ then the frame
is degenerate, but the matrix is symmetric hyperbolic nonetheless.
Beware that
this is a poor toy example for a gauge theory
like general relativity.
}}

Going back to \degnontechs,
the most naive proposal is to
perturb the zero solution $0 \in \Sol(\Ein{})$.
The associated frame vanishes,
that is, the frame is fully degenerate.
While
perturbations certainly exist\footnote{%
Take the ray through any nontrivial
element of $\Sol(\Ein{})$, say Minkowski spacetime.},
they cannot obviously be obtained from perturbation theory about zero.
The problem is that
the derivative of $\Ein{1}\to \Ein{2},
\unk \mapsto \eb{\unk}{\unk}$ vanishes at zero.

A more serious proposal is to perturb a $\unkp \in \Sol(\Ein{})$
whose frame
satisfies,
at each point of the \mfd,
 the equivalent conditions:
\begin{itemize}
\item 
The image of the past cone is
disjoint from the image
of the future cone.
\item The kernel
is a spacelike
subspace of \modelspace.
\end{itemize}
These are natural conditions to impose
and one could guess that they suffice to do
perturbation theory\toyex~about $\unkp$,
but this is not obviously the case.

The derivative of $\Ein{1}\to \Ein{2},
\unk \mapsto \eb{\unk}{\unk}$ at $\unkp$
is two times the $k=1$ instance of the map
$\dec{\dd} = \eb{\unkp}{\,\cdot\,}: \Ein{k}\to \Ein{k+1}$,
but it is good to define this map for all $k$ right away.
The graded Lie algebra axioms imply\differentialfootnote~that
this is a differential,
$(\dec{\dd})^2 = 0$. Its cohomologies
play an important role in perturbation theory,
see for example Gerstenhaber \cite{Gerstenhaber}.
In particular, the 2nd cohomology
\[
\ker(\dec{\dd}|_{\Ein{2}})\,\big/\,\image(\dec{\dd}|_{\Ein{1}})
\]
is called the space of obstructions.
If this 2nd cohomology is zero, then
every finite-order perturbation extends to one of
all orders, i.e.~to a formal power series perturbation,
by a standard argument.

In this paper we do not consider all such $\unkp$,
only a certain class.
For this class,
the 2nd cohomology of $\dec{\dd}$ is nonzero as shown in \sref{obstrexample}.
Perturbations may well exist,
but  they cannot obviously
be obtained from perturbation theory about $\unkp$,
much like for the zero solution $0 \in \Sol(\Ein{})$.

We emphasize that these are issues encountered at the level of
formal power series
perturbations, before even broaching the issue of convergence.


\subsection{Goal of this paper and open questions}\label{sec:goalopen}

In this paper we construct a filtration of $\Ein{}$
that can be used to set up a filtered expansion,
a refinement of the above $\unkp$-proposal.
We call $\unkp$ the naive leading term.
It is only a piece of an
object $\unk_0 = \unkp \oplus \ldots$ that we call without reservation
the leading term;
the direct sum is defined later in this introduction.
Whereas $\unkp$ has a degenerate frame,
$\unk_0$ already describes the infinitesimal opening-up to
a nondegenerate frame.

This $\unk_0$ is an element not of $\Ein{}$,
but of another graded Lie algebra
that has the same `size' as $\Ein{}$, and that is
 defined using the filtration.

This $\unk_0$ defines a differential $\dd_0$
whose 2nd cohomology is the new space of obstructions.
We do not prove that this space is zero,
but it seems possible that it is.
The 1st cohomology
is the space of non-equivalent solutions to the linearized equations.
It would be interesting to calculate these cohomologies.
If things pan out, one
would have a concrete mathematical object
for the heuristic idea of \BKL,
at the preliminary level of
formal power series perturbations\footnote{%
It would be very interesting to look systematically
for other interesting filtrations of $\Ein{}$.}.

By \BKL~we mean the basic building block of
an inspiring but very heuristic conjecture
due to Belinskii, Khalatnikov, Lifshitz \cite{LK,BKL},
which would involve sticking together (cf.~\sref{fakepar})
an infinite sequence of bounces.
Here we do not even attempt to summarize this conjecture in words;
we have used the idea of the \degnontech~as
the pedagogical motivation for this paper,
because it can be read and understood on its own.

Technically, what we do in this paper is to construct a filtration that:
\begin{itemize}
\item Encodes a class of $\unk_0 = \unkp \oplus \ldots$
(\sref{sksapjsl} and \sref{ljfldfjfdss}).
\item Imposes
a lower triangular structure on
perturbation theory (\sref{lowertriangularity}).
\item Is by design amenable to gauge-fixing
to hyperbolic equations,
even though the construction of the filtration
does not involve any gauge-fixing (\sref{symhypxxx}).
\end{itemize}
Choosing filtrations
involves conceptual issues,
specific to general relativity.

The primary objects in this paper are filtrations.
By contrast,
$\unk_0 = \unkp \oplus \ldots$
only appears in this introduction and again briefly at the end of this
paper, where we write down explicitly
the equations that $\unk_0$ has to satisfy (\sref{leadingtermeqs}).



\subsection{What is a filtration?}\label{sec:whatisfilt}

A filtration is an increasing sequence of subspaces of $\Ein{}$:
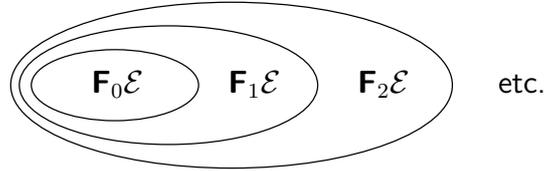
\begin{center}
\input{filintro.pstex_t}
\end{center}%

The basic algebraic requirements for a filtration
indexed by $\filind \in \Z_{\geq 0}$ are:
\begin{itemize}
\item 
$\fil{\filind}{\Ein{}} = \bigoplus_k \fil{\filind}{\Ein{k}}$
where $\fil{\filind}{\Ein{k}}\subset \Ein{k}$ are linear subspaces\footnote{%
Vector subspaces,
or maybe even submodules over the ring of smooth real functions
on the \mfd. All actual constructions in this paper yield
submodules.
}.
\item
$\fil{\filind}{\Ein{}}
\subset \fil{\filind+1}{\Ein{}}$.

\item $\eb{\fil{\filind}{\Ein{}}}{\fil{\otherfilind}{\Ein{}}}
\subset \fil{\filind+\otherfilind}{\Ein{}}$.
\item $\exists p: \fil{p}{\Ein{}} = \Ein{}$.
\end{itemize}
That is, the filtration is by graded subspaces;
is increasing; respects the bracket;
exhausts\footnote{
To be exhaustive simply means $\bigcup_p \fil{p}{\Ein{}} = \Ein{}$.
We use the stronger condition
$\exists p: \fil{p}{\Ein{}} = \Ein{}$
since it is more concrete, and it suffices for this paper.
} at a finite $\filind$.
Note in particular that $\fil{0}{\Ein{}}$ is a subalgebra.

\newcommand{\TSP}{\rule{0pt}{13pt}}
To appreciate these abstract conditions,
it is good to recall some of the various roles of
the graded Lie algebra bracket
$\eb{\,\cdot\,}{\,\cdot\,}: \Ein{k}\times \Ein{\ell} \to \Ein{k+\ell}$:\\
{\centering
\begin{tabular}{l@{\quad\;\;\;}l}
\rule{0pt}{20pt}%
$\Ein{0}\times \Ein{0}\to \Ein{0}$ &
bracket of the `infinitesimal gauge group'
Lie algebra $\Ein{0}$\\
\TSP
$\Ein{0}\times \Ein{k}\to \Ein{k}$ &
action of the infinitesimal gauge group on $\Ein{}$\\
\TSP
$\Ein{1}\times \Ein{1}\to \Ein{2}$ &
used to state the \eve\\
\TSP
$\Ein{1}\times \Ein{2}\to \Ein{3}$ &
used to state the key identity
$\forall \unk \in \Ein{1}: \eb{\unk}{\eb{\unk}{\unk}} = 0$
\end{tabular}
}


\subsection{Filtered expansions and the Rees algebra
}\label{sec:REES}

The basic idea is this:
One looks for formal power series
solutions in a variable $s$,
with the coefficient of $s^{\filind}$ drawn from $\fil{\filind}{\Ein{1}}$.
That is, the filtration specifies
at what stage $\filind$ the different degrees of
freedom can kick in.

To implement this idea,
 introduce the so-called Rees algebra
\[
\REES\;=\;\textstyle\bigoplus_{\filind\geq 0} s^{\filind}\,\fil{\filind}{\Ein{}}
\]
understood as a subspace of $\Ein{}[[s]]$,
the formal power series in $s$ with coefficients in $\Ein{}$.
Note that $\Ein{}[[s]]$ is a graded Lie algebra over $\R[[s]]$.

The first three bullets in \sref{whatisfilt} can now be
restated as:
\begin{itemize}
\item $\REES = \bigoplus_k \REES^k$ where
$\REES^k = \bigoplus_{\filind\geq 0} s^{\filind}\,\fil{\filind}{\Ein{k}}$.
\item $\REES \to \REES, x \mapsto sx$ is an injective map.
\item $\REES$ is itself a graded Lie algebra over $\R[[s]]$,
a subalgebra of $\Ein{}[[s]]$.
\end{itemize}

The set of formal power
series solution to the \eve,
filtered by $\fil{}{}$, is then concisely given by
$\Sol(\REES)
 = \{ \unk \in \REES^1 \mid\eb{\unk}{\unk} = 0\}$
where the bracket is understood to be the bracket in $\REES$.

We informally picture the Rees algebra as
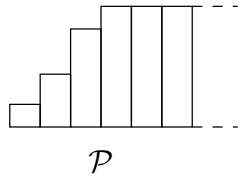
\begin{center}
\begin{picture}(0,0)%
\includegraphics{rees.pstex}%
\end{picture}%
\setlength{\unitlength}{4144sp}%
\begingroup\makeatletter\ifx\SetFigFont\undefined%
\gdef\SetFigFont#1#2#3#4#5{%
  \reset@font\fontsize{#1}{#2pt}%
  \fontfamily{#3}\fontseries{#4}\fontshape{#5}%
  \selectfont}%
\fi\endgroup%
\begin{picture}(1374,1075)(5389,-3644)
\put(5941,-3571){\makebox(0,0)[lb]{\smash{{\SetFigFont{12}{14.4}{\rmdefault}{\mddefault}{\updefault}{\color[rgb]{0,0,0}$\mathclap{\REES}$}%
}}}}
\end{picture}%

\end{center}%

and in drawing the picture we have assumed that
$\fil{p}{\Ein{}} =\Ein{}$
if and only if $p \geq 3$.

\newcommand{\KERFOOTNOTE}{\footnote{
$\ker(\REES/s^{\filind}\REES \leftarrow \REES/s^{\filind+1}\REES)
= s^\filind\REES/s^{\filind+1}\REES$
with canonical isomorphism
$\leftarrow
\GO, s^{\filind}x \mapsfrom x$.
}}

\newcommand{\QUOTIENTFOOTNOTE}{\footnote{%
At the level of formal power series, the basic object of study is the quotient
\[
\big\{\unk \in \Sol(\REES) \;\big|\;
\unk = \unk_0 \bmod s\REES^1
\big\}\,\big/\,\exp([s\REES^0,\,\cdot\,])
\]
The quotient is in the sense of a group action.
The denominator is the group associated to
the Lie algebra $s\REES^0$
by the Baker-Campbell-Hausdorff formula.}}

\subsection{%
The role of $\GO$
}\label{sec:stepbystep}

The quotient $\GO$, a standard object in algebra,
plays an outstanding role for filtered expansions.
It has the same `size' as $\Ein{}$.

Suppose we want to construct
a $\unk \in \Sol(\REES)$.
One can try to break the problem into pieces
and solve $\eb{\unk}{\unk}=0$ first modulo $s\REES$,
then modulo $s^2\REES$, etc.
In the following pictures, the darker parts are divided out:
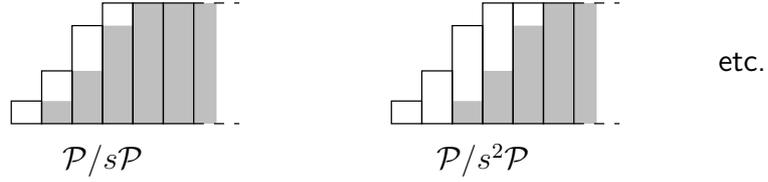
\begin{center}
\input{reesdiv.pstex_t}
\end{center}%

The $ s\REES
\supset s^2\REES \supset  \ldots$ are
 ideals,
$\eb{\REES}{s^{\filind}\REES} \subset s^{\filind}\REES$.
Hence one has a sequence of graded Lie algebra morphisms
$0
\leftarrow \REES/s\REES
\leftarrow \REES/s^2\REES
\leftarrow\ldots$ whereby:
\begin{itemize}
\item Each of these morphisms is surjective.\\
Each of these morphisms 
has kernel canonically isomorphic\KERFOOTNOTE~to $\GO$.
\item
Each $x \in \REES$ corresponds
to a sequence of elements
$0\mapsfrom x_0 \mapsfrom x_1 \mapsfrom
\ldots$
and in particular if $\unk \in \REES^1$
corresponds to 
$0 \mapsfrom \unk_0 \mapsfrom \unk_1 \mapsfrom 
\ldots$
then
\[
\unk \in \Sol(\REES)
\qquad
\Longleftrightarrow
\qquad
\forall \filind: 
\unk_{\filind} \in \Sol(\REES/s^{\filind +1}\REES)
\]
\end{itemize}
When one tries to construct the $\unk_{\filind}$ in succession,
the quotient
$\GO$ plays an outstanding role
as the kernel of the morphisms;
because of the leading term $\unk_0 \in \Sol(\GO)$;
and because of the differential $\dd_0$ defined below.

Explicitly
\begin{align*}
\GO
\;&=\; \textstyle\bigoplus_{\filind \geq 0} s^{\filind}
(\fil{\filind}{\Ein{}}/\fil{\filind-1}{\Ein{}})
\end{align*}
It
is a graded Lie algebra relative to 
the
$k$-grading,
whose bracket respects
both the $\filind$-grading and the $k$-grading.
The $\filind$-grading has only finitely many nonzero components
because the filtration exhausts at a finite $\filind$.

\subsection{The lower triangular differential $\dd_0$}\label{sec:lowertriangularity}

Let $\unk_0 \in \Sol(\GO)$, the leading term\QUOTIENTFOOTNOTE. Define
\[
\dd_0 = \eb{\unk_0}{\,\cdot\,} \;\;:\;\; \GO \to \GO
\]
which is a differential,
$(\dd_0)^2 = 0$.
The role of the cohomologies of $\dd_0$
has been alluded to in \sref{dkhflho4hire}, \sref{goalopen}.
This differential raises the $k$-grading by one:
\[
0\ax\sx{0}\ax\sx{1}\ax
\;\cdots\;
\ax\sx{4}\ax 0
\]
In the $\filind$-grading it is lower-triangular:
\[
\dd_0\;=\;\begin{pmatrix}
\dec{\dd} & 0 & 0 & \cdots\\
\ast & \dec{\dd} & 0 & \cdots \\
\ast & \ast & \dec{\dd} & \cdots\\
\vdots & \vdots & \vdots & \ddots
\end{pmatrix}
\]
This is the block decomposition of $\dd_0$
by the $\filind$-grading of $\GO$.
The rows and columns are indexed by $\filind \geq 0$,
and there are only finitely many of them.

The $\filind$-th entry on the main diagonal is the differential
\[
\dec{\dd} = \eb{\unkp}{\,\cdot\,} \;\;:\;\;
\fil{\filind}{\Ein{}}/\fil{\filind-1}{\Ein{}} \;\to\;
\fil{\filind}{\Ein{}}/\fil{\filind-1}{\Ein{}}
\]
where $\unkp \in \Sol(\fil{0}{\Ein{}})$ is
the first direct summand of $\unk_0 = \unkp \oplus \ldots$
and is called the naive leading term.
Hence each entry of the main diagonal is a complex:
\[
0\ay\sy{0}\ay\sy{1}\ay\;\cdots\;\ay\sy{4}\ay 0
\]

\newcommand{\quot}{X}
By contrast, the entries of the subdiagonals are determined by 
the remaining summands in $\unk_0 = \unkp \oplus \ldots$
and do not individually
correspond to \twofootnotes{complexes}{%
As a toy example, let
$m = (\begin{smallmatrix} a & 0\\
b & c
\end{smallmatrix})$ where $a,c$ are real square matrices,
$b$ a real rectangular matrix,
and $m^2 = 0$.
Then $a^2=ba+cb=c^2=0$.
Hence $b$ descends to a map
$H(a) \to H(c)$
where $H(a) = \ker a / \image a$.
There is a canonical vector space isomorphism 
\[
H(m)/\quot\; \oplus\; \quot \;\; \simeq \;\; H(0 \to H(a) \xrightarrow{b} H(c) \to 0)\]
where $\quot \subset H(m)$
is the subspace of elements that have a representative of the form
$(\begin{smallmatrix} 0 \\ \ast \end{smallmatrix})$.
This is a spectral sequence \cite{Chow}.
Note that
$H(m) \simeq H(m)/\quot \oplus \quot $
but non-canonically so.
}{%
The subdiagonals may play a crucial role
for the filtration $\bkl{}{}$ constructed in this paper.
An observation in \sref{degobs} suggests that
there are degeneracies in the diagonal entries that
may be resolved by the subdiagonals.
See also the closely related \sref{obstrexample}.
}.

\subsection{Notational overview}\label{sec:dfdlshs}

In this paper we first construct a specific
$\Z_{\geq 0}$-indexed
filtration
$\fil{}{}$.
We then take two copies of this filtration,
$\filp{}{}$ and $\filpp{}{}$,
that differ only by a relative rotation of \modelspace.
We then define a $\Z_{\geq 0}\times \Z_{\geq 0}$-indexed filtration:
\[
\bkl{\filind'\filind''}{\Ein{}}
\;=\; \filp{\filind'}{\Ein{}}\cap \filpp{\filind''}{\Ein{}}
\]
This is the filtration intended for \BKL:
\begin{center}
\input{filintro2.pstex_t}
\end{center}%

While filtered expansions based on $\fil{}{}$ involve
a single formal variable $s$,
those based on $\bkl{}{}$
involve two formal variables $s',s''$.
While the naive leading term is a
$\unkp \in \Sol(\fil{0}{\Ein{}})$
in one case,
it is a
$\unkp \in \Sol(\bkl{00}{\Ein{}})$ in the other.
And so forth.
Modulo such modifications, the previous discussion stays intact.

\subsection{Degeneracy of the frame encoded in $\fil{0}{}$ and $\bkl{00}{}$}\label{sec:sksapjsl}

Recall that a
frame is, over each point of the \mfd, a linear
map
with domain a fixed \modelspace~with fixed conformal
inner product\footnote{%
This \modelspace~is the dual of
what is called $\VVreal$ in \cite{RT} and in this paper.
}:
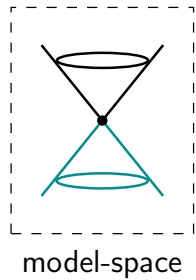
\begin{center}
\begin{picture}(0,0)%
\includegraphics{ocollapse.pstex}%
\end{picture}%
\setlength{\unitlength}{4144sp}%
\begingroup\makeatletter\ifx\SetFigFont\undefined%
\gdef\SetFigFont#1#2#3#4#5{%
  \reset@font\fontsize{#1}{#2pt}%
  \fontfamily{#3}\fontseries{#4}\fontshape{#5}%
  \selectfont}%
\fi\endgroup%
\begin{picture}(1104,1660)(5839,-3869)
\put(6391,-3796){\makebox(0,0)[lb]{\smash{{\SetFigFont{12}{14.4}{\rmdefault}{\mddefault}{\updefault}{\color[rgb]{0,0,0}$\mathclap{\textsf{\modelspace}}$}%
}}}}
\end{picture}%

\end{center}%

The filtration $\fil{}{}$ constructed in this paper has the following
property:
\begin{itemize}
\item There is a fixed 2-dim spacelike subspace of \modelspace~that
is contained in the kernel of the frame of every element of
$\fil{0}{\Ein{1}}$.
\end{itemize}
The copies
$\filp{}{}$, $\filpp{}{}$ are
rotated relative to each other
 such that the associated pair of 2-dim
subspaces span only a 3-dim spacelike subspace:
\begin{itemize}
\item There is a fixed 3-dim spacelike subspace of \modelspace~that is contained in the kernel of the frame of every element of
$\bkl{00}{\Ein{1}}$.
\end{itemize}
Hence the frame of each
$\unkp \in \bkl{00}{\Ein{1}}$ has rank 1,
if no extra degeneracy occurs.
The associated tangent space direction is $\unkp$-dependent, not enshrined in
$\bkl{00}{\Ein{1}}$.

This discussion gives some geometric intuition
for $\fil{}{}$ and $\bkl{}{}$,
but it is not a full description of these filtrations.

\newcommand{\kasnerfootnote}{\footnote{The quotients
$p_i^{\pm \infty} = a_i^{\pm \infty}/
(a_1^{\pm\infty}+a_2^{\pm \infty} + a_3^{\pm\infty})$
are known as Kasner parameters in the literature.
They satisfy
$p_1 + p_2 + p_3 = 1$ and $(p_1)^2 + (p_2)^2 + (p_3)^2=1$
for both $p_i = p_i^{\pm \infty}$.
}}

\subsection{%
Relation to BKL
}\label{sec:ljfldfjfdss}

The relation to BKL is
most apparent in \sref{diz38hro}
where we point out that,
up to innocuous simplifications also discussed there,
$\unkp \in \Sol(\bkl{00}{\Ein{}})$
if and only if
\begin{align*}
\axider{\aunkn{1}} & = \squarethis{\aunkn{6}}\\
\axider{\aunkn{2}} & = -\squarethis{\aunkn{6}}\\
\axider{\aunkn{3}} & = \squarethis{\aunkn{6}} \displaybreak[0]\\
\axider{\aunkn{6}} & = \aunkn{2}\aunkn{6}\\
0 & = \mathrlap{
\aunkn{2}\aunkn{3}
+\aunkn{3}\aunkn{1}
+\aunkn{1}\aunkn{2} - \squarethis{\aunkn{6}}}
\end{align*}
Here $\unkp$ is parametrized by
a vector field $\axider{}$ (the frame) and by real functions
$\aunkn{1},\aunkn{2},\aunkn{3},\aunkn{6}$ on the \mfd;
the details are in \sref{diz38hro}.

These are four ordinary differential equations
along $\axider{}$,
and one equation without derivatives.
Suppose for concreteness that we are on $\R^4$
and that $\axider{}$ is the first partial derivative.
Then $\aunkn{6}$ goes to zero along
each integral curve of $\axider{}$,
both towards the past and future.
The other three functions approach past limits
$a^{-\infty}$ and future limits $a^{+\infty}$
that can differ from integral
curve to integral curve.
These limits satisfy
\[
0 \;=\;
\aunkn{2}^{\pm\infty}\aunkn{3}^{\pm\infty}
+\aunkn{3}^{\pm\infty}\aunkn{1}^{\pm\infty}
+\aunkn{1}^{\pm\infty}\aunkn{2}^{\pm\infty}
\]
On integral curves on which $\aunkn{6}$
is not identically zero,
one has
$\aunkn{2}^{-\infty} > 0$ and\kasnerfootnote
\[
a^{+\infty}
=
\begin{pmatrix}
1 & 2 & 0\\
0 & -1 & 0\\
0 & 2 & 1
\end{pmatrix}
a^{-\infty}
\]

This $\axider{}$ 
is the only direction of differentiation of 
the first order operator $\dec{\dd}$,
but this is not true for entries of the subdiagonals
of the matrix of $\dd_0$ in \sref{lowertriangularity}.

\subsection{%
Rees algebra associated to $\bkl{}{}$
}\label{sec:hjdjd}
The filtration $\bkl{}{}$,
indexed by $\alpha \in \Z_{\geq 0}\times \Z_{\geq 0}$,
has the following algebraic properties:
\begin{itemize}
\item 
$\bkl{\alpha}{\Ein{}} = \bigoplus_k \bkl{\alpha}{\Ein{k}}$
where $\bkl{\alpha}{\Ein{k}}\subset \Ein{k}$ are linear subspaces.
\item
$\bkl{\alpha}{\Ein{}}
\subset \bkl{\beta}{\Ein{}}$ whenever $\alpha \leq \beta$.
\item $\eb{\bkl{\alpha}{\Ein{}}}{\bkl{\beta}{\Ein{}}}
\subset \bkl{\alpha+\beta}{\Ein{}}$.
\item $\exists \alpha: \bkl{\alpha}{\Ein{}} = \Ein{}$.
\end{itemize}
It has the additional property\footnote{%
Explicitly
$\bkl{\filind'\filind''}{\Ein{}}
\cap \bkl{\otherfilind'\otherfilind''}{\Ein{}}=
\bkl{\min(\filind',\otherfilind')\min(\filind'',\otherfilind'')}{\Ein{}}$.
}:
\begin{itemize}
\item 
$\bkl{\alpha}{\Ein{}} \cap \bkl{\beta}{\Ein{}} \subset 
\sum_{\gamma \leq \alpha\text{ and }\gamma \leq \beta} \bkl{\gamma}{\Ein{}}$.
\end{itemize}

Define the Rees algebra
$\REES = \textstyle\bigoplus_{\alpha}
s^{\alpha}\,\bkl{\alpha}{\Ein{}}$
where $s = (s',s'')$ is a pair of formal variables.
One can develop things much as in
\sref{REES},
\sref{stepbystep},
\sref{lowertriangularity},
as discussed in \aref{reestwo}.
The upshot is that now the basic space
is the graded Lie algebra
\begin{align*}
\GO
\;& \DEF\; \REES/(s'\REES + s''\REES)\\
\;& \rule{1pt}{0pt}=\; \textstyle\bigoplus_{\alpha}
s^{\alpha}\,(\bkl{\alpha}{\Ein{}}/\bkl{<\alpha}{\Ein{}})
\end{align*}
where $\bkl{<\alpha}{\Ein{}} = \sum_{\beta \leq \alpha\,\text{and}\,\beta \neq \alpha}
\bkl{\beta}{\Ein{}}$.


\subsection{Reflections}

Recall the 2-dim spacelike subspaces of \modelspace~encountered
in \sref{sksapjsl}.
Point-reflections within these subspaces,
leaving their orthogonal complement fixed,
yield
graded Lie algebra automorphisms
$\fundi',\fundi'' \in \Aut(\Ein{})$
with $(\fundi')^2 = (\fundi'')^2 = \mathbbm{1}$.
Because of the particular
geometric arrangement,
$\fundi'\fundi'' = \fundi''\fundi'$.

Hence $\Ein{}$ decomposes into a direct sum of four pieces,
schematically
\[
\even'\even'' \oplus \even'\odd'' \oplus
\odd'\even'' \oplus \odd'\odd''
\]
and the bracket respects this grading.
Each component of $\filp{}{},\filpp{}{}$
and $\bkl{}{}$
is by construction invariant under both $\fundi'$, $\fundi''$,
hence also decomposes into four pieces.

Having introduced these reflections, we can now
also state\footnote{%
This is a complete description of the filtration,
except for $\filp{0}{\Ein{}}$.
}:
\begin{align*}
\filp{0}{\Ein{}} & \;\subset\; \text{($\even'$ part of $\Ein{}$)}\\
\filp{1}{\Ein{}} & \;=\; \filp{0}{\Ein{}} \oplus \text{($\odd'$ part of $\Ein{}$)}\\
\filp{2}{\Ein{}} & \;=\; \Ein{}
\intertext{%
Analogous for $\filpp{}{}$. One can then 
draw various conclusions about $\bkl{}{}$, for example:}
\bkl{00}{\Ein{}} & \;\subset\; \text{($\even'\even''$ part of $\Ein{}$)}\\
\bkl{22}{\Ein{}} & \;=\; \Ein{}
\end{align*}


\subsection{Symmetric hyperbolicity}\label{sec:symhypxxx}

The basic algebraic conditions on a filtration
(\sref{whatisfilt}
or
\sref{hjdjd})
are very permissive:
many uninteresting filtrations pass.
In this paper we formulate additional algebraic conditions\footnote{%
In the technical jargon of this paper:
Each filtration component is a \ecube~and has a
representative that satisfies the $234$-condition.
See \sref{deff}.
Both $\fil{}{}$ and $\bkl{}{}$
satisfy these conditions.
}
that are specific to $\Ein{}$;
that are rather restrictive;
and that we think are natural conditions.
We now summarize some of their implications.

\newcommand{\exarr}{\;\xrightarrow{\;\cre\;}\;}
\newcommand{\exterm}[1]{\bkl{\alpha}{\Ein{#1}}}
Creation operators are
certain linear maps
$\cre: \Ein{k}\to \Ein{k+1}$ with $\cre^2 = 0$.
Coming with
$\bkl{}{}$ is a set of `admissible' creation operators for which
\[
0\exarr\exterm{0}\exarr\exterm{1}\exarr\exterm{2}\exarr\exterm{3}
\exarr\exterm{4}\exarr 0
\]
is an exact sequence for all $\alpha$.
For example, if $\unkp\in \Sol(\bkl{00}{\Ein{}})$
and if $f$ is a time function\footnote{%
A real function on the \mfd~is a time function for $\unkp$ if
its derivative along any vector in the
(image of \modelspace's) future cone is positive.
},
then 
$\cre = \eb{\unkp}{f\,\cdot\,} - f\eb{\unkp}{\,\cdot\,}$
is an admissible creation operator.

The exact sequence above is a by-product
of the gauge-fixing algorithm \sref{genalg},
a minor generalization of
an algorithm in \cite{RT}.
For every choice of a gauge\footnote{%
The word gauge has here a technical meaning \cite{RT}.}
$\gauge$
it returns a graded subspace
$\bklG{\alpha} \subset \bkl{\alpha}{\Ein{}}$ that `splits'
\[
\bkl{\alpha}{\Ein{}} = \bklG{\alpha} \oplus \cre(\bklG{\alpha})
\]
where
$\cre : \bklG{\alpha} \to \bkl{\alpha}{\Ein{}}$
is injective, for all admissible creation operators $\cre$.

Let
$\dec{\dd} = \eb{\unkp}{\,\cdot\,}$, for all instances of this map.
By design of the algorithm, $\dec{w}
= (\dec{\dd} : \bklG{\alpha} \to \bkl{\alpha}{\Ein{}}/\bklG{\alpha})$
is symmetric hyperbolic, and the complex
$\dec{\dd}:
\bkl{\alpha}{\Ein{}} \to \bkl{\alpha}{\Ein{}}$
has a deformation retraction\footnote{%
Provided the geometry associated to $\unkp$ is globally
hyperbolic, so that
one can solve the initial value for symmetric hyperbolic systems
based on $\unkp$. Then $\dec{w}$ is surjective.
} to
$\dec{\dd}:
\ker \dec{w} \to \ker \dec{w}$, in
particular the cohomologies are the same \cite{RT}.
The point is that, while
elements of $\bkl{\alpha}{\Ein{}}$
live in four dimensions,
elements of $\ker \dec{w}$ live in only three,
since they are homogeneous solutions to a
 linear symmetric hyperbolic system.

We emphasize that there is no gauge-fixing going
on in this paper; we only use amenability to gauge-fixing
in the sense of \sref{genalg}
as an algebraic condition.



\newcommand{\stdim}{d_{\rR}}
\section{Language}\label{sec:langag}
This is a summary of parts of \cite{RT},
with an emphasis on the things we need.

Let $\rR$ be a commutative ring
with a distinguished subring isomorphic to the reals
$\R \subset \rR$,
such that their multiplicative units coincide.
Suppose $\Der(\rR)$\footnote{%
The space of $\R$-linear derivations on $\rR$}
is a \ff~$\rR$-module, and set
$\stdim = \rank_{\rR}\Der(\rR)$.
Example: $\rR$ the set of real smooth
functions on a manifold diffeomorphic to $\R^{\stdim}$.

Denote by
$\rC = \rR \oplus i\rR$ the
complexification of the ring,
$i^2 = -1$. Let $V$ be a \ff~$\rC$-module of rank 2.
Denote by $\cc{V}$ the complex conjugate module.
Let $\Lang$ be the free $\rC$-algebra generated by $V$ and $\cc{V}$,
\[
\Lang
= \rC \oplus V \oplus \cc{V}
\oplus VV \oplus V\cc{V}
\oplus \cc{V}V \oplus \cc{V}\cc{V} \oplus VVV \oplus \ldots
\]
where juxtaposition means tensor product over $\rC$.
Let $\DERIV$ be the set of derivations on $\Lang$
that are $\C$-linear
and that preserve the grading\footnote{%
Hence each element of $\DERIV$ maps
$\rC \to \rC$ and $V \to V$ and $\cc{V} \to \cc{V}$ and so on.
Each element of $\DERIV$ is determined by its restriction
to $\rC \oplus V \oplus \cc{V}
\to \rC \oplus V \oplus \cc{V}$.
}. Elements of $\DERIV$ need not be $\rC$-linear.
We sometimes write
$\DERIVVERT = \{\delta \in \DERIV \mid \delta(\rC)=0\}$.

Set $\DC{} = (\bigwedge \VV)\DERIV$,
and again juxtaposition means tensor product over $\rC$.
Then $\DC{}$ is a complex graded Lie algebra with bracket defined by
\[
\db{\omega\delta}{\omega'\delta'}
= (\omega\wedge\omega')[\delta,\delta']
+ (\omega \wedge \delta(\omega'))\delta'
- (\delta'(\omega) \wedge \omega')\delta
\]
for all $\omega,\omega' \in \bigwedge \VV$
and all $\delta,\delta' \in \DERIV$, with $[\delta,\delta']$
the commutator of derivations.
See \aref{sampleeval} for a sample evaluation of the bracket.

\subsection{Conjugation}\label{sec:hdkhkdhiooao333}
There is a canonical $\rC$-antilinear involution on $\Lang$,
taking $\rC \to \rC$
($f+ig \mapsto f-ig$ for all $f,g\in\rR$)
and $V \to \cc{V}$ and $\cc{V}\to V$ and
$V\cc{V}\to \cc{V}V$ and so on. 
It induces a canonical $\rC$-antilinear involution $\DERIV
\to \DERIV, \delta \mapsto \cc{\delta}$
by $\cc{\delta}(\cc{a}) = \cc{\delta(a)}$ for all $a \in \Lang$.

By conjugation on $\VV$
we mean conjugation followed by exchange of factors:
 $\VV \to \VV, v\cc{w} \mapsto w\cc{v}$. We denote by
$\VVreal \subset \VV$ the real submodule.

These conjugations induce a conjugation on $\DC{}$
that we sometimes denote by
 $\CONJ: \DC{} \to \DC{}, \omega\delta \mapsto \cc{\omega}\cc{\delta}$.
We denote 
$\RE = \tfrac{1}{2}(\mathbbm{1}+\CONJ)$
and $\IM = \tfrac{1}{2i}(\mathbbm{1}-\CONJ)$.

\subsection{The real graded Lie algebra $\Ein{}$}
Let $\D{} \subset \DC{}$ be the
real elements, a real graded Lie algebra and an $\rR$-module.

Set $\IWeyl{0}=\IWeyl{1}=0$.
Let $\IWeyl{2}\subset \D{2}$ be the elements $x$ that
satisfy $x(\rC)=x(V\wedge V)=0$ and
for which $x(V)$ is contained in:
the submodule of $V\cc{V}V\cc{V}V$ that is symmetric in the three $V$'s.
Set $\IWeyl{k} = \VVreal \wedge \IWeyl{k-1}$ recursively for $k\geq 3$.
It is shown in \cite{RT}
that $\IWeyl{} = \bigoplus_k \IWeyl{k}$ is an ideal,
$\db{\D{}}{\IWeyl{}}\subset\IWeyl{}$.

Set
$\Ein{} = \D{}/\IWeyl{}$, and denote the
induced bracket on $\Ein{}$ by $\eb{\,\cdot\,}{\,\cdot\,}$.
Since $\VV$ has rank 4,
we have $\D{k}=0$
and $\Ein{k} = 0$ whenever $k > 4$.
The equation $\eb{\unk}{\unk}=0$ for $\unk \in \Ein{1}$
is equivalent to
the \eve,
if $\rR$ are the smooth real functions on a manifold
diffeomorphic to $\R^4$,
and provided the frame $\unk|_{\rR}$
is nondegenerate. See \cite{RT}.

\newcommand{\vtrsf}{t}
\subsection{Vertical gauge transformations}\label{sec:vgt}
Associated to every $\vtrsf \in \Aut_{\rC}(V)$
are automorphisms on various other spaces:
\begin{itemize}
\item $\vtrsf \in \Aut_{\rC}(\cc{V})$ by $\vtrsf(\cc{v}) = \cc{\vtrsf(v)}$
for all $v \in V$.
\item $\vtrsf \in \Aut_{\rC}(\Lang)$ by $\vtrsf(ab) = \vtrsf(a)\vtrsf(b)$ for all $a,b \in \Lang$.
\item $\vtrsf \in \Aut_{\rC}(\DC{})$ by $(\vtrsf(x))(a) = \vtrsf(x(\vtrsf^{-1}(a)))$
for all $x\in \DC{}$ and $a \in \Lang$.
\end{itemize}
This is a graded Lie algebra automorphism
of $\DC{}$ that commutes with conjugation, and
therefore we have the additional graded Lie algebra automorphisms:
\begin{itemize}
\item $\vtrsf \in \Aut_{\rR}(\D{})$
which in particular satisfies $\vtrsf(\IWeyl{}) \subset \IWeyl{}$.
\item $\vtrsf \in \Aut_{\rR}(\Ein{})$.
\end{itemize}
The so defined maps
$\Aut_{\rC}(V) \to \Aut(\ldots)$
are group homomorphisms.

\subsection{Canonical conformal inner product on $\VVreal$}
By definition, a volume form on $V$ is
an antisymmetric nondegenerate $\rC$-bilinear map
$\varepsilon: V \times V \to \rC$.
Define a symmetric $\rC$-bilinear
map $\varepsilon\cc{\varepsilon}: \VV \times \VV \to \rC$ by
\[
\varepsilon\cc{\varepsilon}(v\cc{w},v'\cc{w'})
\;=\; \varepsilon(v,v')
\cc{\varepsilon(w,w')}
\]
Its restriction to an $\rR$-bilinear
$\VVreal \times \VVreal \to \rR$
has signature ${+}{-}{-}{-}$,
which is of course equivalent
to one with signature ${-}{+}{+}{+}$.

The choice of $\varepsilon$ is unique up to multiplication
by an invertible element of $\rC$.
Some constructions in this paper
refer to a volume form $\varepsilon$,
but this is only for convenience.
Results depend on it either not at all or only in obvious ways.
All objects that we define scale
homogeneously in $\varepsilon$ and $\cc{\varepsilon}$,
but we never sum objects unless they have the same homogeneity
in $\varepsilon$ and $\cc{\varepsilon}$.

\subsection{Creation operators}
For every $c \in \VV$
we define the $\rC$-linear map
$\cre_c : \DC{k} \to \DC{k+1}$ by
\[
\cre_c(\omega \delta) = (c \wedge \omega)\delta
\]
for all $\omega \in \bigwedge \VV$ and all $\delta \in \DERIV$.
Note that
$\cre_c\cre_{c'}+\cre_{c'}\cre_c= 0$ for all $c,c'$.
	
If $c \in \VVreal$ then we get a map
$\cre_c : \D{k}\to \D{k+1}$.
Since $\cre_c\IWeyl{}\subset \IWeyl{}$ by definition of the ideal,
we also get a map $\cre_c : \Ein{k}\to  \Ein{k+1}$.

\subsection{Annihilation operators}

For every $c \in \VV$
we define the $\rC$-linear map
$\ann_c : \DC{k} \to \DC{k-1}$ by requiring that
\[
\ann_c\cre_{c'} + \cre_{c'}\ann_c \;=\;
\tfrac{1}{2}\varepsilon\cc{\varepsilon}(c,c')\,\mathbbm{1}
\]
for all $c' \in \VV$.
This definition is recursive in $k$, with base case
$\ann_c\DC{0} = 0$. Uniqueness of $\ann_c$ is clear,
but one has to check existence.
The map
$c \mapsto \ann_c$ is $\rC$-linear.
Note that
$\ann_c$ scales homogeneously like
$\varepsilon\cc{\varepsilon}$ as a function of $\varepsilon$.

If $c \in \VVreal$
then we get a map
$\ann_c : \D{k} \to \D{k-1}$.
In general $\ann_c\IWeyl{} \not\subset \IWeyl{}$.


\subsection{The anticommutation relations}

For all $c,c' \in \VV$ we have
\begin{align*}
\cre_c\cre_{c'} + \cre_{c'}\cre_c & \;=\; 0\\
\ann_c\ann_{c'} + \ann_{c'}\ann_c & \;=\; 0\\
\ann_c\cre_{c'} + \cre_{c'}\ann_c & \;=\;
\tfrac{1}{2}\varepsilon\cc{\varepsilon}(c,c')\,\mathbbm{1}
\end{align*}
which we refer to as the anticommutation relations.

\subsection{Projection operators}\label{sec:projs}

For all $c \in \VV$
such that $\varepsilon\cc{\varepsilon}(c,c) \in \rC$ is invertible we define
\[
\pro_c = (\tfrac{1}{2}\varepsilon\cc{\varepsilon}(c,c))^{-1}\,\ann_c\cre_c
\]
which maps $\DC{k}\to \DC{k}$
and which satisfies $(\pro_c)^2 = \pro_c$.

If $\varepsilon\cc{\varepsilon}(c,c')=0$
then
$\pro_c\pro_{c'} = \pro_{c'}\pro_c$.


\nextpart{Cubes}


\section{Two points of view}
Recall that $\Ein{} = \D{}/\IWeyl{}$.
Given a submodule $\Dp \subset \D{}$,
there are two interpretations
of what the quotient of $\Dp$ by $\IWeyl{}$ should be:
\begin{itemize}
\item $(\Dp+\IWeyl{})/\IWeyl{}$
\item $\Dp/(\Dp\cap \IWeyl{})$
\end{itemize}
There is of course a canonical module isomorphism between these two.
We use both points of view in this paper.
The first in particular when we use the bracket in $\Ein{}$.
The second in particular for the gauges algorithm in \sref{genalg}.


\subsection{%
Lemma
}
\label{sec:downup}
Suppose $\Dp,\Dpp \subset \D{}$ are submodules.
Equivalent are\footnote{%
This holds for any three subgroups of an Abelian group.
We show
$\Uparrow$.
Given $x \in (\Dp + \IWeyl{})\cap (\Dpp+\IWeyl{})$,
i.e.~$x = s' + i' = s''+i''$ for some $s' \in \Dp$, $s'' \in \Dpp$
and $i',i'' \in \IWeyl{}$.
Then $s'-s''=i''-i' \in (\Dp+\Dpp) \cap \IWeyl{}$.
By the hypothesis, $s'-s'' = i''-i' = j'+j''$ with $j' \in \Dp \cap \IWeyl{}$
and $j'' \in \Dpp \cap \IWeyl{}$. 
Then $x = a + b$ with $a = s'-j' = s''+j''$
and $b = i'+j' = i''-j''$. Since $a \in \Dp\cap \Dpp$ and $b \in \IWeyl{}$
we get $x \in (\Dp\cap \Dpp) + \IWeyl{}$, as required.
}:
\begin{itemize}
\item $(\Dp + \IWeyl{}) \cap (\Dpp + \IWeyl{}) \;\;\subset\;\; (\Dp\cap \Dpp) + \IWeyl{}$
\item $(\Dp+\Dpp) \cap \IWeyl{} \;\;\subset\;\; (\Dp \cap \IWeyl{}) + (\Dpp \cap \IWeyl{})$
\end{itemize}
and the opposite $\supset$ holds trivially in both bullets.

In the way we plan to use this,
the first bullet,
which says that some intersection in $\Ein{}$
is equivalent to an intersection in $\D{}$,
is the one we are interested in,
but the second bullet is sometimes easier to check.


\section{\Dcubes~and \Ecubes}\label{sec:cubesbig}

\Dcubes~are submodules of $\D{}$ that have particularly simple
structure: they are representations of
a set of anticommutation relations.
\Ecubes~are submodules of $\Ein{}$ of the form
$(\text{\dcube}+\IWeyl{})/\IWeyl{}$.
We discuss these notions in detail, since
we use them a lot throughout this paper.

\subsection{%
Collections of creation
and annihilation operators}\label{sec:reg}

Let $\reg = \VVreal$
be a
submodule of rank $0 \leq n \leq 4$
with the following property:
there exists a basis of $\VVreal$,
orthogonal with respect to its canonical conformal inner product,
such that $\reg$ is the span of the first $n$ basis elements.

In detail,
there exist $c_1,c_2,c_3,c_4$ with
\[
\begin{aligned}
\VVreal &\;=\;\rR c_1 \oplus \rR c_2 \oplus \rR c_3 \oplus \rR c_4\\
\reg &\;=\; \rR c_1 \oplus \ldots \oplus \rR c_n
\end{aligned}
\]
such that
\[
\varepsilon\cc{\varepsilon}(c_i,c_j)
= \begin{cases}
0 & \text{if $i\neq j$}\\
\text{(invertible element of $\rR$)} & \text{if $i=j$}
\end{cases}
\]
We often abbreviate $\cre_i = \cre_{c_i}$, $\ann_i = \ann_{c_i}$,
$\pro_i = \pro_{c_i}$.
The projections $\pro_i$ exist
and commute pairwise by \sref{projs}.
Beware that $\reg$ does not come with
a preferred basis. 

Sometimes we
require $c_1 \in \VVpos
\DEF
\{v\cc{v}+w\cc{w}\mid \text{$v,w$ a basis for $V$}\}$
and $n \geq 1$;
we always say explicitly when we do.

We denote by $\cre_\reg$
the span of all creation operators
of all elements of $\reg$.
Similar for $\ann_{\reg}$. They map an element
of $\D{}$ to a subset of $\D{}$,
or more generally a subset to a subset.
Note that
$(\cre_\reg)^n \neq 0$ whereas
$(\cre_\reg)^k = 0$ for all $k > n$.

\subsection{Generated submodules}

Let $\reg$ be as in \sref{reg}.
For every graded submodule\footnote{%
By definition, a submodule $S \subset \D{}$
is a graded submodule if and only if $S = \bigoplus_k (S \cap \D{k})$.
} $\svac \subset \D{}$ set
\[
\fock{\svac}{\reg}
\;=\; \svac + \cre_\reg \svac + \ldots + (\cre_\reg)^n \svac
\]
Then
$\fock{\svac}{\reg} \subset \D{}$
is itself a graded submodule.
In general one cannot replace $+$ by
 $\oplus$ in this definition,
but we will discuss conditions that allow one to do so.

\newcommand{\wedgeall}{\wedge_{\text{tot}}}
\subsection{Bracket estimate}\label{sec:BRACKETing}
Suppose $\svac,\svac' \subset \D{}$ are graded submodules.
Denote by
$W,W' \subset \bigwedge \VVreal$
the submodules
$W = \wedgeall(\svac(\reg))$ and
$W' = \wedgeall(\svac'(\reg))$.
Then one has
\[
\db{\fock{\svac}{\reg}}{\fock{\svac'}{\reg}}
\;\subset\;
 \fock{\wedgeall (W\svac')}{\reg}
+\fock{\wedgeall (W'\svac)}{\reg}
+\fock{\db{\svac}{\svac'}}{\reg}
\]

To see this, use the identity
$\db{x}{\cre_c x'}
= \wedgeall(x(c)x') + (-1)^{|x|} \cre_c\db{x}{x'}$
repeatedly to pull all $\cre_{\reg}$
out of the second argument of
$\db{\fock{\svac}{\reg}}{\fock{\svac'}{\reg}}$,
and then proceed similarly for the first argument.

\subsection{$\reg$-oneway-\dcubes~and $\reg$-\dcubes}\label{sec:fdkhfdksaaa}

We say that a graded submodule $\svac\subset \D{}$ yields
\begin{itemize}
\item an $\reg$-oneway-\dcube~if and only if
$\cre_1\cdots \cre_n: \svac \to \D{}$ is injective\footnote{%
This condition uses a basis
$c_1,\ldots,c_n$ as in \sref{reg}, but is basis-independent.
Since $\svac$ is graded, this condition is equivalent to:
$\cre_1\cdots \cre_n: \svac \cap \D{k}
\to \D{k+n}$ is injective for all $k$.
}.
\item an $\reg$-\dcube~if and only if
$\ann_{\reg}\svac = 0$.
\end{itemize}
In the case of oneway-\dcubes,
different $\svac$ can yield the same $\fock{\svac}{\reg}$.
By contrast, for \dcubes~there
is always a unique $\svac$ that yields a given $\fock{\svac}{\reg}$,
as we will see in \sref{dddhypc}.
In this case we refer to $\fock{\svac}{\reg}$
itself as an $\reg$-\dcube.

Every $\reg$-\dcube~is an $\reg$-oneway-\dcube, since
then $\ann_1\cdots \ann_n$ is a left-inverse
of $\cre_1\cdots \cre_n: \svac \to \D{}$,
up to multiplication by an invertible element of $\rR$.

Given an $\reg$-oneway-\dcube,
the composition of any subset
of $\cre_1,\ldots,\cre_n$ is an injective
$\svac \to \D{}$ map, and the direct sum of the images
of these $2^n$ maps is precisely $\fock{\svac}{\reg}$,
a direct sum of $2^n$ isomorphic copies of $\svac$;
this follows from\footnote{%
One has to show, assuming say $n=2$,
that for all $x,x',x'',x''' \in \svac$
one has
$(x+\cre_1x'+\cre_2x''+\cre_1\cre_2x'''=0)
\Longrightarrow (x=x'=x''=x'''=0)$.
Apply $\cre_1\cre_2$ and use anticommutation relations
to get $\cre_1\cre_2x=0$, hence $x=0$ by injectivity.
Then apply $\cre_2$ to show $x'=0$ etc.
}
$\cre_i\cre_j + \cre_j\cre_i = 0$.
For say $n=3$ the picture is:
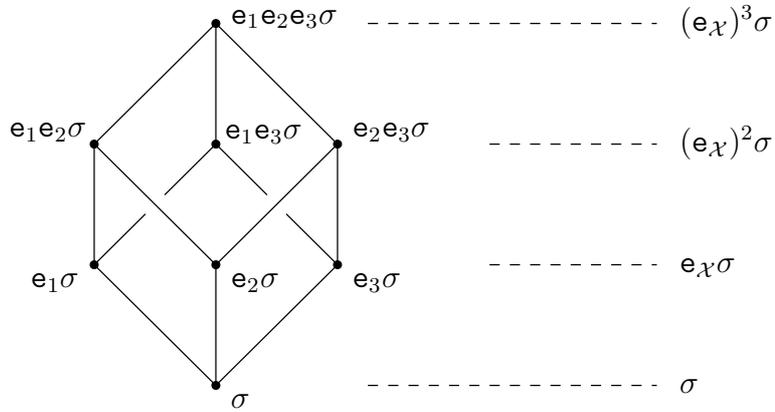
\begin{center}
\input{fock.pstex_t}
\end{center}%

Along each edge of the cube one
can move in the \emph{upward} direction by using one of
$\cre_1,\cre_2,\cre_3$.
Such a move is an  isomorphism between the respective bullets.

In the case of an $\reg$-\dcube~one can in addition
move in the \emph{downward} direction using one of $\ann_1,\ann_2,\ann_3$;
by anticommutation relations these $\ann$-moves are
the inverses of the $\cre$-moves, up to multiplication by an invertible
element of $\rR$.
Any round-trip from a bullet back to itself is
also a multiple of the identity.

This discussion implies, for $\reg$-oneway-\dcubes~and hence for $\reg$-\dcubes,
the coarser but basis-independent decomposition
$\fock{\svac}{\reg}
= \svac \oplus \cre_\reg \svac \oplus \ldots \oplus (\cre_\reg)^n \svac$.


\subsection{Equivalent characterization of $\reg$-\dcubes}\label{sec:dddhypc}

For every graded submodule $S \subset \D{}$ the following are equivalent:
\begin{itemize}
\item $\cre_\reg S \subset S$ and $\ann_{\reg} S \subset S$.
\item $S$ is an $\reg$-\dcube:
$S = \fock{\svac}{\reg}$
for a graded submodule $\svac \subset \D{}$
with
$\ann_\reg \svac = 0$.
\end{itemize}
Furthermore,
if either condition holds, then $\svac = \{x \in S\mid \ann_\reg x = 0\}$.

Direction $\Uparrow$ follows from the discussion in
\sref{fdkhfdksaaa}.
For $\Downarrow$ we use a standard argument.
Pick a basis as in \sref{reg}.
The hypothesis implies $\pro_iS \subset S$.
The commuting projections $\pro_1,\ldots,\pro_n$ and
$\mathbbm{1}-\pro_1,\ldots,\mathbbm{1}-\pro_n$
decompose $S$ into a direct sum of $2^n$ spaces.
Denote one of these spaces by
\[
\svac = \image(\pro_1\cdots \pro_n: S \to S)
= \image(\ann_1\cdots\ann_n\cre_1\cdots \cre_n: S \to S)
\]
It is graded and satisfies $\ann_\reg \svac = 0$
and $S = \fock{\svac}{\reg}$, and we are done with $\Downarrow$.
It is not difficult to see that this particular
choice of $\svac$ is forced.
It is equivalent to $\svac = \{x \in S\mid \ann_\reg x = 0\}$,
and also to
$\svac = S \cap \gvac{\reg}$ using \sref{odhdld}.

Note that
$\svac \subset 
\D{0} \oplus \ldots \oplus \D{4-n}$.

\subsection{Remark}
By \sref{dddhypc}, the intersection
of two $\reg$-\dcubes~is again an $\reg$-\dcube,
as is the span of two $\reg$-\dcubes.
Assuming $\ann_\reg \svac = \ann_\reg \svac' = 0$
 these operations are equivalently given by
$\fock{\svac}{\reg} \cap \fock{\svac'}{\reg} = \fock{\svac \cap \svac'}{\reg}$
and $\fock{\svac}{\reg} + \fock{\svac'}{\reg} = \fock{\svac + \svac'}{\reg}$.


\subsection{%
The vacuum $\gvac{\reg}$ 
}\label{sec:odhdld}
We can view $\D{}$ itself as an $\reg$-\dcube,
$\D{} = \fock{\gvac{\reg}}{\reg}$ with
\[
\gvac{\reg} \;=\; \{x \in \D{}\mid \ann_{\reg}x=0\}
\]
The bigger $\reg$ the smaller $\gvac{\reg}$.
We always have
$\D{0}\subset \gvac{\reg} \subset \D{0} \oplus \ldots \oplus \D{4-n}$.

We have a bijection
\begin{align*}
\mathllap{\{\text{graded submodules of $\gvac{\reg}$}\}}
 & \;\to\; \mathrlap{\{\text{$\reg$-\dcubes}\}}\\
\svac & \;\mapsto\; \fock{\svac}{\reg}
\end{align*}

Conveniently,
$\gvac{\reg}$
is itself given by\footnote{%
Special case of:
$\gvac{\reg} = \fock{\gvac{\reg \oplus \mathcal{Y}}}{\mathcal{Y}}$
when $\mathcal{X}\perp \mathcal{Y}$.
}
\[
\gvac{\reg} \;=\; \fock{\D{0}}{\regperp}
\]
where by definition
$\regperp = \rR c_{n+1}\oplus \ldots \oplus \rR c_4$.
The notation $\regperp$ is justified since $\regperp$
is determined by $\reg$;
neither comes with a preferred basis.


\subsection{The $234$-condition} \label{sec:twothreefour}

For every $\reg$-\dcube~$S$ we define
\[
\text{$(\reg,S)$
satisfies the $234$-condition}
\qquad
\Longleftrightarrow
\qquad
S \cap \IWeyl{}\; \subset \;\fock{S \cap \IWeyl{2}}{\reg}
\]

The inclusion $\supset$
is automatic, since
$S$ and $\IWeyl{}$ are closed under $\cre_\reg$.
The $234$-condition
is necessary to run the gauges algorithm \sref{genalg}.
Trivially
$(\VVreal,\D{})$ satisfies the $234$-condition;
it is equivalent to the definition of
$\IWeyl{}$ in terms of $\IWeyl{2}$.

\subsection{Maximal $\reg$-\dcube}\label{sec:maximalcube}

Recall that the span and the
intersection of $\reg$-\dcubes~are again $\reg$-\dcubes.

To every graded submodule $\Ep \subset \Ein{}$ we can associate the
biggest $\reg$-\dcube~for which
$(\text{$\reg$-\dcube}\,+\,\IWeyl{})/\IWeyl{}\subset \Ep$.
We denote it by $\maxcube_{\reg}\Ep \subset \D{}$.

Note that:
\[
\maxcube_{\reg}(\Ep\cap \Epp )
= \maxcube_{\reg}\Ep \cap \maxcube_{\reg}\Epp
\]


\subsection{$\reg$-\ecube}

By definition, an $\reg$-\ecube~is a subset
of $\Ein{}$ of the form
$(\text{$\reg$-\dcube}\,+\,\IWeyl{})/\IWeyl{}$.

One $\reg$-\ecube~can arise from more than one $\reg$-\dcubes.
Hence
every $\reg$-\ecube~is an equivalence class of $\reg$-\dcubes.
Every $\reg$-\ecube~is closed under $\cre_{\reg}$.

For every graded submodule $\Ep \subset \Ein{}$ we have
\[
\text{$\Ep$ is an $\reg$-\ecube}
\qquad
\Longleftrightarrow
\qquad
(\maxcube_{\reg}\Ep + \IWeyl{})/\IWeyl{}=\Ep
\]
The representative returned by $\maxcube_{\reg}$
will be called the maximal representative.


\subsection{Lemma}\label{sec:ebucintersection}

For all $\reg$-\ecubes~$\Ep,\Epp \subset \Ein{}$ the following
statements are equivalent\footnote{Clearly $\text{3rd}\Rightarrow \text{2nd}$.
We have $\text{2nd} \Rightarrow \text{1st}$
because then one of the bullets in \sref{downup} holds,
hence the first does, hence
$(\Dp +\IWeyl{})\cap(\Dpp+\IWeyl{})
=  (\Dp\cap \Dpp)+\IWeyl{}$,
and since $\Dp \cap \Dpp$
is an intersection of $\reg$-\dcubes~and hence itself an $\reg$-\dcube,
we are done.
We have $\text{1st}\Rightarrow \text{3rd}$
because we can then check the first bullet in \sref{downup}:
$
((\maxcube_{\reg}\Ep+\IWeyl{})/\IWeyl{})\cap
((\maxcube_{\reg}\Epp+\IWeyl{})/\IWeyl{})
=
\Ep\cap \Epp
=
(\maxcube_{\reg}(\Ep\cap \Epp)+\IWeyl{})/\IWeyl{}
= ((\maxcube_{\reg}\Ep \cap \maxcube_{\reg}\Epp)+\IWeyl{})/\IWeyl{}$;
the last step uses \sref{maximalcube}.}:
\begin{itemize}
\item $\Ep \cap \Epp$ is an $\reg$-\ecube.
\item There exist $\reg$-\dcubes~$\Dp,\Dpp$
with $(\Dp+\IWeyl{})/\IWeyl{} = \Ep$
and $(\Dpp+\IWeyl{})/\IWeyl{} = \Epp$
such that $\Dp,\Dpp$ satisfy one
of the bullets in \sref{downup}.
\item
$\Dp = \maxcube_{\reg}\Ep$ and $\Dpp = \maxcube_{\reg}\Epp$
satisfy one of the bullets in \sref{downup}.
\end{itemize}
Furthermore,
if the 2nd bullet holds then
$\Ep\cap \Epp$ has representative $\Dp\cap \Dpp$,
and if the 3rd bullet holds
then $\Ep\cap \Epp$  has maximal representative
$\max_{\reg}\Ep \cap \max_{\reg} \Epp$.


\section{Gauges algorithm in a \dcube}\label{sec:genalg}

\newcommand{\sreg}{\reg}
\newcommand{\sregpos}{\reg_{\text{positive}}}
This \sref{genalg} is not used in the rest of this paper.
However it motivates our working with
\ecubes~and \dcubes~throughout this paper,
as well as the $234$-condition.

This \sref{genalg} must be read together
with the gauges algorithm in \cite{RT}.
It concerns a straightforward
generalization of the algorithm in \cite{RT},
that instead of $\D{}$ works entirely within a \dcube,
provided this \dcube~satisfies
conditions that we state below,
including the $234$-condition.

To see how this \sref{genalg} is applied, we refer to \cite{RT}.


\subsection{Input}\label{sec:subalginp}
The algorithm takes as input a positive definite Hermitian form
$\gauge$ as in the original algorithm in \cite{RT},
satisfying conditions \ixherm, \ixker, \ixpos~stated there.

It additionally takes as input:
\begin{itemize}
\item An $\sreg \subset \VVreal$
that satisfies \sref{reg}
with 
$n\geq 1$ and $c_1 \in \VVpos$.
\item A graded submodule
$\sub{\D{}}\subset \D{}$
such that:
\begin{itemize}
\item $\sub{\D{}}$ is an $\sreg$-\dcube, see \sref{dddhypc}.
\item
The pair $(\sreg,\sub{\D{}})$
satisfies the $234$-condition, see \sref{twothreefour}.
\end{itemize}
\end{itemize}
Set
$\sub{\IWeyl{}} = \IWeyl{} \cap \sub{\D{}}$
and
$\sub{\Ein{}} = \sub{\D{}}/\sub{\IWeyl{}}$.
We additionally require:
\begin{itemize}
\item
$\sub{\D{k}}$ is \ff~and has \ff~complement in $\D{k}$.\\
$\sub{\IWeyl{k}}$ 
is \ff~and has \ff~complement in $\sub{\D{k}}$.\\
$\sub{\Ein{k+1}}/\cre_1(\sub{\Ein{k}})$ is
 \ff, where $\cre_1 = \cre_{c_1}$.
\end{itemize}

Observe that
the conditions that $\gauge$ must satisfy
are completely separate from
the conditions that $\sreg$ and $\sub{\D{}}$
must satisfy.


\subsection{Output}\label{sec:subalgout}

The output is very similar to \cite{RT}, but here everything refers to
$\sub{\Ein{}}$ and $\sreg$.

Set $\sub{\EinG{-1}} = 0$.

The algorithm returns sequentially for $k=-1,0,1,2,\ldots$
an $\rR$-bilinear map
\[
\sub{\bil{k}} \;\;:\;\; \sub{\EinG{k}} \;\times\; \sub{\Ein{k+1}} \;\to\; \rR
\]
such that we have:
\begin{itemize}
\item[\kk{\suboxsp}{k}]
$\sub{\EinG{k}} \times \sub{\EinG{k}} \to \rR,
\;(x,y) \mapsto \sub{\bil{k}}(x,\cre_cy)$ is symmetric
for all $c \in \sreg$, and positive definite for all
$c \in  \VVpos \cap \sreg$.
\item[\kk{\suboxds}{k}]
Define $\sub{\EinG{k+1}}\subset\sub{\Ein{k+1}}$ by
\[ \sub{\bil{k}}(\sub{\EinG{k}},\sub{\EinG{k+1}}) = 0 \]
Then for all $c \in
 \VVpos \cap \sreg$,
 \[ \sub{\Ein{k+1}} = \cre_c(\sub{\EinG{k}}) \oplus \sub{\EinG{k+1}} \]
is an internal direct sum decomposition,
and $\cre_c: \sub{\EinG{k}} \to \sub{\Ein{k+1}}$ is injective.
\item[\kk{\suboxff}{k}] $\sub{\EinG{k+1}}$ is \ff.
\end{itemize}

\subsection{The generalized algorithm (outline)}

The algorithm is completely analogous to
the one in \cite{RT}, except that everything takes
place in the  $\sreg$-\dcube~$\sub{\D{}}$.
One always restricts to $c \in \sreg$
so that one stays in the \dcube~at all intermediate steps.
The $234$-condition is used in key places to show that
things descend from $\sub{\D{}}$ to $\sub{\Ein{}}$.

Note that \kk{\suboxsp}{k} implies \kk{\suboxds}{k},
and \kk{\suboxds}{k-1} and \kk{\suboxds}{k} imply \kk{\suboxff}{k}.
Hence one only has to
show \kk{\suboxsp}{k} at each step.

The map $\km$ in \cite{RT} satisfies 
$\km(\sub{\IWeyl{3}}) \subset \sub{\D{1}}$
by the $234$-condition and since we have an
$\sreg$-\dcube.
Similar to \cite{RT} we denote by $\sub{\DG{k}}\subset \sub{\D{k}}$
the preimage of $\sub{\EinG{k}}\subset \sub{\Ein{k}}$ under
$\sub{\D{k}}\to \sub{\Ein{k}}$.
Therefore $\sub{\IWeyl{k}} \subset \sub{\DG{k}}$
and $\sub{\EinG{k}} =
\sub{\DG{k}}/\sub{\IWeyl{k}}$. 
Set $\sub{\EinG{-1}}=0$ and $\sub{\bil{-1}}=0$.

Set $\sub{\bil{0}} = \BG{0}$
on $\sub{\DG{0}} \times \sub{\D{1}}$.
It descends to $\sub{\EinG{0}}\times \sub{\Ein{1}}$
and satisfies \kk{\suboxsp}{0}.

Set $\sub{\bil{1}} = \BG{1}$
on $\sub{\DG{1}} \times \sub{\D{2}}$.
It descends to $\sub{\EinG{1}}\times \sub{\Ein{2}}$
and satisfies \kk{\suboxsp}{1}.

Define $\sub{\auxx{2}}: \sub{\DG{2}}\to \sub{\D{0}}$ by
$\BG{1}(\sub{\D{1}},(\je\,\sub{\auxx{2}}+\mathbbm{1})(\sub{\DG{2}}))=0$.
To show the existence of a unique such map, one uses
the fact that $\VVpos \cap \sreg$ is nonempty.
Set $\sub{\Auxx{2}} = \je\,\sub{\auxx{2}} + \mathbbm{1}: \sub{\DG{2}} \to (\VV)^{\otimes 2}\DERIV$.
Then $\sub{\Auxx{2}}=\mathbbm{1}$ on $\sub{\IWeyl{2}}$.
Set $\sub{\bil{2}} = \BG{2}(\sub{\Auxx{2}}(\,\cdot\,),\,\cdot\,)$
on $\sub{\DG{2}} \times \sub{\D{3}}$. 
For all $x,y \in \sub{\DG{2}}$ and $c \in \sreg$
we have
$\sub{\bil{2}}(x,\cre_cy)=\BG{2}(\sub{\Auxx{2}}(x),c \sub{\Auxx{2}}(y))$;
one instance where having a \dcube~is essential.
It descends
to $\sub{\EinG{2}}\times \sub{\Ein{3}}$;
one instance where the $234$-condition is essential.
It satisfies 
\kk{\suboxsp}{2}.

Define $\sub{\auxx{3}}: \sub{\DG{3}}\to \sub{\D{1}}$ by
$\BG{2}(
\je\,\sub{\D{0}}\oplus \sub{\D{2}},(\je\,\sub{\auxx{3}}+\mathbbm{1})(\sub{\DG{3}}))=0$.
Set $\sub{\Auxx{3}} = \je\,\sub{\auxx{3}} + \mathbbm{1}:
\sub{\DG{3}} \to (\VV)^{\otimes 3}\DERIV$.
Then $\sub{\Auxx{3}}=\je\km + \mathbbm{1}$ on $\sub{\IWeyl{3}}$;
one of several instances where
$\km(\sub{\IWeyl{3}}) \subset \sub{\D{1}}$ is used.
Set $\sub{\bil{3}} = \BG{3}(\sub{\Auxx{3}}(\,\cdot\,),\,\cdot\,)$
on $\sub{\DG{3}} \times \sub{\D{4}}$.
For all $x,y \in \sub{\DG{3}}$ and $c \in \sreg$
we have 
$\sub{\bil{3}}(x,\cre_cy)=\BG{3}(\sub{\Auxx{3}}(x),c \sub{\Auxx{3}}(y))$.
It descends to
$\sub{\EinG{3}}\times \sub{\Ein{4}}$.
It satisfies \kk{\suboxsp}{3}.

We show that $\sub{\EinG{4}} = 0$.
Pick a $c \in \VVpos \cap \sreg$.
We have\footnote{%
By $\cre_c(\sub{\D{4}})=0$
and anticommutation relations and having a \dcube.
}
$\cre_c(\sub{\D{3}}) = \sub{\D{4}}$. 
Hence $\cre_c(\sub{\Ein{3}}) = \sub{\Ein{4}}$.
But we also have
$\cre_c(\sub{\Ein{3}}) = \cre_c(\sub{\EinG{3}})$
by \kk{\suboxds}{2}.
Hence $\sub{\Ein{4}} = \cre_c(\sub{\EinG{3}})$,
and now \kk{\suboxds}{3} implies
$\sub{\EinG{4}} = 0$.


\section{Filtration by \ecubes}

In this section we use the letter $\bkl{}{}$ to denote
an unspecified filtration.
The same letter is going to stand for a particular
filtration, later in this paper.


\subsection{Definition of a filtration by $\reg$-\ecubes}\label{sec:deff}

Suppose
\begin{itemize}
\item $\reg$ satisfies \sref{reg} with $n\geq 1$ and $c_1 \in \VVpos$.
\end{itemize}
We say that a collection $(\bkl{\alpha}{\Ein{}})_{\alpha}$,
where the index $\alpha$ runs over a power of $\Z_{\geq 0}$,
is a filtration by $\reg$-\ecubes~if and only if the following conditions hold:
\begin{itemize}
\item $\bkl{\alpha}{\Ein{}} \subset \Ein{}$ is a graded submodule.
\item $\bkl{\alpha}{\Ein{}} \subset \bkl{\beta}{\Ein{}}$
if $\alpha \leq \beta$.
\item
$\eb{\bkl{\alpha}{\Ein{}}}{\bkl{\beta}{\Ein{}}}
\subset \bkl{\alpha+\beta}{\Ein{}}$.
\item $\exists \alpha: \bkl{\alpha}{\Ein{}} = \Ein{}$.
\item $\bkl{\alpha}{\Ein{}} \cap \bkl{\beta}{\Ein{}}
\subset \sum_{\gamma \leq \alpha\text{ and }\gamma \leq \beta}
\bkl{\gamma}{\Ein{}}$.
\item $\bkl{\alpha}{\Ein{}}$ is an $\reg$-\ecube.
\item $\bkl{\alpha}{\Ein{}}$ has a representative that
satisfies, with $\reg$, the $234$-condition\footnote{%
One may want to consider a choice of such representatives as being
part of the filtration.
}.
\item $\unk(\rR) \subset \reg$ for all $\unk \in \bkl{0}{\Ein{1}}$.
\end{itemize}
The last three are motivated by \sref{genalg}.


\subsection{Remarks: Constructing such a filtration}\label{sec:rem1}

Here is a rough strategy for constructing a filtration as in \sref{deff}.
There is of course no guarantee that this strategy works,
it is just a strategy.

Choose $\reg$. Calculate $\gvac{\reg} = \fock{\D{0}}{\regperp}$.

Choose $\bkl{\alpha}{\gvac{\reg}}
\subset \gvac{\reg}$ and set $\bkl{\alpha}{\D{}}
= \fock{\bkl{\alpha}{\gvac{\reg}}}{\reg}$.

Check that $\db{\bkl{\alpha}{\D{}}}{\bkl{\beta}{\D{}}}
\subset \bkl{\alpha+\beta}{\D{}} + \IWeyl{}$;
one can use the bracket estimate \sref{BRACKETing}.

Check that $(\reg,\bkl{\alpha}{\D{}})$ satisfies the $234$-condition.

Check that $\bkl{\alpha}{\Ein{}} = (\bkl{\alpha}{\D{}}+\IWeyl{})/\IWeyl{}$
satisfies all conditions in \sref{deff}.


\subsection{Remarks: Intersecting two such filtrations}\label{sec:rem2}
Suppose $\filp{}{},\filpp{}{}$ are two filtrations
as in \sref{deff} relative to the same $\reg$.
Set
\[
\bkl{\alpha'\alpha''}{\Ein{}}
\;=\; \filp{\alpha'}{\Ein{}}\cap \filpp{\alpha''}{\Ein{}}
\]
where $\alpha'\alpha''$ is a shorthand for a pair of indices\footnote{%
$(\alpha'\alpha'' \leq \beta'\beta'')
\Longleftrightarrow
(\alpha'\leq\beta' \text{ and } \alpha''\leq\beta'')$
and 
$\alpha'\alpha'' + \beta'\beta''
=
(\alpha'+\beta')(\alpha''+\beta'')$
}. Then $\bkl{}{}$ automatically
satisfies many, but not necessarily all,
properties in \sref{deff} relative to $\reg$.

In particular, $\bkl{\alpha'\alpha''}{\Ein{}}$
is not obviously an $\reg$-\ecube.
The
 intersection of a representative
of $\filp{\alpha'}{\Ein{}}$ with a representative of
$\filpp{\alpha''}{\Ein{}}$ is not obviously
a representative of $\bkl{\alpha'\alpha''}{\Ein{}}$.
See \sref{ebucintersection} for details.


\newcommand{\contdisc}{%
Given is a decomposition $V = V_- \oplus V_+$. See \eqref{dxsul}.}

\nextpart{Structure 
associated to a decomposition of $V$}

Suppose a direct sum decomposition
of $V$ is given,
\begin{equation}\label{dxsul}
V = V_- \oplus V_+
\end{equation}
where both summands are \ff~$\rC$-modules of rank $1$.

In \sref{xsdfil}
we construct a filtration by \ecubes~associated
to \eqref{dxsul}.
The sections leading up to \sref{xsdfil}
are to get acquainted with \eqref{dxsul}.


\section{Associated decomposition of $\D{}$} \label{sec:decddd}
\contdisc

\subsection{Notation}\label{sec:grdnotation}

We introduce gradings on various $\rC$-modules, using the notation
\[
\text{(module)}
\;=\;\textstyle\bigoplus_{i,j\in \half\Z}
\grd{i,j}{
\text{(module)}}
\]
In all figures below, $i$ is constant on vertical lines
and increases towards the right in steps of $\half$,
whereas the index $j$ is constant on horizontal lines
and increases towards the top.
No bullet means that the corresponding component
vanishes. The presence of a bullet means that the corresponding
component has rank $\geq 1$.

We also define coarser gradings:
for all $\ell \in \Z$ we set
$\grd{\ell}{} = \bigoplus_{i-j=\ell}\grd{i,j}{}$
and we set
$\grd{\text{$\even$ ($\odd$)}}{}
= \bigoplus_{\text{$\ell$ even (odd)}}\grd{\ell}{}$.

Whenever defined, conjugation $\CONJ$ maps $\grd{i,j}{}\to \grd{j,i}{}$.

The grading on a tensor product is always the tensor product grading.

\subsection{Grading on $\rC$ and $V$ and $\cc{V}$}
Set
\begin{align*}
\grd{0,0}{\rC} & = \rC & 
\grd{-\half,0}{V} & = V_- &
\grd{0,-\half}{\cc{V}} & = \cc{V_-}\\
&&\grd{\half,0}{V} & = V_+ &
\grd{0,\half}{\cc{V}} & = \cc{V_+}
\end{align*}
and set all other components to zero:
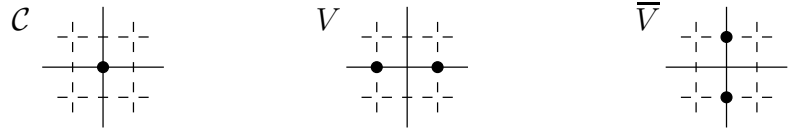
\begin{center}
\input{grstart.pstex_t}
\end{center}%

\subsection{Grading on the free algebra $\Lang$}
The grading on $\rC \oplus V \oplus \cc{V}$ induces
a grading on $\Lang$.

For example,
$\VV \subset \Lang$ has
nontrivial components:
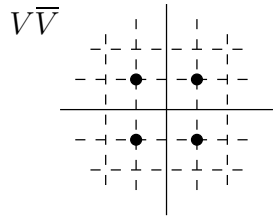
\begin{center}
\begin{picture}(0,0)%
\includegraphics{grvv.pstex}%
\end{picture}%
\setlength{\unitlength}{4144sp}%
\begingroup\makeatletter\ifx\SetFigFont\undefined%
\gdef\SetFigFont#1#2#3#4#5{%
  \reset@font\fontsize{#1}{#2pt}%
  \fontfamily{#3}\fontseries{#4}\fontshape{#5}%
  \selectfont}%
\fi\endgroup%
\begin{picture}(1287,1284)(10606,-4033)
\put(10621,-2941){\makebox(0,0)[lb]{\smash{{\SetFigFont{12}{14.4}{\rmdefault}{\mddefault}{\updefault}{\color[rgb]{0,0,0}$\mathllap{V\cc{V}}$}%
}}}}
\end{picture}%

\end{center}%

For example,
the lower right bullet is $\grd{\half,-\half}{\VV} = V_+\cc{V_-}$.

\subsection{Grading on $\bigwedge \VV$}\label{sec:vvgrad}
Nontrivial components:
\begin{center}
\begin{picture}(0,0)%
\includegraphics{grext.pstex}%
\end{picture}%
\setlength{\unitlength}{4144sp}%
\begingroup\makeatletter\ifx\SetFigFont\undefined%
\gdef\SetFigFont#1#2#3#4#5{%
  \reset@font\fontsize{#1}{#2pt}%
  \fontfamily{#3}\fontseries{#4}\fontshape{#5}%
  \selectfont}%
\fi\endgroup%
\begin{picture}(1287,1284)(10606,-4033)
\put(10621,-2941){\makebox(0,0)[lb]{\smash{{\SetFigFont{12}{14.4}{\rmdefault}{\mddefault}{\updefault}{\color[rgb]{0,0,0}$\mathllap{\textstyle\bigwedge V\cc{V}}$}%
}}}}
\end{picture}%

\end{center}%

For example, the center bullet and the rightmost bullet are given by
\begin{align*}
\grd{0,0}{(\textstyle\bigwedge \VV)}
& = 
\rC \oplus (V_+\cc{V_+} \wedge V_-\cc{V_-})
\oplus (V_+\cc{V_-} \wedge V_-\cc{V_+})
\oplus (\textstyle\bigwedge^4 \VV)\\
\grd{1,0}{(\textstyle\bigwedge \VV)} & = V_+\cc{V_-} \wedge V_+\cc{V_+}
\end{align*}

\subsection{Grading on $\DERIV$}\label{sec:ouefdldf}
Define
\[
\grd{i,j}{\DERIV}
\; = \; \big\{
\delta \in \DERIV \mid 
\forall i',j':\;
\delta(\grd{i',j'}{\Lang}) \subset \grd{i+i',j+j'}{\Lang} \big\}
\]
Then
$\DERIV = \bigoplus_{i,j \in \half\Z} \grd{i,j}{\DERIV}$.
Nontrivial components:
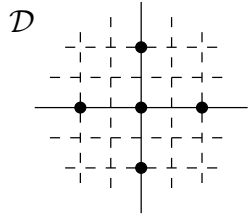
\begin{center}
\begin{picture}(0,0)%
\includegraphics{grderiv.pstex}%
\end{picture}%
\setlength{\unitlength}{4144sp}%
\begingroup\makeatletter\ifx\SetFigFont\undefined%
\gdef\SetFigFont#1#2#3#4#5{%
  \reset@font\fontsize{#1}{#2pt}%
  \fontfamily{#3}\fontseries{#4}\fontshape{#5}%
  \selectfont}%
\fi\endgroup%
\begin{picture}(1287,1284)(10606,-4033)
\put(10621,-2941){\makebox(0,0)[lb]{\smash{{\SetFigFont{12}{14.4}{\rmdefault}{\mddefault}{\updefault}{\color[rgb]{0,0,0}$\mathllap{\DERIV}$}%
}}}}
\end{picture}%

\end{center}%

The four outer bullets
are all in  $\DERIVVERT$. 

Center bullet:
$\delta \in \grd{0,0}{\DERIV}$
if and only if
$\delta(\rC) \subset \rC$,
$\delta(V_{\pm}) \subset V_{\pm}$,
$\delta(\cc{V_{\pm}}) \subset \cc{V_{\pm}}$,
hence $\rank_{\rC}\grd{0,0}{\DERIV} = 4 + \stdim$.
Rightmost bullet:
$\delta \in \grd{1,0}{\DERIV}$
if and only if $\delta(\rC)=\delta(V_+) = 
\delta(\cc{V_\pm}) = 0$ and
$\delta(V_-) \subset V_+$,
hence $\rank_{\rC} \grd{1,0}{\DERIV} = 1$.

\subsection{Grading on the complex graded Lie algebra $\DC{}$}\label{sec:hajahkhsk}

By combining \sref{vvgrad}
and \sref{ouefdldf} we get:
\begin{center}
\begin{picture}(0,0)%
\includegraphics{grgla.pstex}%
\end{picture}%
\setlength{\unitlength}{4144sp}%
\begingroup\makeatletter\ifx\SetFigFont\undefined%
\gdef\SetFigFont#1#2#3#4#5{%
  \reset@font\fontsize{#1}{#2pt}%
  \fontfamily{#3}\fontseries{#4}\fontshape{#5}%
  \selectfont}%
\fi\endgroup%
\begin{picture}(2052,1824)(6601,-4303)
\put(6616,-2761){\makebox(0,0)[lb]{\smash{{\SetFigFont{12}{14.4}{\rmdefault}{\mddefault}{\updefault}{\color[rgb]{0,0,0}$\mathllap{\DC{}}$}%
}}}}
\end{picture}%

\end{center}%

One actually has a 3-grading,
\[
\DC{} = \textstyle\bigoplus_k\bigoplus_{i,j\in \half\Z} \grd{i,j}{\DC{k}}
\]
Hence the figure is really an overlay of five figures,
one for each of $\DC{0},\ldots,\DC{4}$.
Each of the four corner bullets has rank 1 and is contained in $\DC{2}$.
Each of the eight corner-neighbors
has rank 2: one from $\DC{1}$, one from $\DC{3}$.

By construction, the grading respects the bracket:
\[
\db{
\grd{i,j}{\DC{}}}{
\grd{i',j'}{\DC{}}}
\;\subset\;
\grd{i+i',j+j'}{\DC{}}
\]

The coarser grading
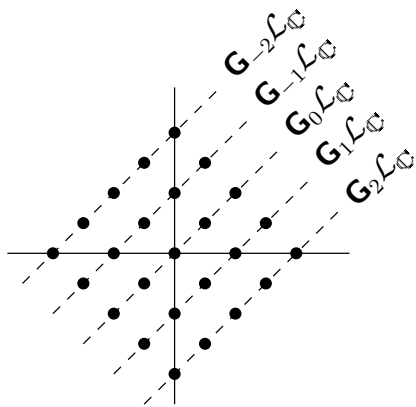
\begin{center}
\input{grglared.pstex_t}
\end{center}%

also respects the bracket,
$\db{
\grd{\ell}{\DC{}}}{
\grd{\ell'}{\DC{}}}
\subset
\grd{\ell+\ell'}{\DC{}}$.

Conjugation, defined in \sref{hdkhkdhiooao333}, reflects about the diagonal:
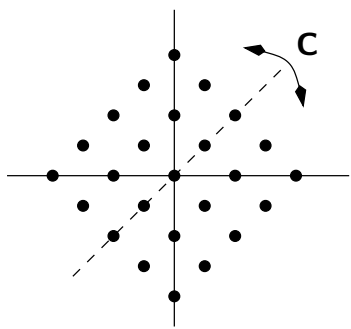
\begin{center}
\begin{picture}(0,0)%
\includegraphics{grconj.pstex}%
\end{picture}%
\setlength{\unitlength}{4144sp}%
\begingroup\makeatletter\ifx\SetFigFont\undefined%
\gdef\SetFigFont#1#2#3#4#5{%
  \reset@font\fontsize{#1}{#2pt}%
  \fontfamily{#3}\fontseries{#4}\fontshape{#5}%
  \selectfont}%
\fi\endgroup%
\begin{picture}(2049,1914)(6649,-4303)
\put(8371,-2671){\makebox(0,0)[lb]{\smash{{\SetFigFont{12}{14.4}{\rmdefault}{\mddefault}{\updefault}{\color[rgb]{0,0,0}$\CONJ$}%
}}}}
\end{picture}%

\end{center}%


\subsection{Grading on $\D{}$\\
that respects the bracket only in a restricted sense}\label{sec:realglagr}
Recall that $\D{}\subset \DC{}$ is the subset of real elements.
Since this is an $\rR$-module, the notation here does
not quite agree with \sref{grdnotation}.
Define
\[
\grd{i,j}{\D{}}
\;=\;
\{
\text{real elements of $
\grd{i,j}{\DC{}}
+ 
\grd{j,i}{\DC{}}$}
\}
\]
Then
$\grd{i,j}{\D{}} = \grd{j,i}{\D{}}$
and
$\D{} = \bigoplus_{i \geq j} \grd{i,j}{\D{}}$.
Hence we restrict figures to $i\geq j$:
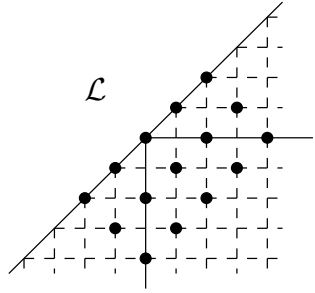
\begin{center}
\begin{picture}(0,0)%
\includegraphics{grrealgla.pstex}%
\end{picture}%
\setlength{\unitlength}{4144sp}%
\begingroup\makeatletter\ifx\SetFigFont\undefined%
\gdef\SetFigFont#1#2#3#4#5{%
  \reset@font\fontsize{#1}{#2pt}%
  \fontfamily{#3}\fontseries{#4}\fontshape{#5}%
  \selectfont}%
\fi\endgroup%
\begin{picture}(1824,1734)(6829,-4303)
\put(7426,-3166){\makebox(0,0)[lb]{\smash{{\SetFigFont{12}{14.4}{\rmdefault}{\mddefault}{\updefault}{\color[rgb]{0,0,0}$\mathllap{\D{}}$}%
}}}}
\end{picture}%

\end{center}%

This grading respects the bracket only in a restricted sense:
\[
\db{\grd{i,j}{\D{}}}{\grd{i',j'}{\D{}}}
\;\;\subset\;\;
\grd{i+i',j+j'}{\D{}}
\;+\;
\grd{i+j',j+i'}{\D{}}
\]
In particular the bracket respects
$\grd{}{\D{}} = \grd{\even}{\D{}}\oplus \grd{\odd}{\D{}}$
as a $\Z_2$-grading.

\section{Associated decomposition of $\IWeyl{}$} \label{sec:oshohssnlls}

\contdisc

In this section we use bases $V_{\pm} = \rC v_{\pm}$.
This is a matter of convenience;
final results are $v_{\pm}$-independent, as we repeatedly point out.

\subsection{Creation and annihilation
operators $\cre_{\pm}$ and $\ann_{\pm}$}\label{sec:dsohlald}
Abbreviate
\begin{align*}
\cre_{+} & = \cre_{v_+\cc{v_+}} & \ann_{+} & = \ann_{v_+\cc{v_+}} \\
\cre_{-} & = \cre_{v_-\cc{v_-}} & \ann_{-} & = \ann_{v_-\cc{v_-}}
\end{align*}
We use these operators in a way that does not
depend on the choice of $v_{\pm}$, nor on the
choice of the auxiliary volume form hidden in $\ann_{\pm}$.
Note that
\begin{align*}
\cre_+\ann_+ + \ann_+\cre_+ \;=\; \cre_-\ann_- + \ann_-\cre_- &\;=\;0\\
\cre_+\ann_- + \ann_-\cre_+ \;=\; \cre_-\ann_+ + \ann_+\cre_- 
&\;=\; 
\text{(an invertible element of $\rC$)}\,\mathbbm{1}
\end{align*}
In addition to these, the anticommutator
of any two creation operators vanishes,
as does the anticommutator of any two
annihilation operators.
The creation and annihilation operators
shift the $\grd{}{}$-grading on $\DC{}$ as follows:
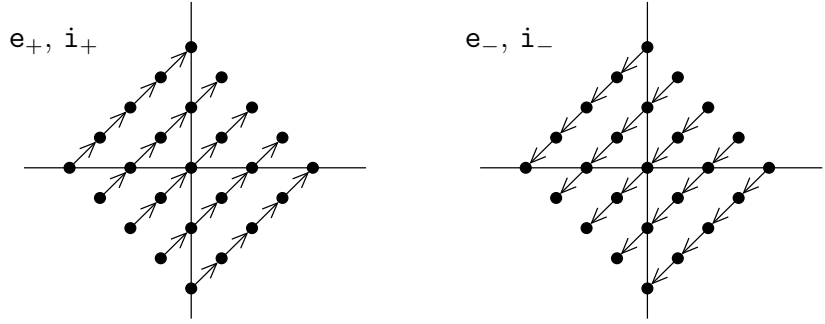
\begin{center}
\input{grglacre.pstex_t}
\end{center}%

These figures do not show
$\cre_{\pm}(\DC{k}) \subset \DC{k+1}$ and $\ann_{\pm}(\DC{k})
\subset \DC{k-1}$.

\subsection{Decomposition of the ideal $\IWeylC{}$}\label{sec:idhksv}
Set $\IWeylC{} = \IWeyl{} \oplus i\IWeyl{}$,
a $\rC$-submodule and ideal of the graded Lie algebra $\DC{}$.
It has an alternative decomposition
into $\rC$-modules,
$\IWeylC{} = \IWeylX{}\oplus \CONJ \IWeylX{}$
with
\[
\IWeylX{} \;=\; \{x \in \IWeylC{}\mid x(\cc{V}) = 0 \}
\]
In this \sref{idhksv} we decompose $\IWeylX{}$.
This decomposition immediately induces a decomposition of $\IWeyl{}$,
via the $\rR$-module isomorphism
$\RE : \IWeylX{} \to \IWeyl{}$.

For each $z \in \C$ abbreviate
$v_z = v_- + zv_+$.

Define $\qq{z} \in \DC{2}$ by $\qq{z}(\rC)=\qq{z}(\cc{V})=0$ and
\[
\qq{z}(v_{w}) \;=\; (w-z)
(v_z\cc{v_+} \wedge v_z \cc{v_-})v_z
\]
To extract $\qq{z}(v_{\pm})$ from this definition,
expand both sides as linear polynomials in $w$
and compare coefficients.
It is immediate that\footnote{%
By definition $\qq{z}(\rC)=\qq{z}(\cc{V})
= 0$, and
$\qq{z}(V)$ is contained in the submodule of $\VV\VV V$
symmetric in the three $V$'s.
To check $\qq{z}(V\wedge V)=0$, note that
$\qq{z}(v_{w}v_{w'})
=
(w+w'-2z)(v_z\cc{v_+} \wedge v_z \cc{v_-})v_zv_z
+
(w-z)(w'-z)(v_z\cc{v_+} \wedge v_z \cc{v_-})(v_zv_++v_+v_z)$
is symmetric in $w,w'$.
}
$\qq{z} \in \IWeylX{2}$.

Since $\qq{z}$ is polynomial in $z$ of degree four,
we can assign names to its coefficients\footnote{%
We are abusing notation here, e.g.~$\qq{-2}$ is not
equal to $\qq{z}$
evaluated at $z=-2$.
},
$\qq{z}=
\qq{-2} + z\qq{-1} + z^2\qq{0} + z^3\qq{1} + z^4 \qq{2}$.
The rank 1 modules
$\QQ{j} = \rC \qq{j}$
are $v_{\pm}$-independent\footnote{
By contrast, $\rR \qq{j}$ does depend on $v_{\pm}$.
The one exception is $j=0$, in fact
$\QQ{0} = \rR \qq{0} \oplus i\rR \qq{0}$
is a $v_{\pm}$-independent
decomposition into two modules of $\rR$-rank 1.}
and we have
\[
\IWeylX{2} \;=\;
\QQ{-2}\,\oplus\,
\QQ{-1}\,\oplus\,
\QQ{0}\,\oplus\,
\QQ{1}\,\oplus\,
\QQ{2}
\]
with $\rank_{\rC} \IWeylX{2} = 5$.
We have:
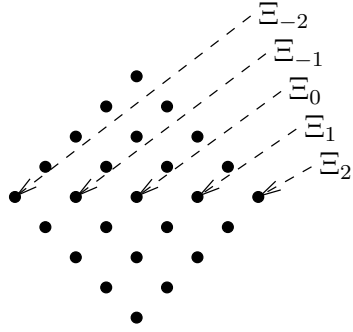
\begin{center}
\input{gri2.pstex_t}
\end{center}%

where an arrow means `contained in'.
Note that the
$\CONJ \QQ{i}$
are on the vertical axis;
that $\QQ{0}
\cap \CONJ \QQ{0} = 0$;
and that $\QQ{\pm 2}$
and $\CONJ \QQ{\pm 2}$
are equal to the four corner bullets.

Note that $\cre_{\pm} \QQ{i} \subset \IWeylX{3}$
and $\cre_+\cre_-\QQ{i} \subset \IWeylX{4}$.
We have
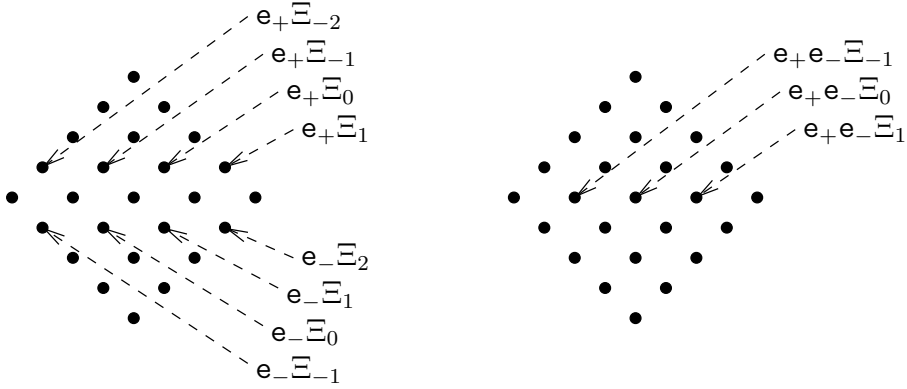
\begin{center}
\input{gri3.pstex_t}
\end{center}%

where an arrow means `contained in'.
Note that $\cre_-\QQ{-2} = \cre_+\QQ{2}=0$.
We have
\begin{align*}
\IWeylX{3} &\;=\;
\cre_-\QQ{-1}\oplus
\cre_-\QQ{0}\oplus
\cre_-\QQ{1}\oplus
\cre_-\QQ{2}
\oplus
\cre_+\QQ{-2}\oplus
\cre_+\QQ{-1}\oplus
\cre_+\QQ{0}\oplus
\cre_+\QQ{1}\\
\IWeylX{4} & \;=\;
\cre_+\cre_-\QQ{-1}\oplus
\cre_+\cre_-\QQ{0}\oplus
\cre_+\cre_-\QQ{1}
\end{align*}
where each summand has rank 1, hence
$\rank_{\rC}\IWeylX{3} = 8$
and
$\rank_{\rC}\IWeylX{4} = 3$.


\subsection{Grading on the complex graded Lie algebra $\EinC{}$}
Set $\EinC{} = \DC{}/\IWeylC{}$.
The nontrivial components of the $\grd{}{}$-grading
on $\EinC{}$ are
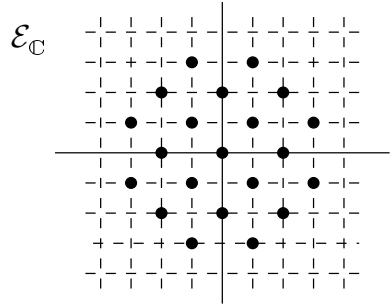
\begin{center}
\begin{picture}(0,0)%
\includegraphics{grglaein.pstex}%
\end{picture}%
\setlength{\unitlength}{4144sp}%
\begingroup\makeatletter\ifx\SetFigFont\undefined%
\gdef\SetFigFont#1#2#3#4#5{%
  \reset@font\fontsize{#1}{#2pt}%
  \fontfamily{#3}\fontseries{#4}\fontshape{#5}%
  \selectfont}%
\fi\endgroup%
\begin{picture}(2052,1824)(6601,-4303)
\put(6616,-2761){\makebox(0,0)[lb]{\smash{{\SetFigFont{12}{14.4}{\rmdefault}{\mddefault}{\updefault}{\color[rgb]{0,0,0}$\mathllap{\EinC{}}$}%
}}}}
\end{picture}%

\end{center}%

Each of the eight corner bullets has rank 1 and is contained in $\EinC{1}$.


\section{Associated filtration by \ecubes}\label{sec:xsdfil}

\contdisc

We construct a filtration as in \sref{deff},
indexed by $\Z_{\geq 0}$.
We loosely follow \sref{rem1}.

\subsection{%
Invariance condition
}\label{sec:breq}

We require that the filtration 
that we are about to construct
be invariant, in the sense of \sref{vgt},
under the subgroup
\[
\big\{
\vtrsf \in \Aut_{\rC}(V) \mid
\vtrsf(V_{\pm}) \subset V_{\pm}\text{ or }
\vtrsf(V_{\pm}) \subset V_{\mp}
\big\}
\]

\subsection{Choice of $\reg$}\label{sec:dkfhlskkka}
Set\footnote{%
The numbering is slightly 
inconsistent with \sref{reg}. This choice
is more in line with standard conventions: Pauli matrices etc.
}
\begin{align*}
\reg \;&=\; (V_-\cc{V_-} \oplus V_+\cc{V_+})_{\text{real}}\\
&=\; \rR c_0 \oplus \rR c_3
\end{align*}
where an explicit orthogonal basis of $\VVreal$ as in \sref{reg} is given by
\begin{align*}
c_0 & \;=\; v_+\cc{v_+} + v_-\cc{v_-}
& c_1 & \;=\; v_+\cc{v_-} + v_-\cc{v_+}\\
c_3 & \;=\; v_+\cc{v_+} - v_-\cc{v_-}
& c_2 & \;=\; i(v_+\cc{v_-} - v_-\cc{v_+})
\end{align*}
Note that $c_0 \in \VVpos$.

Then $\cre_{\reg}
= \rR\cre_-+\rR \cre_+$ and $\cre_{\regperp}
= \rR\cre_1 + \rR \cre_2$
 are $v_{\pm}$-independent.

\subsection{The vacuum $\gvac{\reg}$}

We have
\begin{align*}
\gvac{\reg}
& = \fock{\D{0}}{\regperp}\\
& = \D{0}
\;\oplus\;
\cre_{\regperp}\D{0}
\;\oplus\;
 (\cre_{\regperp})^2\D{0}
\end{align*}
Recall from \sref{realglagr} the grading of $\D{}$:
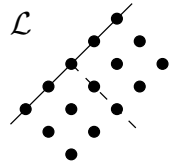
\begin{center}
\begin{picture}(0,0)%
\includegraphics{omegat.pstex}%
\end{picture}%
\setlength{\unitlength}{4144sp}%
\begingroup\makeatletter\ifx\SetFigFont\undefined%
\gdef\SetFigFont#1#2#3#4#5{%
  \reset@font\fontsize{#1}{#2pt}%
  \fontfamily{#3}\fontseries{#4}\fontshape{#5}%
  \selectfont}%
\fi\endgroup%
\begin{picture}(952,951)(7459,-2800)
\put(7606,-2041){\makebox(0,0)[lb]{\smash{{\SetFigFont{12}{14.4}{\rmdefault}{\mddefault}{\updefault}{\color[rgb]{0,0,0}$\mathllap{\D{}}$}%
}}}}
\end{picture}%

\end{center}%

The three components of $\gvac{\reg}$ are then graded as follows:
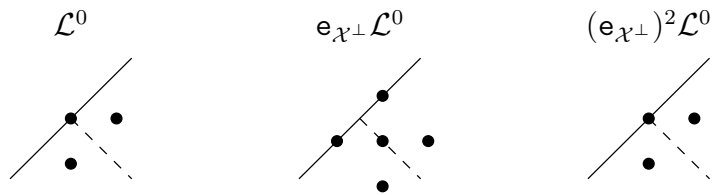
\begin{center}
\input{omega.pstex_t}
\end{center}%

Note that $\cre_{\regperp}$ shifts parallely to the dashed line.
Exchanging $V_-$ and $V_+$ corresponds to
reflection about the dashed line.

\newcommand{\dshort}{m}
The $\rR$-ranks are given by:
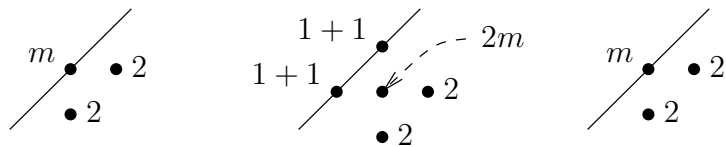
\begin{center}
\input{omegaranks.pstex_t}
\end{center}%

where $\dshort = 4 + \stdim$. Somewhat informal remarks:
\begin{itemize}
\item The three bullets of rank $\dshort,2\dshort,\dshort$
have invariant submodules of ranks $4,8,4$ respectively,
obtained by intersecting with $\DERIVVERT$.
These vertical submodules
have further invariant decompositions.
\item The six bullets of rank 2
are contained in $\DERIVVERT$
and are irreducible.
\item The bullets of rank $1+1$
are contained in $\DERIVVERT$ and have
invariant decomposition
$\rR \RE \tju_{\pm} \oplus \rR \IM \tju_{\pm}$
with $\tju_{\pm} \in \DC{1}$ given by
$\tju_{\pm}(\rC)=\tju_{\pm}(v_{\pm})=\tju_{\pm}(\cc{V})=0$
and $\tju_+(v_-) = (v_-\cc{v_+})v_+$
and $\tju_-(v_+) = (v_+\cc{v_-})v_-$.
\end{itemize}


\subsection{Choice of $\fil{0}{\D{}}$}\label{sec:ch0}

Set $\fil{0}{\D{}} = \fock{\fil{0}{\gvac{\reg}}}{\reg}$
where $\fil{0}{\gvac{\reg}} \subset \gvac{\reg}$ is the direct sum of:
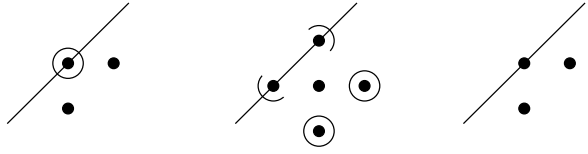
\begin{center}
\begin{picture}(0,0)%
\includegraphics{omegafilt00.pstex}%
\end{picture}%
\setlength{\unitlength}{4144sp}%
\begingroup\makeatletter\ifx\SetFigFont\undefined%
\gdef\SetFigFont#1#2#3#4#5{%
  \reset@font\fontsize{#1}{#2pt}%
  \fontfamily{#3}\fontseries{#4}\fontshape{#5}%
  \selectfont}%
\fi\endgroup%
\begin{picture}(3444,874)(4039,-2723)
\end{picture}%

\end{center}%

A circle around a bullet means that we include that entire
bullet in $\fil{0}{\gvac{\reg}}$.
The semicircles mean that we include $\rR \RE \tju_{\pm}$
but not $\rR \IM \tju_{\pm}$.

This choice is consistent with  \sref{breq}.
Note that $\fil{0}{\D{}}\subset \grd{\even}{\D{}}$.

We now check that $\fil{0}{\D{}}$ is a subalgebra modulo the ideal,
\[
\db{\fil{0}{\D{}}}{\fil{0}{\D{}}}
\subset \fil{0}{\D{}} + \IWeyl{}
\]
By the bracket estimate in \sref{BRACKETing} it suffices to
check two things:
\begin{itemize}
\item 
$\wedgeall(x(\reg)) \subset \bigwedge \reg$
for all $x \in \fil{0}{\gvac{\reg}}$.
\item 
$\db{\fil{0}{\gvac{\reg}}}{\fil{0}{\gvac{\reg}}}
\subset
\fock{\fil{0}{\gvac{\reg}}}{\reg} + \IWeyl{}$.
\end{itemize}
The first is by direct calculation\footnote{%
By contrast, this does not hold
for all $x \in \rR \IM \tju_{\pm}$.}. We now discuss the second.

We decompose
$\fil{0}{\gvac{\reg}} = a\oplus b \oplus c$ with local abbreviations
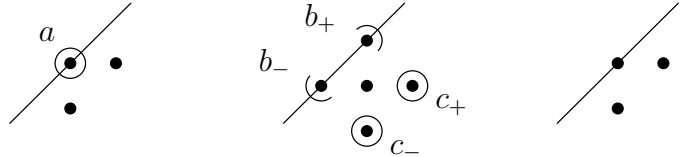
\begin{center}
\input{omegafilt0.pstex_t}
\end{center}%

and $b = b_-\oplus b_+$ and $c = c_- \oplus c_+$.
In particular $b_{\pm} = \rR \RE \tju_{\pm}$.

Then $\db{a}{a}\subset a$,
$\db{a}{b}\subset b$,
$\db{a}{c}\subset c$.
Also $\db{b}{c} \subset
\bigoplus_{i-j=2} \grd{i,j}{\D{}}
\subset \fock{c}{\reg}$,
by gradings alone.
Finally, by direct calculation\footnote{%
The last of these four is
 the sample calculation in \aref{sampleeval}.
Minor notational mismatch:
here $c_{\pm}$ are $\rR$-modules of rank $2$;
there $c_{\pm}$ are elements with parameters $\lambda_{\pm} \in \rC
= \rR \oplus i\rR$.
}:
\begin{align*}
\db{b_{\pm}}{b_{\pm}} & \;\subset\; \fock{b_{\pm}}{\reg}
&
\db{b_-}{b_+} & \;\subset\; \fock{a}{\reg}
+ \fock{b}{\reg} + \IWeyl{2}\\
\db{c_{\pm}}{c_{\pm}} & \;\subset\; \fock{b_{\pm}}{\reg}
&
\db{c_-}{c_+} & \;\subset\; \fock{a}{\reg} + \IWeyl{2}
\end{align*}
and there is in particular,
and crucially, no $\rR \IM \tju_{\pm}$ on the right hand sides.

\subsection{%
Choice of $\fil{\filind}{\D{}}$ for $p > 0$}\label{sec:sdoohds63768}

Set
\begin{align*}
\fil{1}{\D{}} & \;=\; \fil{0}{\D{}} \oplus \grd{\odd}{\D{}}\\
\fil{2}{\D{}} & \;=\; \D{}\\
\fil{3}{\D{}} & \;=\; \D{}\\
&\;\;\;\vdots
\end{align*}
This choice is consistent with \sref{breq}.
We have
\[
\db{\fil{\filind}{\D{}}}{\fil{\otherfilind}{\D{}}}
\;\subset\; \fil{\filind+\otherfilind}{\D{}} + \IWeyl{}
\]
using \sref{ch0}, including $\fil{0}{\D{}}\subset \grd{\even}{\D{}}$. 

The definition is equivalent to
$\fil{\filind}{\D{}}
= \fock{\fil{\filind}{\gvac{\reg}}}{\reg}$
where:
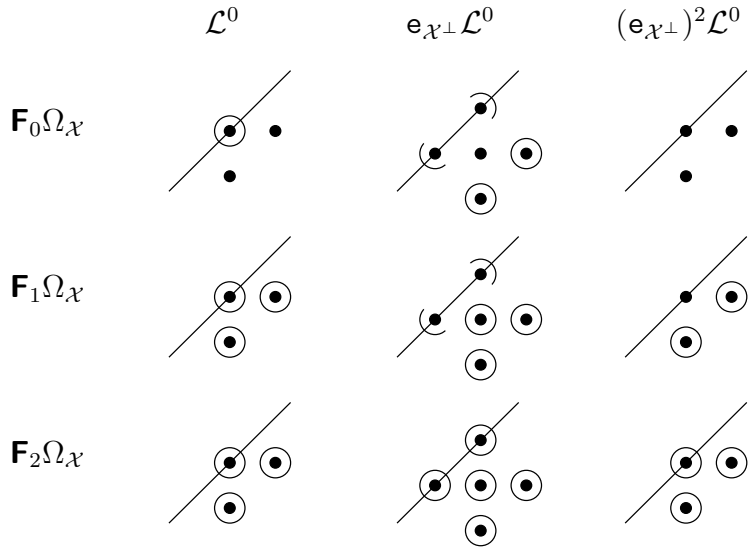
\begin{center}
\input{omegafilt.pstex_t}
\end{center}%


\subsection{%
The $234$-condition}\label{sec:djsjdojls2}

By inspection,
\begin{align*}
\fil{0}{\D{}}\cap \IWeyl{} &\;=\;
\fock{\textstyle\bigoplus_{2\leq |i|\leq 2}\RE \QQ{i}}{\reg}\\
\fil{1}{\D{}}\cap \IWeyl{} &\;=\;
\fock{\textstyle\bigoplus_{1\leq |i| \leq 2}\RE \QQ{i}}{\reg}\\
\fil{2}{\D{}}\cap \IWeyl{} &\;=\;
\fock{\textstyle\bigoplus_{0\leq |i|\leq 2}\RE \QQ{i}}{\reg}
\;=\;\IWeyl{}
\end{align*}
Hence
$(\reg,\fil{\filind}{\D{}})$ satisfies the $234$-condition
for all $p$.

\subsection{Filtration of $\Ein{}$}

Set $\fil{\filind}{\Ein{}}
= (\fil{\filind}{\D{}}+\IWeyl{})/\IWeyl{}$.
It has all the properties in \sref{deff}.


\section{Remarks that may be helpful when choosing a gauge $\gauge$}\label{sec:helpwithgauges}

\contdisc

This section is not used in the rest of this paper.

This section uses notation from \cite{RT}.
Briefly, by definition, a gauge 
is a positive definite $\rC$-Hermitian map
$\gauge:(\cc{V}\DERIV/\KERN) \times (\cc{V}\DERIV/\KERN) \to \rC$.
Here $\KERN \subset \cc{V}\DERIV$ is the subset of $x$ for which
$x(\rC)=x(V)=x(\cc{V}\wedge\cc{V})=0$
and for which $x(\cc{V}) \subset \cc{V}\cc{V}$
is symmetric in the two $\cc{V}$'s.

\subsection{Grading on $\cc{V}\DERIV$}\label{sec:dihdkhkkka22}
Nontrivial components:
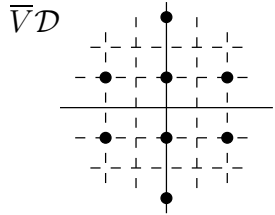
\begin{center}
\begin{picture}(0,0)%
\includegraphics{grvd.pstex}%
\end{picture}%
\setlength{\unitlength}{4144sp}%
\begingroup\makeatletter\ifx\SetFigFont\undefined%
\gdef\SetFigFont#1#2#3#4#5{%
  \reset@font\fontsize{#1}{#2pt}%
  \fontfamily{#3}\fontseries{#4}\fontshape{#5}%
  \selectfont}%
\fi\endgroup%
\begin{picture}(1287,1284)(9526,-4033)
\put(9541,-2941){\makebox(0,0)[lb]{\smash{{\SetFigFont{12}{14.4}{\rmdefault}{\mddefault}{\updefault}{\color[rgb]{0,0,0}$\mathllap{\cc{V}\DERIV}$}%
}}}}
\end{picture}%

\end{center}%

The six outer bullets have rank $1$.
The
bullets in the center have
rank $5 + \stdim$.


\subsection{Decomposition of $\KERN \subset \cc{V}\DERIV$}

The decomposition of $\KERN$
is obtained similarly to the decomposition of $\IWeylC{}$ in \sref{idhksv}.

Define
$\jj{z} \in \cc{V}\DERIV$ by $\jj{z}(\rC)=\jj{z}(V)=0$ and
\[
\jj{z}(\cc{v_{z'}}) \;=\; \cc{(z'-z)}\, \cc{v_z}\cc{v_z}
\]
Then\footnote{%
Note that
$\jj{z}(\cc{v_{z'}}\cc{v_{z''}})
=
\cc{(z'+z''-2z)} \cc{v_z}\cc{v_z}\cc{v_z}
+
\cc{(z'-z)(z''-z)} \cc{v_z}(\cc{v_z}\cc{v_+} + \cc{v_+}\cc{v_z})$.
Since this is symmetric in $z',z''$
we get $\jj{z}(\cc{V}\wedge \cc{V})=0$,
one of the requirements for $\jj{z}\in \KERN$.
} $\jj{z} \in \KERN$.
Denote $\jj{z}
= \jj{-\thalf} + \jj{-\half}\cc{z} +
\jj{\half}\cc{z}^2 + \jj{\thalf}\cc{z}^3$.
The rank 1 modules $\JJ{i} = \rC \jj{i}$
are $v_{\pm}$-independent. We have
\[
\KERN \;=\; \JJ{-\thalf}\oplus \JJ{-\half} \oplus \JJ{\half}\oplus \JJ{\thalf}
\]

\subsection{Complement of $\KERN \subset \cc{V}\DERIV$}\label{sec:indcompl}
It turns out that invariantly associated to \eqref{dxsul}
is a complement of $\KERN \subset \cc{V}\DERIV$.

In the following figure, an arrow means `contained in':
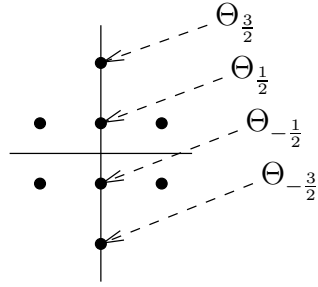
\begin{center}
\input{grj.pstex_t}
\end{center}%

The bullets on the vertical axis are
\begin{align*}
\grd{0,+\thalf}{(\cc{V}\DERIV)} & \;=\; \JJ{+\thalf}\\
\grd{0,+\half}{(\cc{V}\DERIV)} & \;=\; \JJ{+\half} \,\oplus\, \cc{V_+} \grd{0,0}{\DERIV}\\
\grd{0,-\half}{(\cc{V}\DERIV)} & \;=\; \JJ{-\half} \,\oplus\, \cc{V_-} \grd{0,0}{\DERIV}\\
\grd{0,-\thalf}{(\cc{V}\DERIV)} & \;=\; \JJ{-\thalf}
\end{align*}
The remaining four bullets are
\begin{align*}
\grd{-1,\pm \half}{(\cc{V}\DERIV)} &\;=\; \cc{V_{\pm}}\grd{-1,0}{\DERIV}
&
\grd{1,\pm \half}{(\cc{V}\DERIV)} &\;=\; \cc{V_{\pm}}\grd{1,0}{\DERIV}
\end{align*}
Combining, we get the internal direct sum decomposition\footnote{%
To be sure,
$\grd{\text{any},0}{\DERIV} =
\grd{-1,0}{\DERIV}\oplus \grd{0,0}{\DERIV}\oplus\grd{1,0}{\DERIV}$.
}
\[
\cc{V}\DERIV\;=\; \KERN\;\oplus\; \KERNcomp
\]

In particular, a gauge $\gauge$
is determined by its restriction to $\KERNcomp$,
and every positive definite $\rC$-Hermitian form on
$\KERNcomp$ determines a $\gauge$.

\subsection{Grading on $\cc{V}\DERIV/\KERN \simeq \KERNcomp$}
\label{sec:dihdkhkkka}
The nontrivial components are:
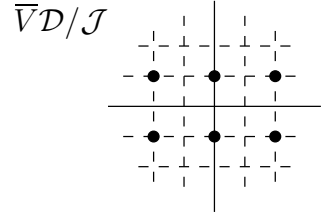
\begin{center}
\begin{picture}(0,0)%
\includegraphics{grvdj.pstex}%
\end{picture}%
\setlength{\unitlength}{4144sp}%
\begingroup\makeatletter\ifx\SetFigFont\undefined%
\gdef\SetFigFont#1#2#3#4#5{%
  \reset@font\fontsize{#1}{#2pt}%
  \fontfamily{#3}\fontseries{#4}\fontshape{#5}%
  \selectfont}%
\fi\endgroup%
\begin{picture}(1287,1284)(9526,-4033)
\put(9541,-2941){\makebox(0,0)[lb]{\smash{{\SetFigFont{12}{14.4}{\rmdefault}{\mddefault}{\updefault}{\color[rgb]{0,0,0}$\mathllap{\cc{V}\DERIV/\KERN}$}%
}}}}
\end{picture}%

\end{center}%

The four corner bullets have rank $1$.
The two center bullets have
rank $4 + \stdim$.

\subsection{Remarks}

The above discussion suggests conditions
that one might impose on $\gauge$,
and that are invariantly associated to a decomposition \eqref{dxsul}.
Here are two examples:
\begin{itemize}
\item One might require that
the six bullets in \sref{dihdkhkkka}
be mutually $\gauge$-orthogonal.
\item One might require
that
$\gauge(\cc{v_-} \grd{0,0}{\D{0}},
\cc{v_-} \grd{0,0}{\D{0}}) \subset \rR$.\\
One might require
that
$\gauge(\cc{v_+} \grd{0,0}{\D{0}},
\cc{v_+} \grd{0,0}{\D{0}}) \subset \rR$.
\end{itemize}
The second bullet is independent of the choice of $v_{\pm}$.


\nextpartxx{Structure associated to two compatible decompositions of $V$}{%
Structure associated to\\
two compatible decompositions of $V$}

\newcommand{\rundi}{\texttt{r}}
Suppose we have two decompositions like \eqref{dxsul},
\begin{subequations}\label{eq:twod}
\begin{equation}\label{eq:twod1}
\begin{aligned}
V & = V_-' \oplus V_+'\\
 & =  V_-'' \oplus V_+''
\end{aligned}
\end{equation}
that are compatible in the sense that
\begin{equation} \label{eq:twod2} 
\rundi' \rundi'' = - \rundi'' \rundi'
\end{equation}
\end{subequations}
where $\rundi',\rundi''\in \Aut_{\rC}(V)$ are the involutions defined by
\begin{align*}
\rundi'|_{V_-'} & = -\mathbbm{1} &
\rundi''|_{V_-''} & = -\mathbbm{1}\\
\rundi'|_{V_+'} & = \mathbbm{1} &
\rundi''|_{V_+''} & = \mathbbm{1}
\end{align*}

\newcommand{\contdiscus}{%
Given are two compatible decompositions $V = V_-'\oplus V_+'
= V_-'' \oplus V_+''$. See \eqref{eq:twod}.}

\newcommand{\contdiscusext}{\contdiscus~We use bases $v_{\pm}'$ and $v_{\pm}''$ that are compatible
 precisely as in \sref{compbases}.}


\section{Reflections}

\contdiscus

\subsection{Equivalent conditions} \label{sec:shd2393}

Condition \eqref{eq:twod2} is separately equivalent
to each of the following:
\begin{itemize}
\item $\rundi'(V_{\pm}'') = V_{\mp}''$.
\item $\rundi''(V_{\pm}') = V_{\mp}'$.
\item There exist bases 
$V_{\pm}' = \rC v_{\pm}'$ and $V_{\pm}'' = \rC v_{\pm}''$ such that
\begin{align*}
v_{\pm}' & = 2^{-1/2}(v_+'' \pm v_-'')
&
v_{\pm}'' & = 2^{-1/2}(v_+' \pm v_-')
\end{align*}
\end{itemize}

\subsection{Compatible bases}\label{sec:compbases}

We say that $V_{\pm}' = \rC v_{\pm}'$ and $V_{\pm}'' = \rC v_{\pm}''$
are compatible if
\begin{align*}
v_{\pm}' & = \rundi''(v_{\mp}')
&
v_{\pm}'' & = \rundi'(v_{\mp}'')\\
v_{\pm}' & = 2^{-1/2}(v_+'' \pm v_-'')
&
v_{\pm}'' & = 2^{-1/2}(v_+' \pm v_-')
\end{align*}
Choosing one basis element determines the other three.

The associated bases for $\VVreal$ defined in \sref{dkfhlskkka} satisfy
\begin{align*}
c_0' & = c_0'' &
c_1' & = c_3'' &
c_2' & = -c_2'' &
c_3' & = c_1''
\end{align*}

\subsection{Associated commuting reflections}

Associated to $\rundi',\rundi'' \in \Aut_{\rC}(V)$
are
automorphisms on various spaces, see \sref{vgt}.
We collectively
denote by $\fundi',\fundi''$
these automorphisms
on spaces on which $-\mathbbm{1} \in \Aut_{\rC}(V)$
is represented as the identity.
Such spaces include $\VV$, $\DERIV$, $\D{}$, $\IWeyl{}$, $\Ein{}$.

Note that $\fundi',\fundi''$ are involutions, and they commute,
\[
\fundi'\fundi'' = \fundi''\fundi'
\]

Note that on spaces on which $-\mathbbm{1} \in \Aut_{\rC}(V)$
is represented as the identity,
both $\pm \rundi'$ are represented as $\fundi'$,
and both $\pm \rundi''$ are represented as $\fundi''$.


\subsection{Projections}

Define the projections
\begin{align*}
\pundievenall & = \tfrac{1}{2}(\mathbbm{1}+ \fundi')&
\pundialleven & = \tfrac{1}{2}(\mathbbm{1}+ \fundi'')\\
\pundioddall & = \tfrac{1}{2}(\mathbbm{1}- \fundi')&
\pundiallodd & = \tfrac{1}{2}(\mathbbm{1}- \fundi'')
\end{align*}
They pairwise commute and we often abbreviate
$\pundievenodd = \pundievenall\pundiallodd$ etc.



\newcommand{\evod}{\text{$\even'$ ($\odd'$)}}
\subsection{Lemma}\label{sec:sii1kns}
Let $\grdp{}{}$ and $\grdpp{}{}$ be the gradings
associated to the decompositions \eqref{eq:twod1}. Then
\begin{align*}
\pundievenall \DC{} & = \grdp{\even}{\DC{}}&
\pundialleven \DC{} & = \grdpp{\even}{\DC{}}\\
\pundioddall \DC{} & = \grdp{\odd}{\DC{}}&
\pundiallodd \DC{} & = \grdpp{\odd}{\DC{}}
\end{align*}
This is not quite as obvious as may first appear.
One can explicitly check that
\begin{align*}
\pundi_{\text{$\even'$ ($\odd'$)}} \VV &
\;=\; \grdp{\text{$\even$ ($\odd$)}}{\VV}\\
\pundi_{\text{$\even'$ ($\odd'$)}} \DERIV &
\;=\; \grdp{\text{$\even$ ($\odd$)}}{\DERIV}
\end{align*}
which implies the claim.


\section{%
Associated decomposition of $\IWeyl{}$}\label{sec:irev}

\contdiscusext

We freely use notation from the discussion of the ideal in \sref{oshohssnlls}.
All the objects defined there come in two copies, tagged $'$ and $''$ respectively.

The discussion is limited to
the $\rC$-module $\IWeylX{}$.
Recall that $\RE: \IWeylX{} \to \IWeyl{}$
is an $\rR$-module isomorphism.
It commutes with $\fundi'$ and $\fundi''$.

\subsection{Lemma}\label{sec:dhdf2222khfs}
We have\footnote{%
One calculates
$\fundi'(\qq{z}'(\fundi'(v_w'))) = \ldots = \qq{-z}'(v_w')$
and
$\fundi''(\qq{z}'(\fundi''(v_w')))
= \ldots = z^4 \qq{1/z}'(v_w')$,
using the definition of $\qq{z}'$ in \sref{oshohssnlls}.
Here $v_z' = v_-' + zv_+'$
satisfies
$\fundi'v_z' = -v_{-z}'$
and
$\fundi''v_z' = z v_{1/z}'$.
}
\begin{align*}
\fundi'\qq{z}' & = \qq{-z}' &
\fundi''\qq{z}' & = z^4 \qq{1/z}'
\end{align*}
and the same result holds
with all $'$ and $''$ exchanged.

\subsection{Lemma}\label{sec:dhsihfdhdf}

We have
$\fundi'\cre_{\pm}' = \cre_{\pm}'\fundi'$ but
$\fundi'' \cre_{\pm}' = \cre_{\mp}'\fundi''$,
as operators on $\DC{}$.
Hence\footnote{%
Here $\cre_0' = \cre_+' + \cre_-'$
and $\cre_3' = \cre_+' - \cre_-'$,
instances of $\cre_i' = \cre_{c_i'}'$
with $c_i'$ defined in \sref{dkfhlskkka}.
}
\begin{align*}
\pundi_{\lambda'\lambda''}\cre_0'
 & =
\cre_0'\pundi_{\lambda'\lambda''}
& 
\pundi_{\lambda'\lambda''}\cre_3'
& =
\cre_3'\pundi_{\lambda'(\lambda''+1)}
\end{align*}
where $\lambda' \in \{\even',\odd'\}$
and $\lambda'' \in \{\even'',\odd''\}$.
Hence for $k=3,4$ we have\footnote{%
Using
$\IWeylX{k} =
\cre_-'\IWeylX{k-1}
+
\cre_+'\IWeylX{k-1}
=
\cre_0'\IWeylX{k-1}
+
\cre_3'\IWeylX{k-1}$ for $k=3,4$.
}
\begin{align*}
\pundi_{\lambda'\lambda''}\IWeylX{k} & =
\cre_0' \pundi_{\lambda'\lambda''}\IWeylX{k-1}
+
\cre_3' \pundi_{\lambda'(\lambda''+1)}\IWeylX{k-1}
\end{align*}
which we use below.

\subsection{Decomposition of $\IWeylX{}$}\label{sec:irevxx}

By \sref{dhdf2222khfs}
and \sref{dhsihfdhdf} 
we have
\begin{align*}
\pundieveneven\IWeylX{2} &
 = \rC(\qq{-2}' + \qq{2}') \oplus \rC \qq{0}'
\\
\pundievenodd\IWeylX{2} &
=\rC(\qq{-2}' - \qq{2}')
\\
\pundioddeven\IWeylX{2} & 
=\rC(\qq{-1}' + \qq{1}')
\\
\pundioddodd\IWeylX{2} &
=\rC(\qq{-1}' - \qq{1}') \displaybreak[0]\\
\rule{0pt}{16pt}
\pundieveneven\IWeylX{3} &
 =
\rC(\cre_+'\qq{-2}' + \cre_-'\qq{2}') \oplus \rC \cre_0'\qq{0}'\\
\pundievenodd\IWeylX{3} &
=
\rC(\cre_+'\qq{-2}' - \cre_-'\qq{2}') \oplus \rC \cre_3'\qq{0}'
\\
\pundioddeven\IWeylX{3} & 
=
\rC\cre_0'(\qq{-1}' + \qq{1}')
\oplus
\rC\cre_3'(\qq{-1}'-\qq{1}')\\
\pundioddodd\IWeylX{3} &
=
\rC\cre_3'(\qq{-1}' + \qq{1}')
\oplus
\rC\cre_0'(\qq{-1}'-\qq{1}')
\displaybreak[0]\\
\rule{0pt}{16pt}
\pundieveneven\IWeylX{4} &
 = 0\\
\pundievenodd\IWeylX{4} &
=\rC \cre_0'\cre_3' \qq{0}'
\\
\pundioddeven\IWeylX{4} & 
=
\rC\cre_0'\cre_3'(\qq{-1}'-\qq{1}')\\
\pundioddodd\IWeylX{4} &
=
\rC\cre_0'\cre_3'(\qq{-1}' + \qq{1}')
\intertext{Same with all $\pundievenodd$ and $\pundioddeven$
exchanged on the left hand sides,
and with all $'$ replaced by $''$ on the right hand sides. In particular,}
\pundieveneven\IWeylX{2}
&=
\rC(\qq{-2}' + \qq{2}') \oplus \rC \qq{0}'\\
&=
\rC(\qq{-2}'' + \qq{2}'') \oplus \rC \qq{0}''
\end{align*}



\subsection{
Lemma
}\label{sec:shkkska2344}

We have\footnote{%
First equation:
It essentially suffices to check that the two summands
have trivial intersection; take
$2\cre_-'\cre_+'
= \cre_0'\cre_3'
= \cre_0''\cre_1''$ which annihilates the first summand
but not the second.
}
\begin{align*}
\pundieveneven\IWeylX{2} & =
\rC(\qq{-2}' + \qq{2}') \oplus \rC(\qq{-2}'' + \qq{2}'')\\
\pundieveneven\IWeylX{3} & = 
\cre_0 \pundieveneven\IWeylX{2}\\
\pundieveneven\IWeylX{4} & = 0
\end{align*}
with $\cre_0 = \cre_0' = \cre_0''$.


\section{Associated filtration by \ecubes}\label{sec:oddklwej3l}

\contdiscusext

We use notation from our discussion of the filtration
in \sref{xsdfil}.
All objects defined there come in two copies, tagged $'$ and $''$ respectively.
For example, $\filp{}{}$ and $\filpp{}{}$.

We construct a filtration $\bkl{}{}$ as in \sref{deff},
indexed by $\Z_{\geq 0} \times \Z_{\geq 0}$.
We use \sref{rem2}.

\subsection{Choice of $\reg$}\label{sec:geomcompxx}

Define $\reg'$ and $\reg''$ as in \sref{dkfhlskkka}. Then define
\begin{align*}
\reg & = \reg' \cap \reg''\\
& = \rR c_0
\end{align*}
with $c_0 = c_0' = c_0'' \in \VVpos$.
It satisfies all conditions
in \sref{reg}.

\subsection{Choice of $\bkl{\filind'\filind''}{\Ein{}}$}

For all $\filind',\filind''\in \Z_{\geq 0}$ set
\[
\bkl{\filind'\filind''}{\Ein{}} \;=\;
\filp{\filind'}{\Ein{}} \cap \filpp{\filind''}{\Ein{}}
\]

Recall that $\filp{\filind'}{\Ein{}}$ is an $\reg'$-\ecubes,
and $\filpp{\filind''}{\Ein{}}$ is an $\reg''$-\ecubes.
Therefore they are 
$\reg$-\ecubes.
But we have yet to see that $\bkl{\filind'\filind''}{\Ein{}}$
is an $\reg$-\ecube.


\subsection{Lemma}\label{sec:sizd7sd}

The 
following spaces are closed under $\fundi'$ and $\fundi''$:
\begin{itemize}
\item The ideal $\IWeyl{}$.
\item The representatives
$\filp{\filind'}{\D{}}$ and $\filpp{\filind''}{\D{}}$ defined in \sref{xsdfil}.
\item Intersections and sums of these spaces.
\end{itemize}
The first follows from \sref{vgt}.
The second follows from  \sref{breq} and \sref{shd2393}.


\subsection{Proof that $\bkl{\filind'\filind''}{\Ein{}}$
is an $\reg$-\ecube}\label{sec:dodlslhfls}

Set
$\Dp{} = \filp{\filind'}{\D{}}$
and
$\Dpp{} = \filpp{\filind''}{\D{}}$,
keeping in mind the implicit dependence on $\filind',\filind''$.
These $\reg$-\dcubes~are representatives of
$\filp{\filind'}{\Ein{}}$ and $\filpp{\filind''}{\Ein{}}$.
We prove that their intersection
$\Dp{}\cap \Dpp{}$ is a representative of $\bkl{\filind'\filind''}{\Ein{}}$.

\newcommand{\STIU}[1]{\textbf{I}_{#1}}
By the 2nd bullet in \sref{ebucintersection}
it suffices to check the 2nd bullet in \sref{downup}, that is
\[
(\Dp{}+ \Dpp{})
\cap \IWeyl{}
\;\subset\;
(\Dp{} \cap \IWeyl{})
+
(\Dpp{}\cap \IWeyl{})
\]
All terms and subterms are closed under
$\pundi_{\lambda'\lambda''}$ by \sref{sizd7sd}, and
therefore it suffices to show
for all $\filind',\filind'',\lambda',\lambda''$
the statement $\STIU{\filind'\filind''\lambda'\lambda''}$ defined by
\[
\STIU{\filind'\filind''\lambda'\lambda''}
\quad
\Longleftrightarrow
\quad
(\pundi\Dp{}+ \pundi\Dpp{})
\cap \pundi\IWeyl{}
\;\subset\;
(\pundi\Dp{} \cap \pundi\IWeyl{})
+
(\pundi\Dpp{}\cap \pundi\IWeyl{})
\]
Here and below $\pundi = \pundi_{\lambda'\lambda''}$,
keeping in mind the implicit dependence on $\lambda',\lambda''$.
We show below that one is always in one of the following three sufficient
situations:
either $\pundi\Dp{} = 0$; or $\pundi\Dpp{}=0$;
or $\pundi \IWeyl{}
\subset (\pundi\Dp{} \cap \pundi\IWeyl{})
+
(\pundi\Dpp{}\cap \pundi\IWeyl{})$.

{\bf Proof of $\STIU{\filind'\filind''\even'\even''}$.}
$\pundi\IWeyl{}
\subset 
\pundi\fock{\RE \QQ{-2}' \oplus \RE \QQ{2}'}{\reg'}
+
\pundi\fock{\RE \QQ{-2}'' \oplus \RE \QQ{2}''}{\reg''}$
by \sref{shkkska2344}.
Since 
$\pundi\fock{\RE \QQ{-2}' \oplus \RE \QQ{2}'}{\reg'}
\subset
\pundi\Dp{} \cap \pundi\IWeyl{}$
by \sref{djsjdojls2}, and similar for $''$, 
we conclude that
$\pundi \IWeyl{}
\subset (\pundi\Dp{} \cap \pundi\IWeyl{})
+
(\pundi\Dpp{}\cap \pundi\IWeyl{})$.

{\bf Proof of $\STIU{\filind'\filind''\odd'\lambda''}$.}
Distinguish two cases:
\begin{itemize}
\item If $\filind' = 0$ then $\pundi \Dp{} = 0$.\\
Use $\Dp{} \subset \grdp{\even}{\D{}}$
from \sref{ch0}, and \sref{sii1kns}.
\item If $\filind' \geq 1$
then $\pundi\IWeyl{} \subset \pundi \Dp{} \cap \pundi\IWeyl{}$.\\
Use $\grdp{\odd}{\D{}} \subset \Dp{}$
from \sref{sdoohds63768}, and
$\pundi \IWeyl{} \subset \pundi \D{}
\subset \pundioddall \D{} \subset \grdp{\odd}{\D{}}$
using \sref{sii1kns}.
\end{itemize}

{\bf Proof of $\STIU{\filind'\filind''\lambda'\odd''}$.} Analogous.


\subsection{Lemma}\label{sec:fd3hl3}

Recall the notation $\fock{\svac}{\reg} = \svac + \cre_0 \svac$
where $\cre_0=\cre_0'=\cre_0''$.

\newcommand{\SHD}{\sigma_{\even'\even''}}
We have, with the abbreviation
$\SHD=
\RE(\rC(\qq{-2}' + \qq{2}')) \subset \pundieveneven\IWeyl{2}$,
\begin{align*}
\filp{0}{\D{}}\cap \filpp{0}{\D{}}
\cap \IWeyl{} & \;=\; 0\\
\filp{0}{\D{}}\cap \filpp{1}{\D{}}
\cap \IWeyl{} & \;=\; 
\fock{\pundievenodd\IWeyl{2}}{\reg}\\
\filp{0}{\D{}}\cap \filpp{2}{\D{}}
\cap \IWeyl{} & \;=\;
\fock{\SHD}{\reg}
\oplus
\fock{\pundievenodd\IWeyl{2}}{\reg}
\\
\filp{1}{\D{}}\cap \filpp{1}{\D{}}
\cap \IWeyl{} & \;=\; 
\fock{\pundievenodd\IWeyl{2}}{\reg}
\oplus
\fock{\pundioddeven\IWeyl{2}}{\reg}
\oplus
\pundioddodd\IWeyl{}\\
\filp{1}{\D{}}\cap \filpp{2}{\D{}}
\cap \IWeyl{} & \;=\; 
\fock{\SHD}{\reg}
\oplus
\fock{\pundievenodd\IWeyl{2}}{\reg}
\oplus
\pundioddeven\IWeyl{}
\oplus
\pundioddodd\IWeyl{}\\
\filp{2}{\D{}}\cap \filpp{2}{\D{}}
\cap \IWeyl{} & \;=\; 
\fock{\pundieveneven\IWeyl{2}}{\reg}
\oplus
\pundievenodd\IWeyl{}
\oplus
\pundioddeven\IWeyl{}
\oplus
\pundioddodd\IWeyl{}
\end{align*}
This follows from the formulas in \sref{djsjdojls2}
and from \sref{irevxx}, \sref{shkkska2344}.


\subsection{Lemma}\label{sec:sohdsl}

If $(\lambda',\lambda'') \neq (\even',\even'')$
then there is a $Y \subset \pundi \IWeyl{3}$ such that
\begin{subequations}\label{eq:zz}
\begin{align}
\pundi \IWeyl{} & \;=\;
\rule{19mm}{0pt}\mathllap{\fock{\pundi \IWeyl{2}}{\reg}}
\oplus Y \oplus \pundi \IWeyl{4}
\label{eq:zz1}
\\
\pundi \D{} & \;=\;
\rule{19mm}{0pt}\mathllap{\fock{\pundi\gvac{\reg,\leq 2}}{\reg}}
\oplus Y \oplus \pundi \IWeyl{4}
\label{eq:zz2}
\end{align}
\end{subequations}
where $\pundi = \pundi_{\lambda'\lambda''}$
and where $\gvac{\reg,\leq 2}
= \gvac{\reg} \cap (\D{0}\oplus \D{1}\oplus \D{2})$.

To check this, first note that
$\pundi
\D{} = \fock{\pundi\gvac{\reg}}{\reg}
= \fock{\pundi\gvac{\reg,\leq 2}}{\reg}
\oplus \fock{Z}{\reg}$
where $Z = \pundi(\gvac{\reg}\cap \D{3})$.
It is easy to see that $\rank_{\rR}Z = 2$
given $(\lambda',\lambda'') \neq (\even',\even'')$.
Hence
$Z = \ann_0(\pundi \IWeyl{4})$,
essentially because $\supset$ and the right hand side has rank
$2$. Hence\footnote{%
Use the anticommutation relation
$\ann_0\cre_0+\cre_0\ann_0 = \text{(invertible)}\mathbbm{1}$.
}
$\fock{Z}{\reg}
= \ann_0(\pundi \IWeyl{4})
\oplus \pundi \IWeyl{4}$.
But $\ann_0(\pundi \IWeyl{4}) \not\subset \IWeyl{3}$ cannot
be taken to be $Y$.
Instead take
$Y = \RE(\rC(\cre_3'\qq{0}'))$
or
$Y = \RE(\rC(\cre_3'(\qq{-1}'\pm\qq{1}')))$,
depending on $\lambda'\lambda''$;
then \eqref{eq:zz1} follows from \sref{irevxx}.
For \eqref{eq:zz2} it essentially suffices,
since $Y \subset \pundi \D{3}$ has rank 2,
to check that it does not intersect
$\fock{\pundi\gvac{\reg,\leq 2}}{\reg}
\cap \D{3}$. But applying $\cre_0'$ to the latter gives zero,
whereas $\cre_0' Y = \pundi \IWeyl{4}$ has rank 2.

Corollary:
If $(\lambda',\lambda'') \neq (\even',\even'')$
then the
$\reg$-\dcube~$S = \fock{\pundi\gvac{\reg,\leq 2}}{\reg}$
satisfies
$\pundi \D{} = S + \pundi \IWeyl{}$,
and $(\reg,S)$ satisfies the $234$-condition.

To see that the $234$-condition holds,
note that
$\pundi \IWeyl{2} \subset \pundi \D{2}
\subset S$.
Hence 
$\fock{\pundi\IWeyl{2}}{\reg} \subset S$,
and then \eqref{eq:zz}
implies
$S \cap \pundi\IWeyl{} = \fock{\pundi\IWeyl{2}}{\reg}$.


\subsection{The $234$-condition}

By \sref{dodlslhfls},
$S_0 = \filp{\filind'}{\D{}}\cap \filpp{\filind''}{\D{}}$ is a representative
of the $\reg$-\ecube~$\bkl{\filind'\filind''}{\Ein{}}$.
We construct another representative $S$ such that $(\reg,S)$
satisfies the $234$-condition:
\begin{itemize}
\item If $\filind' = 0$ or $\filind''=0$
then we can take $S = S_0$ by \sref{fd3hl3}.
\item If $\filind' = \filind''=1$ then 
$\pundioddodd S_0 
= \pundioddodd \D{}$. Set
\[
S = \pundieveneven
S_0 \oplus \pundievenodd S_0 \oplus
\pundioddeven S_0 \oplus \fock{\pundioddodd\gvac{\reg,\leq 2}}{\reg}
\]
using notation from \sref{sohdsl}.
This is a representative,
$S_0+\IWeyl{} = S+\IWeyl{}$,
because
$\pundi S_0 + \pundi\IWeyl{}
= \pundi S + \pundi\IWeyl{}$
with $\pundi = \pundioddodd$,
by the corollary in \sref{sohdsl};
both sides are equal to $\pundi \D{}$.
Use the same corollary to check the
$\pundi$-part of the $234$-condition
for $(\reg,S)$; for the other three parts use
\sref{fd3hl3}.
\item
If $\filind'=1$
and $\filind'' = 2$
then
$\pundioddall S_0 = \pundioddall \D{}$.
In this case set $S$ equal to
\[
\pundieveneven
S_0 \oplus \pundievenodd S_0
\oplus \fock{\pundioddeven\gvac{\reg,\leq 2}}{\reg}
\oplus \fock{\pundioddodd\gvac{\reg,\leq 2}}{\reg}
\]
\item If $\filind'=\filind''=2$ then
$S_0 = \D{}$. In this case set $S$ equal to
\[
\pundieveneven
S_0 
\oplus \fock{\pundievenodd\gvac{\reg,\leq 2}}{\reg}
\oplus \fock{\pundioddeven\gvac{\reg,\leq 2}}{\reg}
\oplus \fock{\pundioddodd\gvac{\reg,\leq 2}}{\reg}
\]
\end{itemize}

\subsection{Filtration of $\Ein{}$}

The filtration $\bkl{}{}$ has all the properties in \sref{deff}.


\nextpart{Some explicit calculations}

The filtration $\bkl{}{}$ in \sref{oddklwej3l}
can be used to set up a filtered expansion, see the introduction \sref{intro}.
The following sections contain the results
of explicit calculations.

\newcommand{\exc}{\texttt{x}}
\newcommand{\Exc}{\texttt{X}}
We use bases $v_{\pm}'$
and $v_{\pm}''$ compatible as in \sref{compbases}.
We abbreviate $v_{\pm} = v_{\pm}'$.


\section{Notation}

\subsection{Involution $\exc$}
Define $\exc \in \Aut_{\rC}(V)$ by, equivalently,
\begin{itemize}
\item $\exc = 2^{-1/2}(\rundi' + \rundi'')$.
\item $v_{\pm}'' = \exc(v_{\pm}')$.
\item $v_{\pm}' = \exc(v_{\pm}'')$.
\end{itemize}
Then $\exc^2 = \mathbbm{1}$
and
$\rundi'' = \exc \rundi' \exc$.

\subsection{Third reflection $\rundi'''$}

Define $\rundi''' \in \Aut_{\rC}(V)$ by
\[
\rundi''' = i\rundi'\rundi''
\]
It follows immediately that:
\begin{itemize}
\item The $\rundi',\rundi'',\rundi'''$ anticommute pairwise.
\item $(\rundi')^2 = (\rundi'')^2 = (\rundi''')^2 = \mathbbm{1}$.\\
$\rundi''\rundi''' = -i\rundi'$ and
$\rundi'''\rundi' = -i\rundi''$ and
$\rundi'\rundi'' = -i\rundi'''$.
\item
The sixteen elements $\lambda \mathbbm{1}$,
$\lambda \rundi'$,
$\lambda\rundi''$,
$\lambda\rundi'''$
with $\lambda = \pm 1,\pm i$
are pairwise distinct and
form a subgroup of $\Aut_{\rC}(V)$.
\end{itemize}

\newcommand{\excp}{\exc'}
\newcommand{\excpp}{\exc''}

\subsection{Involutions $\exc',\exc''$}

Define $\excp,\excpp\in \Aut_{\rC}(V)$ by
\begin{align*}
\excp &\;=\;2^{-1/2}(\rundi' + \rundi''')\\
\excpp &\;=\;2^{-1/2}(\rundi'' - \rundi''')
\end{align*}
Then
$\excp\rundi' = \rundi'''\excp$
and
$\excpp\rundi'' = -\rundi'''\excpp$ and these are involutions,
$(\excp)^2 = (\excpp)^2 = \mathbbm{1}$.
We have chosen signs such that
$\exc \excp = \excpp \exc$.

\subsection{Associated automorphisms}

Associated to
$\rundi',\rundi'',\rundi''',\exc,\exc',\exc'' \in \Aut_{\rC}(V)$
are
automorphisms on various other spaces, see \sref{vgt}.
We collectively
denote by  $\fundi',\fundi'',\fundi''',\Exc,\Exc',\Exc''$
these automorphisms
on spaces on which $-\mathbbm{1} \in \Aut_{\rC}(V)$
is represented as the identity.


\section{On the naive leading term $\unkp$}
\label{sec:diz38hro}

Part of this section appears again in \sref{leadingtermeqs}.
Calculations are not written out explicitly.
Recall that we abbreviate $v_{\pm} = v_{\pm}'$.

\subsection{Lemma}\label{sec:naivepa}

The following map is an $\rR$-module isomorphism:
\begin{align*}
\Der(\rR) \oplus \rR^6 & \;\to\; \bkl{00}{\Ein{1}}\\
{\axider{}} \oplus \mathrlap{a}\phantom{\rR^6} & \;\mapsto\; \unkp
\end{align*}
where
\NAIVELEADINGTERM
One can check that
$\fundi'\unkp=\fundi''\unkp = \unkp$.
Applying $\Exc$
on the right hand side of the map corresponds to
exchanging $\aunkn{3},\aunkn{1}$ on the left hand side of the map.


\subsection{Lemma} \label{sec:naiveeq}

For every $\unkp \in \bkl{00}{\Ein{1}}$ we have
$\eb{\unkp}{\unkp}=0$
if and only if
\aeqsnA
and
\aeqsnB


\subsection{An explicit solution}\label{sec:stdsol}

Let $\rR$ be the smooth real functions on $\R^4$.
Let $\p_1,\p_2,\p_3,\p_4$ be the partial derivatives.
Let $\tau$ be the first coordinate,
so that $(\p_1\tau,\p_2\tau,\ldots) = (1,0,0,0)$.
Let $u \in \rR$ be any
function that has a multiplicative inverse $u^{-1}\in \rR$,
and that depends only on the second coordinate,
$(\p_1 u,\p_2u,\ldots) = (0,\ast,0,0)$.
Then
\begin{align*}
\axider{} & = \p_1 &
\aunkn{1} & = -u + \tanh \tau &
\aunkn{4} & = 0\\
&&
\aunkn{2} & = -\tanh \tau &
\aunkn{5} & = 0\\
&&
\aunkn{3} & = -u^{-1} + \tanh \tau &
\aunkn{6} & = 1/(\cosh \tau)
\end{align*}
solves all the equations in \sref{naiveeq}.

Informally, the above solution is general up to exceptional cases,
with the understanding that one can obtain other solutions
from the one above by changing coordinates
or rescaling $v_{\pm}$.
We make no attempt to make this precise.


\subsection{Nonzero 2nd cohomology\\
for the explicit solution in \sref{stdsol}}
\label{sec:obstrexample}

\newcommand{\syzzzz}[1]{\Ein{#1}}
\newcommand{\ayyyy}{\;\xrightarrow{\;\dec{\dd}\;}\;}

Define the differential
$\dec{\dd} = \eb{\unkp}{\,\cdot\,} : \Ein{}\to\Ein{}$:
\[
0
\ayyyy\syzzzz{0}
\ayyyy\syzzzz{1}
\ayyyy\syzzzz{2}
\ayyyy\syzzzz{3}
\ayyyy\syzzzz{4}
\ayyyy
0
\]
Then the 2nd cohomology is nonzero,
in fact one has the stronger statement:
there are two solutions
to the linearized equations,
$x,y \in \ker(\dec{\dd}|_{\Ein{1}})$,
whose bracket is nonzero in the 2nd cohomology,
$\eb{x}{y} \notin \image(\dec{\dd}|_{\Ein{1}})$.

\newcommand{\thirdcoord}{f}
By direct calculation,
these $x,y \in \D{1}$ 
solve the linearized \twofootnotes{equations}{As an
aside,
$x \in \bkl{01}{\Ein{1}}$
and
$\fundi'x = - \fundi''x = x$,
and $y \in \bkl{20}{\Ein{1}}$
and
$\fundi'y = \fundi'' y = y$.
}{%
As an aside, note that the set of
solutions to the linearized equations,
$\ker(\dec{\dd}|_{\Ein{1}})$,
is closed under multiplication by
elements of
$\rR$ that are annihilated by $\p_1$.}:
\begin{itemize}
\item
$x|_{\rR} = \exp(-\tfrac{1}{2}u^{-1}\tau) (v_+\cc{v_+} - v_-\cc{v_-}) \p_3$
and
$x(v_{\pm})=0$.
\item
$y|_{\rR} = 0$
and
\begin{align*}
y(v_-)
& \;=\;
\exp(-u\tau) (1 - iu \cosh \tau)(v_+\cc{v_+}v_- + v_-\cc{v_+}v_+)\\
& \qquad
+ \exp(-u\tau) (v_-\cc{v_-}v_- + v_+\cc{v_-}v_+)\\
\rule{0pt}{13pt}
y(v_+) & \;=\;
 \mathrlap{\text{(defined such that $\fundi''y = y$)}}
\end{align*}
\end{itemize}
To check that $\eb{x}{y}$ is nonzero in the 2nd cohomology,
let $\thirdcoord \in \rR$ denote the third coordinate,
$(\p_1\thirdcoord,\p_2\thirdcoord,\ldots) = (0,0,1,0)$.
We use the following lemma:
\begin{itemize}
\item $(\cre_-\cre_+\db{\unkp}{\D{1}})(\thirdcoord) = 0$,
 where $\cre_{\pm}$ are defined as in \sref{dsohlald}.
\end{itemize}
On the other hand,
$(\cre_-\cre_+\db{x}{y})(\thirdcoord) \neq 0$
by direct verification.
Since all elements of $\IWeyl{2}$ annihilate $\rR$,
we conclude that
$\eb{x}{y} \notin \image(\dec{\dd}|_{\Ein{1}})$.


\section{On the leading term $\unk_0$}\label{sec:leadingtermeqs}

We dump the result
of computer calculations of the
equations
$\eb{\unk_0}{\unk_0}=0$ that the leading term $\unk_0 \in \REES^1/s\REES^1$ has to satisfy.
Here $\REES$ is the Rees algebra of $\bkl{}{}$.

We do not suggest that one should explicitly
work with the equations as presented here.
The discussion is incomplete in many ways,
in particular we do not consider $\REES^k/s\REES^k$ for $k=0,3,4$,
which would be necessary to discuss the symmetries
and identities enjoyed by these equations.

Even so, there are interesting things to see here.
For example in \sref{degobs}
we point out certain degeneracies that are very closely
related to \sref{obstrexample}.

\subsection{Ranks of the components of $\GO$}
The $\rR$-ranks of the components of $\GO$ are as follows:
\begin{equation*}
\begin{array}{c | c | c c c c c}
 & \text{multiplicity} & \Ein{0} & \Ein{1} & \Ein{2} & \Ein{3} & \Ein{4} \\
\hline
\bkl{00}{}             & 1 & \stdim+2 & \stdim+6 & 5 & 1 & 0 \\
\bkl{10}{}/\bkl{<10}{} & 2 & 2 & \stdim+7 & \stdim+6 & 1 & 0 \\
\bkl{20}{}/\bkl{<20}{} & 2 & 0 & 1 & 1 & 0 & 0 \\
\bkl{11}{}/\bkl{<11}{} & 1 & 2 & \stdim+8 & 2\stdim+10 & \stdim+4 & 0 \\
\bkl{21}{}/\bkl{<21}{} & 2 & 0 & 1 & \stdim+4 & \stdim+3 & 0 \\
\bkl{22}{}/\bkl{<22}{} & 1 & 0 & 0 & 1 & \stdim+3 & \stdim+2
\end{array}
\end{equation*}
Here rank $m\stdim + n$ means that the respective component is $\simeq \Der(\rR)^m \oplus \rR^n$.
For example,
$\bkl{10}{\Ein{1}}/\bkl{<10}{\Ein{1}}
\simeq \Der(\rR) \oplus \rR^7$, and the multiplicity reminds us
that there is also the isomorphic $\bkl{01}{\Ein{1}}/\bkl{<01}{\Ein{1}}$.


\subsection{Summary of conventions}

We abbreviate $v_{\pm} = v_{\pm}'$. Hence
\begin{align*}
v_{\pm}' & = v_{\pm}\\
v_{\pm}'' & = 2^{-1/2}(v_+ \pm v_-)
\intertext{We use the following definitions, consistent
with the rest of this paper:}
\rundi'(v_{\pm}) & = \pm v_{\pm}\\
\rundi''(v_{\pm}) & = v_{\mp}\\
\exc & = 2^{-1/2}(\rundi'+\rundi'')\\
\exc' & = 2^{-1/2}(\rundi'+i\rundi'\rundi'')\\
\exc'' & = 2^{-1/2}(\rundi'+i\rundi''\rundi')
\end{align*}
We denote by $\fundi',\fundi'',\Exc,\Exc',\Exc''$ the
associated elements of $\Aut(\Ein{})$.


\subsection{The auxiliary map $\auxA$}

Define
\begin{align*}
\Der(\rR) \oplus \rR^5 &\;\to\; \pundieveneven\Ein{1}\\
{\xxider{}} \oplus \mathrlap{x}\phantom{\rR^5} & \;\mapsto\; \auxA_{{\xxider{}},x} = \auxA
\end{align*}
by
\auxAdata
One can check that applying
$\Exc$ on the right hand side of the map corresponds to
exchanging $x_3,x_1$ on the left hand side of the map.
Applying $\Exc'$ corresponds to exchanging $x_1,x_2$.
Applying $\Exc''$ corresponds to exchanging $x_2,x_3$.


\subsection{The auxiliary map $\auxB$}

Define
\begin{align*}
\Der(\rR) \oplus \rR^7 &\;\to\; \pundievenodd\Ein{1}\\
{\xxider{}} \oplus \mathrlap{x}\phantom{\rR^7} & \;\mapsto\; \auxB_{{\xxider{}},x} = \auxB
\end{align*}
by
\auxBdata
One can check that applying
$\Exc''$ on the right hand side of the map corresponds to
multiplying
${\xxider{}},x_1,x_2,x_3,x_6,x_7$
by $-1$ on the left hand side.


\subsection{The auxiliary map $\auxD$}

Define
\begin{align*}
\rR &\;\to\; \pundieveneven\Ein{1}\\
\mathrlap{x}\phantom{\rR} & \;\mapsto\; \auxD_{x} = \auxD
\end{align*}
by
\auxDdata
One can check that $\Exc''\auxD = \auxD$.


\subsection{The auxiliary map $\auxG$}

Define
\begin{align*}
\rR &\;\to\; \pundievenodd\Ein{1}\\
\mathrlap{x}\phantom{\rR} & \;\mapsto\; \auxG_{x} = \auxG
\end{align*}
by
\auxGdata
One can check that $\Exc''\auxG = -\auxG$.


\subsection{Parametrization of $\REES^1/s\REES^1$}\label{sec:29jojewl}
The map
\[
({\axider{}},a,{\bxider{}},b,{\cxider{}},c,d,{\exider{}},e,f,g,h)
\;\mapsto\; \unk_0
\]
where ${\axider{}}, {\bxider{}}, {\cxider{}}, {\exider{}}
\in \Der(\rR)$ and $a \in \rR^6$, $b,c \in \rR^7$, $e \in \rR^8$,
$d,f,g,h \in \rR$ and
\auxGOparametrization
is a parametrization of $\REES^1/s\REES^1$.

Some remarks about this parametrization:
\begin{itemize}
\item The first eight letters of the alphabet are used
as identifiers for the components of the grading
of $\REES^1/s\REES^1$, as in the following table:
\[
\begin{array}{c |c c c}
(s'')^2 & f & h & -\\
s'' & {\cxider{}},\,c & {\exider{}},\,e & g\\
1 & {\axider{}},\,a & {\bxider{}},\,b & d\\
\hline
& 1 & s' & (s')^2
\end{array}
\]
\item
The naive leading term is
$\unkp =
\auxA_{\axider{},\aunkn{1},\ldots,\aunkn{5}} + \Exc'\auxD_{\aunkn{6}}$,
consistent with
\sref{diz38hro}.
\item Note that
\begin{align*}
\unk_0|_{\rR}\;& =\;(\Gw\Gcw+\Gv\Gcv){\axider{}}\\
& \qquad +s'(\Gw\Gcv+\Gv\Gcw){\bxider{}}\\
& \qquad +s''(\Gw\Gcw-\Gv\Gcv){\cxider{}}\\
& \qquad +s's''i(\Gw\Gcv-\Gv\Gcw){\exider{}}
\end{align*}
If $\rR$ are the smooth real functions on a 4-dim manifold,
then informally, the frame $\unk_0|_{\rR}$
can be nondegenerate at order $s's''$ but no earlier.
\item The way in which $\auxB,\auxD,\auxG$ appear
is made more transparent by throwing in an artificial third
variable $s'''$, as discussed in \sref{fakepar}.
\end{itemize}

In \sref{firsteqs} \ldots \sref{lasteqs} we explicitly
write down the equation $\eb{\unk_0}{\unk_0} = 0$,
organized by the grading
of $\REES^2/s\REES^2$. That is, by the powers of $s',s''$.
The bracket on $\GO$ respects this grading,
hence only certain terms can appear.


\newcommand{\eqwithoutder}[1]{and the equations obtained by eliminating
applications of ${\axider{}}$:}
\newcommand{\onlyprod}[1]{The grading only allows products of type #1.}

\subsection{Equation component $\bkl{00}{\Ein{2}}$}\label{sec:firsteqs}
\aeqsnA
\eqwithoutder{a}
\aeqsnB
We already saw these equations in \sref{diz38hro}.

\subsection{Equation component $s'(\bkl{10}{\Ein{2}}/\bkl{<10}{\Ein{2}})$}
\beqsnA
\eqwithoutder{b}
\beqsnB
\onlyprod{$ab$}

\subsection{Equation component $(s')^2(\bkl{20}{\Ein{2}}/\bkl{<20}{\Ein{2}})$}
\deqsnA
\deqsnB
\onlyprod{$ad,b^2$}

\subsection{Equation component $s's''(\bkl{11}{\Ein{2}}/\bkl{<11}{\Ein{2}})$}
\eeqsnA
\eqwithoutder{e}
\eeqsnB
\onlyprod{$ae, bc$}

\subsection{Equation component $(s')^2s''(\bkl{21}{\Ein{2}}/\bkl{<21}{\Ein{2}})$}\label{sec:dohfz3hr}
\geqsnA
\eqwithoutder{g}
\geqsnB
\onlyprod{$ag, be, cd$}

\subsection{Equation component $(s')^2(s'')^2(\bkl{22}{\Ein{2}}/\bkl{<22}{\Ein{2}})$}
\ieqsnA
\ieqsnB
\onlyprod{$bh, cg, df, e^2$}

\subsection{Omitted equation components} \label{sec:lasteqs}

We have omitted the equations components $s'',(s'')^2,s'(s'')^2$.
They can be obtained from the analogous $s',(s')^2,(s')^2s''$ components.


\subsection{Informal remark about \sref{firsteqs} \ldots \sref{lasteqs}}\label{sec:degobs}

Each of these sections contains:
\begin{itemize}
\item Equations involving
applications of ${\axider{}}$.
\item Equations obtained by eliminating
applications of ${\axider{}}$.
\end{itemize}
This organization leaves considerable ambiguity,
in particular equations in the 1st set
can be rewritten by adding
multiples of the equations in the 2nd set.

It is interesting to
consider the 2nd sets of equations,
assuming that 
${\axider{}}$ and $a$, which constitute the naive leading term,
have already been fixed.
Consider for example in \sref{dohfz3hr} the equation:
\[
\commuttor{\exider{}}{\bxider{}}+\eunkn{3}\bxider{}-\eunkn{4}\bxider{}-\bunkn{3}\exider{}-\bunkn{4}\exider{}+\dunkn{1}\cxider{} = 0
\mod \rR {\axider{}}
\]
There are only products of type $be,cd$ in this equation,
 none of type $ag$,
which is a kind of degeneracy.
See the closely related \sref{obstrexample}.


\subsection{Informal remark about the parametrization
in \sref{29jojewl}}\label{sec:fakepar}

The parametrization becomes more transparent by throwing in
a third variable:
\auxGOparametrizationFAKE
This parametrization reduces
 to the one in \sref{29jojewl}
if one sets $s'''=1$.

This expression is of some interest if
one is willing to speculate, with \cite{BKL}, beyond \BKL.
It could conceivably be used to
asymptotically match (stick together)
a BKL-bounce constructed using $s',s''$ with $s'''=1$
to one constructed using say $s'',s'''$ with $s'=1$.
It shows which parametrization components would rotate
into which other components when matching expansions.



\appendix
\addtocontents{toc}{\vskip 2mm}

\section{Rees algebra for a
$\Z_{\geq 0}\times \Z_{\geq 0}$-indexed filtration}\label{app:reestwo}

We elaborate on \sref{hjdjd}.
Here $\REES$ satisfies, among others:
\begin{itemize}
\item
$\REES \to \REES, x \mapsto s'x$
and
$\REES \to \REES, x \mapsto s''x$
are injective maps.
\item $m_1\REES \cap m_2\REES \subset \lcm(m_1,m_2)\REES$
for any two monomials in $s',s''$.\\
Here $\lcm$ is the least common multiple,
for example $\lcm(s',s'') = s's''$.
\end{itemize}

Suppose $M$ is a set of monomials in $s',s''$,
containing almost all monomials,
and with $s'M \subset M$ and $s''M \subset M$.
Set $M\REES = \sum_{m \in M} m\REES$,
an effectively finite sum, and an ideal of $\REES$.
Suppose $M_- \subset M$ is such that $s'M \subset M_-$
and $s''M \subset M_-$, in particular
the set difference $M\setminus M_-$ is finite. \begin{samepage}Then
\[
\ker(\REES/M\REES \leftarrow \REES/M_-\REES)
= M\REES/M_-\REES
\;\;
\xleftarrow{\;\text{vec sp iso}\;}
\;\;
\textstyle\bigoplus_{m \in M\setminus M_-} \REES/(s'\REES+s''\REES)
\]
with the isomorphism given by $\sum_m m x_m \mapsfrom
\bigoplus_{m \in M\setminus M_-} x_m$.\end{samepage}

Example 1: $M$ the set of monomials of total degree $\geq \filind$,
and $M_-$ the set of monomials of total degree $\geq \filind+1$.

\begin{samepage}
Example 2: Here $M_-$ is obtained by removing a single monomial from $M$:
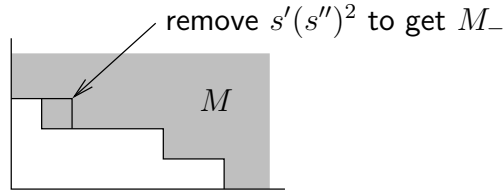
\begin{center}
\input{rees2.pstex_t}
\end{center}%

\end{samepage}

\section{Sample evaluation of the bracket}\label{app:sampleeval}

This example uses a basis,
 $V = \rC v_- \oplus \rC v_+$.

{\bf Goal.}
Let $\lambda_{\pm} \in \rC$ be parameters.
Define $c_{\pm} \in \D{1}$ by $c_{\pm}(\rC)=0$ and by
\begin{align*}
\TWO{c_-(v_-)}{0}{
     c_+(v_-)}{\lambda_+(v_+\cc{v_-})v_+}\\
\TWO{c_-(v_+)}{\lambda_-(v_-\cc{v_+})v_-}{
     c_+(v_+)}{0}
\end{align*}
Then the reality condition coming with $\D{1}$ implies
\begin{align*}
\TWO{c_-(\cc{v_-})}{0}{
     c_+(\cc{v_-})}{\cc{\lambda_+}(v_-\cc{v_+})\cc{v_+}}\\
\TWO{c_-(\cc{v_+})}{\cc{\lambda_-}(v_+\cc{v_-})\cc{v_-}}{
     c_+(\cc{v_+})}{0}
\end{align*}
Our goal is to evaluate
$\db{c_-}{c_+}$.

{\bf Reformulation.} The above definition is equivalent to
$c_{\pm} = (\mathbbm{1}+\CONJ) x_{\pm}$ where $x_{\pm}\in \DC{1}$ are given by
$x_{\pm}(\rC)=x_{\pm}(\cc{V})=0$ and
\begin{align*}
\TWO{x_-(v_-)}{0}{
     x_+(v_-)}{\lambda_+(v_+\cc{v_-})v_+}\\
\TWO{x_-(v_+)}{\lambda_-(v_-\cc{v_+})v_-}{
     x_+(v_+)}{0}
\end{align*}
and accordingly
$(\CONJ x_{\pm})(\rC) = (\CONJ x_{\pm})(V)=0$ and
\begin{align*}
\TWO{(\CONJ x_-)(\cc{v_-})}{0}{
     (\CONJ x_+)(\cc{v_-})}{\cc{\lambda_+}(v_-\cc{v_+})\cc{v_+}}\\
\TWO{(\CONJ x_-)(\cc{v_+})}{\cc{\lambda_-}(v_+\cc{v_-})\cc{v_-}}{
     (\CONJ x_+)(\cc{v_+})}{0}
\end{align*}
Clearly
\[
\db{c_-}{c_+}
\;=\;
\db{x_-}{x_+}
+
\db{x_-}{\CONJ x_+}
+
\db{\CONJ x_-}{x_+}
+
\db{\CONJ x_-}{\CONJ x_+}
\]
which are four instances of the 
$\DC{1}\times \DC{1}\to \DC{2}$
bracket.
We evaluate each of the four terms; only the first will
require an actual calculation.

{\bf The term $\db{x_-}{x_+}$.} Note that $x_{\pm} = \omega_{\pm}\delta_{\pm}$
where $\omega_- = \lambda_-(v_-\cc{v_+})$
and $\omega_+ = \lambda_+(v_+\cc{v_-})$ and where $\delta_{\pm} \in \DC{0}$
are given by $\delta_{\pm}(\rC)=\delta_{\pm}(\cc{V})=0$
and
\begin{align*}
\TWO{\delta_-(v_-)}{0}{
     \delta_+(v_-)}{v_+}\\
\TWO{\delta_-(v_+)}{v_-}{
     \delta_+(v_+)}{0}
\end{align*}
Then
\begin{align*}
\delta_-(\omega_+) & = \lambda_+(v_-\cc{v_-})\\
\delta_+(\omega_-) & = \lambda_-(v_+\cc{v_+})
\end{align*}
and $\delta = [\delta_-,\delta_+] \in \DC{0}$ is given by
$\delta(\rC)=\delta(\cc{V})=0$ and
\begin{align*}
\delta(v_-) & =  v_-\\
\delta(v_+) & = -v_+
\end{align*}
Therefore we have
\begin{align*}
\db{x_-}{x_+} & = \db{\omega_-\delta_-}{\omega_+\delta_+}\\
&= (\omega_-\wedge\omega_+)[\delta_-,\delta_+]
+ (\omega_- \wedge \delta_-(\omega_+))\delta_+
- (\delta_+(\omega_-) \wedge \omega_+)\delta_-\\
& =
\lambda_-\lambda_+ \big[(v_-\cc{v_+}\wedge v_+\cc{v_-})\delta
+(v_-\cc{v_+}\wedge v_-\cc{v_-})\delta_+
-(v_+\cc{v_+}\wedge v_+\cc{v_-})\delta_-\big]
\end{align*}
which is as explicit as it gets.

{\bf The term $\db{x_-}{\CONJ x_+}$.}
This term vanishes,
 by a calculation analogous to the one we just did.
A more systematic way to see this is to introduce gradings
compatible with the bracket; see \sref{decddd}.

{\bf Combining.}
Using $\db{\CONJ\,\cdot\,}{\CONJ\,\cdot\,}
= \CONJ\db{\,\cdot\,}{\,\cdot\,}$ as well as $\CONJ^2 = \mathbbm{1}$ we
see that
$\db{\CONJ x_-}{\CONJ x_+} = \CONJ \db{x_-}{x_+}$
and $\db{\CONJ x_-}{x_+} = \CONJ \db{x_-}{\CONJ x_+} = 0$, hence
\[
\db{c_-}{c_+}
 \;=\; \db{x_-}{x_+} + \CONJ\db{x_-}{x_+}
\]
The first term annihilates $\rC \oplus \cc{V}$,
the second annihilates $\rC \oplus V$.
One can think of it like this:
$\db{c_-}{c_+}|_V = \db{x_-}{x_+}|_V$,
an then $\db{c_-}{c_+}|_{\cc{V}}$ is determined by reality.

{\bf Final observation.}
Set $x' = \lambda_-\lambda_+ (v_-\cc{v_-}\wedge v_+\cc{v_+})\delta
\in \DC{2}$ and add-subtract:
\[
\db{x_-}{x_+}
= x' + (-x' + \db{x_-}{x_+})
\]
One verifies that the second term
is in $\IWeyl{2} \oplus i\IWeyl{2}$, hence
\[
\db{c_-}{c_+} \;\in\; (\mathbbm{1}+\CONJ)x' + \IWeyl{2}
\]


\section*{Acknowledgments}
M.R.~thanks Jim Stasheff for
an interesting and helpful conversation about deformation theory.
M.R.~enjoyed support from the (US) National Science Foundation.

\mostimportant{This material is based upon work supported by the National Science
Foundation under agreement No.~DMS-1128155. Any opinions,
findings and conclusions or recommendations expressed
in this material are those of the author(s) and
do not necessarily reflect the views of the National Science Foundation.}

\end{document}